{
\documentclass[useAMS,usenatbib]{mn2e}

\usepackage{times,graphicx,epsfig}
\newcommand{\bq}{\begin{equation}}
\newcommand{\eq}{\end{equation}}
\newcommand{\bqn}{\begin{eqnarray}}
\newcommand{\eqn}{\end{eqnarray}}
\newcommand{\dd}{\mbox{\rm d}}

\newcommand{\Ei}{\mbox{\rm Ei}}
\newcommand{\apj}{ApJ}
\newcommand{\apjl}{ApJL}
\newcommand{\mnras}{MNRAS}
\newcommand{\aap}{A\&A}
\newcommand{\aj}{AJ}
\newcommand{\pasj}{PASJ}
\newcommand{\araa}{ARAA}
\def\onema{1\!-\!\eta}
\def\twoma{2\!-\!\eta}
\def\threema{3\!-\!\eta}

\def\onepa{1\!+\!\eta}

\def\amone{\eta\!-\!1}
\def\amthree{\eta\!-\!3}
\def\pc{{\rm pc}}
\def\kpc{{\rm kpc}}
\def\msol{{\rm M}_\odot}
\def\kms{{\rm km/s}}
\def\Myr{{\rm Myr}}

\title[Dynamical friction of massive objects]
{Dynamical friction of massive objects in galactic centres}
\author[A. Just et al.]
{A. Just$^{1}$\thanks{E-mail: just@ari.uni-heidelberg.de},
F. M. Khan$^{1,2}$, P. Berczik$^{3,4,1}$, A. Ernst$^{1}$,
and R. Spurzem$^{3,5,1}$\\
$^{1}$Astronomisches Rechen-Institut, Zentrum f\"ur Astronomie der Universit\"at
Heidelberg (ZAH), M\"onchhofstr. 12-14,
D-69120 Heidelberg, Germany\\
$^{2}$Department of Physics, Government College University (GCU), 54000  
Lahore, Pakistan\\
$^{3}$National Astronomical Observatories of China (NAOC), Chinese
Academy of Sciences (CAS), Datun Lu 20A, Chaoyang District,
Beijing 100012, China \\
$^{4}$Main Astronomical Observatory (MAO), National Academy of
Sciences of Ukraine (NASU), Akademika Zabolotnoho 27, 03680 Kyiv,
Ukraine \\
$^{5}$Kavli Institute for Astronomy and Astrophysics, Peking University, China
}
\begin{document}

\date{\today}

\pagerange{\pageref{firstpage}--\pageref{lastpage}} \pubyear{2004}

\maketitle

\label{firstpage}

\begin{abstract}
Dynamical friction leads to an orbital decay of massive objects like young
compact star clusters or Massive Black Holes in central regions of
galaxies.  The dynamical friction
force can be well approximated by Chandrasekhar's standard formula, but recent
investigations show, that corrections to the Coulomb logarithm
are necessary.
With a large set of N-body simulations we show that the improved formula for the Coulomb logarithm fits the orbital decay very well for circular and eccentric orbits. 
The local scale-length of the background density distribution serves as the maximum impact parameter for a wide range of power-law indices of -1\dots -5. 
For each type of code the numerical resolution must be compared to  the effective minimum impact parameter in order to determine the Coulomb logarithm.  We also quantify the correction factors by using self-consistent velocity distribution functions instead of the standard Maxwellian often used. These factors enter directly the decay timescale and cover a range of 0.5\dots 3 for typical orbits.
The new Coulomb
logarithm combined with self-consistent velocity distribution functions in the Chandrasekhar formula provides a significant improvement of orbital decay times with correction up to one order of magnitude compared to the standard case. We suggest the general use of the improved formula in parameter studies as well as in special applications.
\end{abstract}

\begin{keywords}
Stellar dynamics -- black hole physics -- Galaxies: kinematics and dynamics -- Galaxy: centre.
\end{keywords}

\section{Introduction}\label{sec-intro}

Dynamical Friction is a subject with two faces. Many researchers think it
is sufficiently well understood and studied since the classical work of
\citet{cha42} and the many follow-ups. There is a duality
between a collective gas-dynamic approach, studying responses in
a continuum through which a test body moves, and a kinetic or
particles approach where we consider a test particle moving through
a sea of light particles. As is known since long \citep{bon44,rep80}
under certain
limits both approaches can yield similar results. In kinetic approaches
(rooted in Chandrasekhar's work) usually an infinite homogeneous system
of field stars is assumed, and the singularity at large impact parameters
is cut off by the use of a so-called Coulomb logarithm $\ln\Lambda$
(see more detail below). The underlying approximations are that interactions
extend to impact parameters large compared to 
$r_{\rm g}=GM_{\rm bh}/3\sigma^2$ ($M_{\rm bh}$ the test mass
particles, $\sigma$ the 1-dimensional field particle velocity dispersion)
and that the
relative velocities involved are not too small. The error made by the
first assumption is usually absorbed by fitting a certain numerical
value of the Coulomb logarithm to results of numerical simulations; such
a method has been very successful in plasma physics \citep{ros57},
star cluster dynamics \citep{spi87,gie94},
galactic dynamics \citep{bin87,wei89,her89,pru92,cor97} and in \citet{spi03}, where all
citations should be seen as exemplary rather than exhaustive.

There are two developments which have caused renewed interest in more
accurate theoretical determinations of dynamical friction. One is the
realisation (both from theoretical structure formation models and
HST observations of galaxy cores) that most cores of galaxies in the standard
 picture of
hierarchical structure formation are embedded in cuspy distributions
of dark matter \citep{nav97,lau95}.
It means that in many galaxies the density profile of the dark matter,
which is the main background for dynamical friction of dwarf galaxies,
star clusters and compact objects is nowhere constant as assumed in
the standard Chandrasekhar theory. This problem has led to the
suggestion of an empirical variation of the Coulomb logarithm
with radius, so as to account for the different efficiency of
dynamical friction \citep{tre76,has03}.
In \citet{jus05} a detailed theoretical investigation of the parameter dependence in Chandrasekhar's dynamical friction formula was given. An improved analytic approximation for the Coulomb logarithm was presented. 

Recently some doubt has been cast on the validity of Chandrasekhar's dynamical friction
formula at all for certain density profiles, such as harmonic potential, constant density
cores. \citet{rea06} claim that while in other density profiles it may work well
(even in flattened systems, such as \citet{pen04} have shown, if
corrections due to the velocity distribution are taken into account), harmonic cores
are super resonant structures where resonances suppress dynamical friction.  
\citet{boy08} compare estimates of merger time scales of galaxies within
dark matter halos obtained from Chandrasekhar's dynamical friction formula with results of
high resolution $N$-body simulations. They find that Chandrasekhar's friction formula
does not work well, though they attribute it to finite size and mass loss effects of the
merging objects rather than resonant processes. Also \citet{ino09} supported the result
that dynamical friction ceases in harmonic cores, but doubt about resonant effects
as the reason without giving further explanation. 

The idea that dynamical friction tends to stop in harmonic cores is a phenomenon not so 
uncommon in other fields of stellar dynamics. 
For example in simulations of 
super-massive black holes in galactic nuclei \citep{dor03,gua08},
dynamical friction brings the black hole into the core, where it then
starts a wandering process attributed to random encounters / Brownian motion. 
Also in star cluster dynamics dynamical friction of heavy mass stars
sinking to the centre will stall if random encounters support enough pressure in the core. We think
it is unclear whether it is a resonant or a random process which stops dynamical friction
in cores, but this problem deserves more detailed analytical and numerical attention.
In numerical simulations it was shown \citep{nak99a,nak99b} that in elliptical galaxies the settling of a SMBH to the centre can form a shallow cusp in the stellar distribution of the core. In these simulations the influence radius of the SMBH was not resolved and there are also very few observations resolving the influence radius. The investigation of \citet{mer06} on the orbital decay of a massive secondary BH points in the same direction creating a shallow inner cusp.

In this paper, we assume Chandrasekhar's dynamical friction ansatz as a working hypothesis to
be valid at least in principle, but after studying the velocity distribution dependence in
earlier papers \citep{jus05,pen04} we focus here on
the correction due to strong local density gradients, i.e. the opposite case to the disputed
harmonic core situation. 
The detailed investigation of shallow cusps is postponed to a forthcoming paper. The generalised formula for dynamical friction is valid for extended objects like satellite galaxies or star cluster and for point-like objects like SMBHs. For extended objects the Coulomb logarithm is small and corrections to the relevant impact parameter regime are more significant than for SMBHs. Additionally the mass loss and the determination of the effective mass for dynamical friction must be taken into account \citep{fuj06,fuj08}. The present investigation is restricted to the orbital evolution of SMBHs.

Super-massive black holes, most likely
to be present in merging galaxies from the early universe onwards
\citep{kor95,hae93,fer01}, will sink to the centres of galactic merger remnants
by dynamical friction and ultimately
coalesce themselves \citep{fuk92,mak96,ber09}.
Numerical simulations to follow this process in a particle-by-particle
approach \citep{ber05,ber06,mak04,hem02,mil01}
are still too computationally expensive for realistic
particle numbers, and so this situation requires another careful
look at dynamical friction. Here not only the question of the
influence of the ambient density gradient is important, but
the test particle (super-massive black hole, single or binary)
will typically violate also the other condition mentioned above,
since it has small velocity (tendency towards equipartition) at
least in the final phase of its approach to the centre. In
\citet{mer01} we find a comprehensive overview of dynamical
friction for such objects in the limit of small velocities.
An overview of how density gradients affect the sinking
time-scales of massive black holes in galactic nuclei due to
dynamical friction has apparently never been carried out.
This is the aim of the present paper.

Dynamical friction time-scales are also important for another
aspect of massive black hole (binary) dynamics, namely the
eccentricity of the binary. Dynamical friction in homogeneous
media tends to circularise initially eccentric binary orbits, since it is most
efficient at low velocity in apo-centre. Density gradients, if treated
with standard (constant) Coulomb logarithm should influence this
effect. \citet{tsu00} provide a nice overview
of how dynamical friction (including effects of density
gradients and anisotropic velocity distribution) affects
orbital shape (eccentricity), but they use the local epicyclic
approximation. \citet{has03} have shown that the circularisation problem can be solved by adopting a position dependent Coulomb logarithm. Recent models of three black holes in
galactic nuclei show that dynamical friction together with three-body
dynamics (e.g. Kozai effect) can induce extremely high eccentricities of the
innermost black hole binary \citep{iwa06,ama10}.
Also, the cosmological growth of massive central black holes from
minor and major merging depends sensitively on dynamical friction
of satellite galaxies and massive black holes in a background of
stars and dark matter \citep{vol03,vol07,dot10}. 

Direct numerical simulations resolving the full range of impact parameters and relative velocities for taking dynamical friction correctly into account are still extremely tedious. Therefore it is very important to improve our theoretical ansatz for dynamical friction in non-standard cases.
In \citet{jus05} an improved analytic approximation for the Coulomb logarithm was presented. In this article we concentrate on dynamical friction of compact objects in stellar cusps covering a wide range of parameters for circular and eccentric orbits. We present a comprehensive numerical analysis to test the general applicability of the Chandrasekhar formula with the new Coulomb logarithm. We take also into account the self-consistent velocity distribution functions entering the friction force.

In $\S$\ref{sec-fric} the cusp models, the cumulative distribution functions and the Coulomb logarithm are discussed. In $\S$\ref{sec-orb} the theory of the orbital decay is presented. In $\S$\ref{sec-code} the $N$-body and semi-analytic codes, which we are using to evolve our 
models, are described. 
 In $\S$\ref{sec-model} a comparison of the numerical results with semi-analytic calculations is given. In $\S$\ref{sec-appl} a brief discussion of some applications is presented 
and finally $\S$\ref{sec-concl} includes concluding remarks.  

\section{Dynamical friction force}\label{sec-fric}

Throughout the paper we use specific forces, i.e. accelerations.
The dynamical friction force of a massive object with mass $M_{\rm bh}$ and
orbital velocity $V_{\rm bh}$ in a sea of lighter particles
is usually determined by adopting locally a homogeneous background density 
$\rho$
and an isotropic velocity distribution function. The result is a drag force
anti-parallel to the motion, given by the standard formula of Chandrasekhar 
\citep{bin87}
\bqn
\dot{V}_{\rm df} &=& \frac{-4\pi G^2\rho M_{\rm bh}}{V_{\rm bh}^2}\chi 
	\ln\Lambda
        \quad \mathrm{with}\quad \chi=\frac{\rho(<V_{\rm bh})}{\rho}
        \label{dynfric}
\eqn
In general the functions $\chi$ and $\Lambda$ depend on the velocity of the massive object and on the properties of the background system.

\subsection{Cusp models}\label{sec-cusp}

In order to quantify the position dependence of the Coulomb logarithm we investigate the orbital decay of a massive object in stellar cusps. We investigate two general scenarios. In the self-gravitating case the gravitational force is dominated by the power-law cusp. Bulges and the cores of galaxies and of star clusters can be described in that way as long as the massive object is outside the gravitational influence radius of a central black hole (or additional mass concentration). The circular speed is coupled to the cusp mass distribution by the Poisson equation.

In the Kepler case the gravitational force is dominated by a centrally concentrated mass $M_{\rm c}$, which can be a central SMBH of the core of a self-gravitating stellar distribution. 
If the galactic nucleus already harbours a SMBH with mass
$M_{\rm c}$ at the centre, the inner part of the cusp inside the influence radius is dominated by the Kepler
potential of the SMBH. In this case the stellar distribution can be described by a Bahcall-Wolf cusp. The orbital evolution of a second BH entering this region
of gravitational influence is quite different to the self-gravitating case. This
will happen, if a small galaxy merges with a larger one, both harbouring a
central BH. In the minor merger process the bulges of the galaxies will relax
first to the new bulge and the massive SMBH will settle to the centre. Finally
the second BH decays to the centre by dynamical friction. 

Also in the outskirts of self-gravitating systems the density distribution may be approximated by a power law and the potential by a point-mass potential, if it is dominated by the mass concentrated at the centre. We investigate two cases with steep power law distributions to test the maximum impact parameter dependence of the Coulomb logarithm. The Plummer sphere with an outer density slope of $-5$ and the Dehnen models with a slope of $-4$ are the most extreme cases.

\subsubsection{Power law profiles}\label{sec-power}

For analytic estimations we use idealised power law distributions for the background particles.
For both cases (self-gravitating and Kepler potential) the density profile of the cusp and the cumulative mass profile can be approximated by the functions
\bqn
M(y)&=&M_{\rm c}+ M_{\rm t} y^{\eta}\quad\mbox{with}\quad y=\frac{R}{R_0} \label{massy}\\
        \rho(y)&=&\rho_0 y^{\amthree}\quad\mbox{with}\quad 
\rho_{0} = \frac{\eta M_{\rm t}}{4\pi R_{0}^{3}} .\label{deny}
\eqn
It is comfortable to normalise all quantities to the initial values of the orbit. The initial position of the object is $R_0=R(t=0)$ leading to $y_0=1$, the enclosed mass is $M_0=M(y_0)$, and the circular velocity at $R_0$ is $V_{\rm c,0}=V_{\rm c}(R_0)=\sqrt{GM_0/R_0}$. For the self-gravitating case we simply set $M_{\rm c}=0$ leading to $M_0=M_{\rm t}$ via Eq. \ref{deny}. In the Kepler case we must distinguish between $M(y)\approx M_{\rm c}=const.$ for the gravitational forces leading to the circular velocity with $V_{\rm c}^2=GM_{\rm c}/R_0\,y^{-1}$ and the local density determined by $M_{\rm t}$. The case $\eta=5/4$ corresponds to the Bahcall-Wolf (BW) cusp of stationary equilibrium with constant radial mass and energy flow \citep{bah76,lig76}.

We normalise all velocities to the local circular velocity $V_{\rm c}$ instead of the velocity dispersion $\sigma$ by
\bq
u=v/V_{\rm c}\quad\mbox{and}\quad U=V_{\rm bh}/V_{\rm c} \label{eq-u}
\eq 
 The velocity dispersion in self-gravitating cusps behaves quite different for different $\eta$ \citep{tre94}. For $\eta>2$ the kinetic pressure $\rho\sigma^2$ converges to a finite value at the centre, which depends on the outer boundary conditions. The transition case with $\eta= 2$ is of special interest, because it corresponds to the $\rho\propto y^{-1}$ cusp as in the standard NFW cusp and in the Hernquist model. In this case the kinetic pressure $\rho\sigma^2\propto (C-\ln y)$, where the constant C is also determined by the outer boundary conditions. In a Kepler potential the isotropic distribution function degenerates for $\eta=5/2$, because it is completely dominated by particles with low binding energy. Dependent on the outer boundary conditions the distribution function can vary between a $\delta$-function at the escape velocity (i.e. at $E=0$) and a power law $\propto |E|^{-1}$ with some cutoff at $E\approx 0$. For a secondary black hole on a bound orbit in an idealised isotropic cusp the $\chi$-value in Eq. \ref{dynfric} becomes very small for $\eta>2$ and tends to zero for $\eta=5/2$ leading to an unrealistically small dynamical friction force. Any change in the outer boundary conditions, small perturbations to the isotropy, or the effect of higher order terms in the dynamical friction force become important.
 
$U$ can be converted to the standard $X$ variable with the normalised circular velocity $X_{\rm c}$ by
\bq
X=X_{\rm c}U \quad\mbox{with}\quad
X_{\rm c}^2\equiv \frac{V_{\rm c}^2}{2\sigma^2} \label{Xc}
\eq
Inserting Eq. \ref{sig2} into Eq.(\ref{Xc}) we find
\bq
X_{\rm c}^2 = \left\{\begin{array}{lcl}
	 2-\frac{\eta}{2} & \eta< 2.5 & \mbox{Kepler}\\
	 2-\eta & \eta< 2 & \mbox{self-grav.} \\
	 \left[2(C-\ln y)\right]^{-1} & \eta =2 & \mbox{self-grav.}\\
	 C'\,y^{(2\eta-4)} & 2< \eta \le 3 & \mbox{self-grav.}
	 \end{array}\right.
	 \label{Xc2}
\eq
which is independent of position $y$ for the Kepler potential and the self-gravitating cusp with $\eta<2$. 
For a shallow self-gravitating cusp with $\eta>2$ the integration constant $C'$ depends on the outer boundary conditions of the realisation of the cusp. The details of the distribution functions are described in App. \ref{appfe}.

\subsubsection{Physical models}\label{sec-phys}

In many simulations of stellar cusps it turned out that the setup of an initial cusp distribution in dynamical equilibrium with an unphysical outer cutoff is not stationary. The density profiles evolves deep into the inner cusp region. Therefore it is necessary to set up initial particle distributions and velocities with a well-defined outer cutoff of the cusp distribution.

For the self-gravitating cusps we use Dehnen models \citep{deh93} with an outer power law slope of $-4$ for the density. These models are identical to the $\eta$-models of \citet{tre94}. Density and cumulative mass are given by
\bqn
M(y)&=&M_{\rm t} \left(\frac{y}{y_{\rm a}+y}\right)^{\eta}  
\hspace*{.3 cm} y_{\rm a}=\frac{a}{R_0} \label{massdehn}\\
        \rho(y)&=&\frac{\rho_0}{y^{3-\eta}(y_{\rm a}+y)^{\onepa}} ,\quad
	\rho_{0} = \frac{\eta M_{\rm t}y_{\rm a}}{4\pi R_{0}^{3}}\nonumber
\eqn
with Eq. \ref{deny} for the conversion for $\rho_0$.
The Jaffe and Hernquist models correspond to $\eta=1$ and $\eta=2$, respectively. Well inside the scale radius $a$ the particles behave asymptotically like in the idealised power law distributions.

For the Kepler potential case we investigate two different scenarios. In the first case we investigate power law cusps in the vicinity of a central SMBH with mass $M_{\rm c}$, i.e. the BW cusp and a shallower Hernquist (He) cusp. There are no exact distribution functions known describing the Kepler potential part inside the influence radius of the SMBH and the transition to a self-gravitating outer regime. 
\citet{tre94} generalised their $\eta$ models by including the gravitational potential of  a central SMBH and derived the power law distribution function   (Eqs. \ref{fE} and \ref{constKep1}) well inside the influence radius, which is comparable to the scale radius $a$. In \citet{mm07} this approximation was adapted to a Plummer model instead of a $\eta$-model. This model has two advantages. Firstly the steeper slope in the outer part saves particles and computation time for simulations of BW cusps. Secondly the power law distribution function of the Plummer sphere is exact also in the inner part. Therefore realisations with smaller SMBH masses relative to the cusp mass are closer to equilibrium.
The cumulative mass and density distribution are given by
\bqn
M(y) &=& M_{t}\left(\frac{y^2}{y_{\rm a}^2+y^{2}}\right)^{\eta/2}, \hspace*{.3 cm} 0 < \eta \leq 3\label{cumm} \\
\rho(y) &=& \frac{\rho_0}{y^{3-\eta}(y_{\rm a}^2+y^2)^{\eta/2 +1}}, \label{density}
\eqn
where $M_{t}$ is the total mass of stars in the cusp. The approximate DF is
\bq
f(E) = f_{0}E^{7/2}(E_{0}^{s} + E^{s})^{-(\eta + 2)/s},\label{edis}
\eq
where
\bqn
E_{0} &=& \left(\frac{f_{1}}{f_{0}}\right)^{-1/(\eta+2)}\\
f_{0}  &=& \frac{\eta M_{t} \Gamma(4-\eta)}{2^{7/2}\pi^{5/2}M_{c}^{3-\eta}\Gamma(5/2-\eta)}\\
f_{1}  &=& \frac{8\sqrt{2}\eta M_{t}}{7\pi^{3}}.
\eqn
Here $E$ is the binding energy and $s = 5$. We used these equations with $\eta=5/4$ for the BW cusp. The constant $f_{0}$ represents the BW cusp, $f_{1}$ the Plummer model and $E_{0}$ is the transition threshold in energy.
The He cusp is realised in a similar way.

The outskirts of Dehnen models (DE) and the Plummer model (PL) can also be approximated by cusps in a Kepler potential using asymptotic expansions in $y$. From an identification of the density slopes for Dehnen and Plummer of -4 and -5, respectively, with $\eta_0-3$ in Eq. \ref{deny}, we find $\eta_0=-1$ and $\eta_0=-2$ for the outskirts 
(here we use the index 0 in order to distinguish it from the parameter $\eta$ in the core).
This leads with Eq. \ref{constKep1} to the correct distribution functions in Eq. \ref{fE} for the Dehnen models and the Plummer sphere in the limit of small energies. 
If we now identify $M_{\rm t}$ in Eq. \ref{massy} with the mass deficiency compared to the total mass $M_{\rm c}$, then Eqs. \ref{massy} and \ref{deny} also hold for these cases. We find for the Dehnen models
\bqn
M(y)&\approx&M_{\rm c} \left(1-\eta y_{\rm a}y^{-1}\right) =
	M_{\rm c} + M_{\rm t}y^{\eta_0}\label{mdehnapp}\\
        \rho(y)&\approx&\rho_0 y^{-4}=\rho_0 y^{\eta_0-3} \nonumber \\
	 \eta_0&=&-1,\qquad M_{\rm t}=-\eta M_{\rm c}y_{\rm a} \nonumber \\
	 \rho_{0}& =& \frac{\eta M_{\rm c}y_{\rm a}}{4\pi R_{0}^{3}} = 
	\frac{\eta_0 M_{\rm t}}{4\pi R_{0}^{3}} \nonumber
\eqn
and similarly for the Plummer sphere
\bqn
M(y)&=&M_{\rm c} \left(\frac{y^2}{y_{\rm a}^2+y^2}\right)^{3/2}
	\approx M_{\rm c} \left(1-\frac{3}{2}y_{\rm a}^2 y^{-2}\right)
	 \label{mplumapp}\\
	&=& M_{\rm c} + M_{\rm t}y^{\eta_0} \nonumber\\
        \rho(y)&=&\rho_0\left(y_{\rm a}^2+y^2\right)^{-5/2}
	\approx\rho_0 y^{-5}=\rho_0 y^{\eta_0-3}  \nonumber \\
	 \eta_0&=&-2,\qquad M_{\rm t}=-\frac{3}{2} M_{\rm c}y_{\rm a}^2 \nonumber \\
	 \rho_{0} &=& \frac{3 M_{\rm c}y_{\rm a}^2}{4\pi R_{0}^{3}} = 
	\frac{\eta_0 M_{\rm t}}{4\pi R_{0}^{3}}\nonumber
\eqn
completely consistent with the power law cusp description.

\subsection{Cumulative distribution functions}\label{sec-chi}

The cumulative distribution function $\chi(U)$ of the normalised 1-dimensional distribution function $F(u)$ measures the fraction of background particles with velocity smaller than $U=V_{\rm bh}/V_{\rm c}$. We are using the self-consistent $\chi$ functions which are derived in appendix \ref{appfe} for 
the different models. They are significantly different to $\chi_{\rm s}(U)$ 
of the standard Maxwellian which is usually adopted for Chandrasekhar's 
formula. In most applications of the standard formula the local velocity dispersion is not known. Instead $X_{\rm c}=1$ as in the singular isothermal sphere is adopted leading to the identification of $X=U$.
\begin{figure}
\centerline{
  \resizebox{0.98\hsize}{!}{\includegraphics[angle=270]{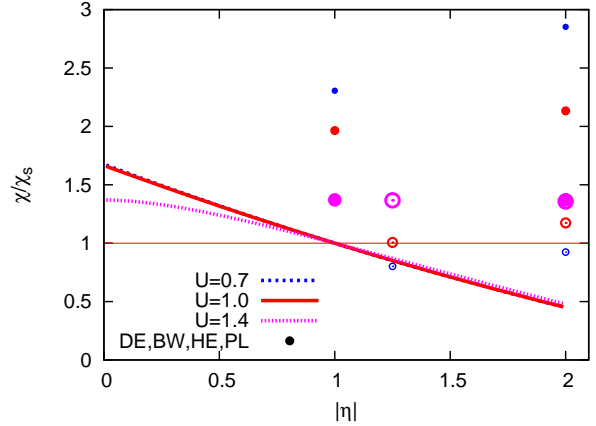}}
  }
\caption[]{
The plot shows the relative variation $\chi(U)/\chi_{\rm s}(U)$ as a function of $\eta$ for
for different values of $U$, where $\chi_{\rm s}$ corresponds to the standard Maxwellian distribution
function (i.e. $\eta=1$). For self-gravitating cusps the full line is
for the circular velocity $U=1$, dotted and dot-dashed lines are for $U=0.7$
and $U=1.4$, respectively.  The circles give the corresponding values for the Kepler potential cases with increasing size for increasing $U$ (open symbols are for positive $\eta$ and full symbols for negative $\eta$).
}\label{figchitot}
\end{figure}
In Fig. \ref{figchitot} 
the correction factors $\chi(U)/\chi_{\rm s}(U)$ entering the friction force formula Eq. \ref{dynfric} are shown. 
The different lines give the
results for the self-gravitating cusps as a function of $\eta$. 
We show the values for the circular velocity $U=1$
(full line) and for $U=0.7, 1.4$ (dotted and dot-dashed
line), typical values for apo- and peri-centre velocities, respectively. In shallow self-gravitating cusps (with
$\eta\ge 1$) the efficiency of dynamical friction is reduced roughly by a
factor of $\eta$ due to the larger fraction of high
velocity particles. In steep cusps dynamical friction is larger compared to the isothermal case, but with systematic deviations from the simple scaling for higher
velocities $U$ during peri-centre passage due to the finite escape velocity. 
The circles in Fig. \ref{figchitot} show $\chi(U)/\chi_{\rm s}(U)$ for the BW and HE cusp (open symbols) and the outskirts of the Plummer (PL) and Dehnen (DE) spheres (full symbols) at the corresponding values for $\eta=5/4,\,2,\,-2,\,-1$, respectively. For the HE case we used the numerically realised values (see also figure \ref{figchiE}.

For circular orbits the orbital decay time varies up to factor larger than two compared to the standard formula due to the self-consistent $\chi$ functions.
For the
evolution of the orbital shape, the relative variation of the friction force
between apo- and peri-centre is also important (see Sect. \ref{sub-ecc}).

\subsection{Coulomb logarithm}\label{sec-coul}

The main uncertainties in the magnitude and parameter dependence of the dynamical friction force is hidden in the Coulomb logarithm $\ln\Lambda$, which gives the effective range of relevant impact parameters and is up to now a weakly determined quantity. Since this formula is applied to wide ranges of parameters, it is very useful to have the explicit parameter dependence of $\ln\Lambda$ instead of fitting a constant value for each single orbit. In \citet{jus05} the effect of the inhomogeneity of the background distribution on the dynamical friction force with Chandrasekhar's approach was discussed. 
The authors derived the approximation
\bq
\ln\Lambda = \ln\frac{b_{\rm max}}{\sqrt{b_{\rm min}^2+a_{\rm 90}^2}}
 \approx \left\{\begin{array}{ll}
	\ln\left(\frac{D_{\rm r}}{b_{\rm min}}\right)
	&\mbox{unres. or ext.}\\
	\\
	\ln\left(\frac{D_{\rm r}}{a_{\rm 90}}\right)
	&\mbox{point-like.}
	\end{array}\right. \label{loglam}
\eq
The Coulomb logarithm depends on the maximum and minimum impact parameter $b_{\rm max}$ and $b_{\rm min}$, resp., and for point-like objects on $a_{\rm 90}$, the typical impact parameter for a $90\degr$-deflection in the 2-body encounters. \citet{jus05} found that the maximum impact parameter is given by the local scale-length $D_{\rm r}$ determined by the density gradient, i.e.
\bq
b_{\rm max} = D_{\rm r} \equiv \frac{\rho}{\left|\nabla\rho\right|}=\frac{R}{\threema}
	\quad \eta\le 2. \label{bmax}
\eq
In an isothermal sphere $b_{\rm max}$ is a factor of 2 smaller than the distance $R$ to the centre. In shallow cusps with $\eta>2$ the local scale length $D_{\rm r}$ exceeds the distance to the centre. 
In that case, the local scale-length may be substituted by $R$ (but  see also \citet{rea06} for additional suppression of dynamical friction in homogeneous cores).

For point-like objects like BHs the effective minimum impact parameter $a_{\rm 90}$ is given by the value for a
$90\degr$-deflection using a typical velocity $v_{\rm typ}$ for the 2-body encounters
\bq
a_{\rm 90} = \frac{GM_{\rm bh}}{v_{\rm typ}^2} 
\approx \frac{GM_{\rm bh}}{2\sigma^2+V_{\rm bh}^2}
= \frac{X_{\rm c}^2}{1+X^2}\frac{M_{\rm bh}}{M(y)}R
= \frac{3\,r_{\rm g}}{2(1+X^2)} \,. \label{a90}
\eq
If the motion of a point-mass is numerically derived by a code, where $a_{\rm 90}$ is not resolved, the minimum impact parameter is determined by the effective spatial resolution of the code. In our simulations we use the direct particle-particle code (PP) $\phi$GRAPE with softening length $\epsilon$ and the Particle-mesh code (PM) {\sc Superbox} with grid cell size $d_{\rm c}$.
We use
\bq
b_{\rm min}=\left\{\begin{array}{ll}
	1.5\,\epsilon
	&\mbox{(PP code)}\\
	\\
	d_{\rm c}/2
	&\mbox{(PM code)}
	\end{array}\right. .\label{bmin}
\eq
These values are a property of the numerical code and do not depend on the application. They are determined by numerical experiments (see also below).
For extended objects like star clusters a good measure of the minimum impact parameter $b_{\rm min}$ is the half-mass radius $r_{\rm h}$. 

The parameter dependence of $b_{\rm max}$ and $a_{\rm 90}$ leads to a position dependence of $\ln\Lambda$, which affects the decay time $\tau_{\rm dec}$ (see Eq. \ref{taudec}) and which also reduces the circularisation of the orbits resolving a longstanding discrepancy between numerical and analytical results. Using the distance to the centre as maximum impact parameter was proposed by different authors \citep{tre76,has03}, but the effect on orbital evolution was never investigated in greater detail or for larger parameter sets.

On circular orbits the local scale-length $D_{\rm r}$ and the deflection parameter $a_{\rm 90}$ are position dependent. On eccentric orbits $a_{\rm 90}$ depends additionally on the velocity. Therefore $\ln\Lambda$ varies systematically during orbital decay and for eccentric orbits along each revolution. This has consequences on the decay time and on the evolution of orbital shape. In eccentric orbits the dynamical friction force varies strongly between apo-and peri-galacticon mainly due to the density variation along the orbit. The variation due to higher peri-centre velocity and to the position dependent Coulomb logarithm weakens the differences.  All these factors depend on the slope of the cusp density. That means that the effective Coulomb logarithm averaged over an orbit depends differently on the eccentricity for different values of $\eta$.

In the case of circular orbits with constant $X_{\rm c}$ the position dependence of the Coulomb logarithm (eq. \ref{loglam}) can be parametrised by
\bq
\ln\Lambda=\ln(\Lambda_0 y^\beta) \,.\label{pdcl}
\eq

Deep in shallow self-gravitating cusps with $\eta> 2$ the contribution from the circular velocity vanishes and $\Lambda$ is also described by eq. \ref{pdcl}.
With Eq. \ref{a90} the Coulomb logarithm is
\bqn
\beta=1&\Lambda_0=\frac{1}{(\threema)}\frac{R_0}{b_{\rm min}} 
&\mbox{extended or unresolved} \label{lam0ex}	 
 \eqn
for extended objects and for point-like objects we find
\bqn
\begin{array}{lll}
\beta=0&\Lambda_0=\frac{(6-\eta)}{(\threema)(4-\eta)}
\frac{M_c}{M_{\rm bh}} &\mbox{Kepler} \\
\beta=\eta&\Lambda_0=\frac{1}{(\twoma)}\frac{M_0}{M_{\rm bh}}
  &\mbox{self-grav.},\,\eta<2 \\
\beta=4-\eta&\Lambda_0=\frac{1}{C'}\frac{M_0}{M_{\rm bh}}
  &\mbox{self-grav.},\,2<\eta<3 
  \end{array}	  \label{lam0} 
 \eqn
We see that the motion of a point-like object in a Kepler potential is also described by a constant Coulomb logarithm, because the linear dependence of $D_{\rm r}$ and $a_{\rm 90}$ cancel.
The standard case corresponds to $\beta=0$ with $R_0$ instead of $D_{\rm r}$ and $r_\mathrm{g}$ instead of $a_{90}$ in equation \ref{loglam} leading to 
\bq
\ln\Lambda_{\rm s} = \left\{\begin{array}{ll}
	\ln\left(\frac{R_0}{b_{\rm min}}\right)
	&\mbox{unres. or ext.}\\
	\\
	\ln\left(\frac{3}{2X_\mathrm{c}^2}\frac{M_0}{M_{\rm bh}}\right)
	&\mbox{point-like.}
	\end{array}\right. \label{loglams}
\eq

In a recent investigation \citet{spi03} determined in a series of N-body calculations quantitatively the value of the Coulomb logarithm. They calculated the orbit of a massive point-like object moving through a background of stars with a power law cusp close to a singular isothermal sphere using different numerical codes (including {\sc Superbox}) and different parameters. They adopted a constant Coulomb logarithm and found that $b_{\rm max}$ is systematically about a factor of two smaller than the initial distance $R_0$ to the centre. 
Additionally they restarted the orbital evolution at smaller initial distances and find a smaller best-fitting $\ln\Lambda$. For $a_{\rm 90}$ they also found a value significantly smaller than the kinematically defined gravitational influence radius $r_{\rm g}$ of the moving BH. 
An inspection of the results of \citet{spi03} shows, in spite of their slightly different interpretation, that they are fully consistent with equations \ref{loglam} -- \ref{lam0}: $\beta\approx 1$ leading to a linear decrease of $\Lambda$ for the resolved and the unresolved cases; $b_{\rm max}=D_{\rm r}\approx R_0/2$; $b_{\rm min}=d_{\rm c}/2$ for the PM code {\sc Superbox}; $a_{\rm 90}=0.75\,r_{\rm g}$.

\section[]{Orbital decay in galactic centres}\label{sec-orb}

We consider the
orbital decay of a massive object in central stellar cusps in detail. 
We investigate the effect of the varying Coulomb logarithm $\ln\Lambda$ and of
a self-consistent distribution function $f(E)$ instead of
the standard Maxwellian on the
decay rate.
The dynamical friction force (Eqs. \ref{dynfric}) depends on the background distribution via the local density of
 slow particles 
$\rho(<V_{\rm bh})=\rho\chi$, the density gradient, i.e. $D_{\rm r}$, and the circular velocity $V_{\rm c}$.

Here we give the explicit solution for a massive object moving on a circular
orbit in a power law density profile. The background
distribution is a self-gravitating stellar cusp with the corresponding self-consistent
distribution function $f(E)$ or a stellar cusp in a Kepler potential. We use $\chi_0=\chi(U=1)$ for the circular speed. The initial decay timescale for angular momentum loss is given by
\bqn
\tau_0 &=& -\frac{V_{\rm c0}}{\dot{V}_{\rm df,0}}
        = \frac{1}{2\pi \eta\chi_0\ln\Lambda_0}
        \frac{M_0^2}{M_{\rm bh}M_{\rm t}}\;T_0 \label{tau0}\\
        &=&\frac{14.9}{\eta\chi_0\ln\Lambda_0} 
        \left[\frac{R_0}{\pc}\frac{M_0}{\msol}\right]^{3/2}
        \left[\frac{M_{\rm t}}{\msol}
	\frac{M_{\rm bh}}{\msol}\right]^{-1}\mathrm{Myr} .\nonumber
\eqn
The first expression is in units of the orbital time $T_0=2\pi R_0/V_{\rm c0}$ and the second expression is in physical units. In the case of a self-gravitating cusp the enclosed cusp mass $M_0$ equals $M_{\rm t}$ and $\tau_0$ is proportional to $M_{\rm t}^{1/2}$. In the Kepler case $M_0$ equals $M_{\rm c}$ and $\tau_0$ is proportional to $M_{\rm c}^{3/2}$.

For the radial evolution $y(t)$ of circular orbits with varying Coulomb logarithm ($\beta\ne 0$) we find the implicit solution (Appendix \ref{apporb})
\bqn
t&=&\tau_{\rm df}\,\frac{\ln z_0}{z_0}\left|\Ei(\ln z_0)-\Ei(\ln z(y))\right|\qquad\kappa, \beta\ne 0\nonumber\\ 
z&=&z_0y^{\kappa}= \Lambda^{\kappa/\beta} \label{ty1}\\
 2\kappa&=&
	\left\{ \begin{array}{lll}
	 3-2\eta & \mbox{Kepler}\\
	 3+\eta & \mbox{self-grav.}\\
	\end{array} \right.
\nonumber
\eqn
and for the  special cases the explicit solutions
\bq
y(t)=\left\{\begin{array}{lll}
\left[1-\frac{t}{\tau_{\rm df}}\right]^{1/\kappa}&\kappa\neq 0, \beta=0\\ \\
\Lambda_0^{\left(\exp(-t/\tau_{\rm df})-1\right)/\beta} & \kappa=0,\beta\neq 0  & \mbox{Kepler}\\ \\
\exp\left(-\frac{t}{\tau_{\rm df}}\right)&\kappa=\beta= 0 & \mbox{Kepler}
\end{array}\right. .\label{yt} 
\eq
The angular momentum evolution can be easily calculated by
\bq
L(t)=\sqrt{GM(y)R}=L_0\times
\left\{\begin{array}{lll}
y^{(1+\eta)/2} & \mbox{self-grav.} \\
y^{1/2} & \mbox{Kepler}
\end{array}\right. \label{Lc} 
\eq
with $L_0=\sqrt{GM_0 R_0}$. We introduced
\bqn
\tau_{\rm df} &=& \tau_0 
\times \left\{ \begin{array}{lll} 
\frac{1+\eta}{3+\eta} & \kappa \ne 0& \mbox{self-grav.}\\ \\
\frac{1}{3-2\eta} & \kappa \ne 0  & \mbox{Kepler}\\ \\
\frac{\ln\Lambda_0}{2\beta} & \kappa = 0,\beta\ne 0 & \mbox{Kepler} \\ \\
\frac{1}{2} & \kappa=\beta = 0 & \mbox{Kepler} \\ \\
\end{array} \right. .\label{taudf}
\eqn
and  $|\tau_{\rm df}|$ measures the decay timescale. 
The first line of eq. \ref{yt} reproduces
the special solution given in \citet{spi03}. 

For shallow cusps with $\eta>3/2$ in a
Kepler potential $\kappa$ becomes negative leading to a negative $\tau_{\rm df}$.
In that case $\ln z_0$ is also negative and there is formally a stalling of the
orbital decay for $\beta>0$ when $\Lambda$ approaches unity. In case of
$\beta=0$ equation \ref{yt} yields an infinite decay time to the centre.

Also for positive $\kappa$ the total decay time with varying $\ln\Lambda$ 
is not well-defined, 
because the approximations in eq. \ref{loglam} breaks down
for $\ln\Lambda<1$. 
In order to get an analytical estimate of the
effective decay time $\tau_{\rm dec}$, we choose the time needed to decrease 
$\ln\Lambda$ from the initial value $\ln\Lambda_0$ to 
$\ln\Lambda=0.3725\beta/\kappa$, where $Ei(\ln z)=0$. For $\ln z_0 \gg 1$
it can be estimated from eq. \ref{taudf} with the help of the asymptotic expansion $\Ei(x)\sim e^x (1/x+1/x^2)$ (8.216) \citep{gra80}
\bq
\tau_{\rm dec}=\left[1+\frac{\beta}{\kappa\ln\Lambda_0}\right]\tau_{\rm df}
   \quad \mbox{for}\quad \frac{\kappa}{\beta}\ln\Lambda_0 \gg 1   \,.\label{taudec}
\eq
The correction factor quantifies the effect due to the position dependence of
the Coulomb logarithm, if $\beta\ne 0$. 
For a negative $\kappa$ (as realised in the Hernquist cusp HE) the last term in eq. \ref{ty1} dominates and diverges as $\Lambda$ approaches unity. Therefore we define a decay timescale for a fixed minimum value $y_\mathrm{dec}$ (e.g. three times the stalling radius) by
\bq
\tau_{\rm dec}=-\frac{\ln z_0}{z_0}\ln|\ln z|\,\tau_{\rm df}
   \quad \mbox{for}\quad |\ln z|=\frac{-\kappa}{\beta}\ln\Lambda \ll 1   \,.\label{taukdec}
\eq
For $\kappa= 0$ and $\beta\ne 0$ the orbital decay of the BH would also stall at $\ln\Lambda=0$.
Only for $\kappa> 0$ and $\beta=0$ there is a finite time 
$\tau_{\rm df}$ to reach the centre. For completeness we mention that for a
negative $\beta$ the total decay time to the centre would be finite due to the
enhanced friction force by an increasing Coulomb logarithm. But for all
realistic cases we find $\beta\ge0$.

The standard case of dynamical friction corresponds to the isothermal sphere and a constant Coulomb logarithm, i.e. initial enclosed
mass $M_0$ at radius $R_0$ with $\eta=1$ and $\beta=0$ in Eq. \ref{taudf}. We use the decay time of the standard case
\bq
\tau_{\rm df0}=
        \frac{1}{\chi_{\rm s}\ln\Lambda_{\rm s}}\frac{M_0}{M_{\rm bh}}
        \frac{R_0}{V_{\rm c0}}
\times \left\{ \begin{array}{lll} 
\frac{1}{2}& \mbox{self-grav.}\\ \\
\frac{M_0}{M_{\rm t}}  & \mbox{Kepler}
\end{array} \right. \label{taudf0}
\eq
as given also in \citet{bin87} (7-26) as normalisation.
For more details of the orbital evolution see App. \ref{apporb}. 

\subsection[]{Self gravitating cusps}\label{sec-selfgcusp}

\begin{figure}
\centerline{
  \resizebox{0.98\hsize}{!}{\includegraphics[angle=270]{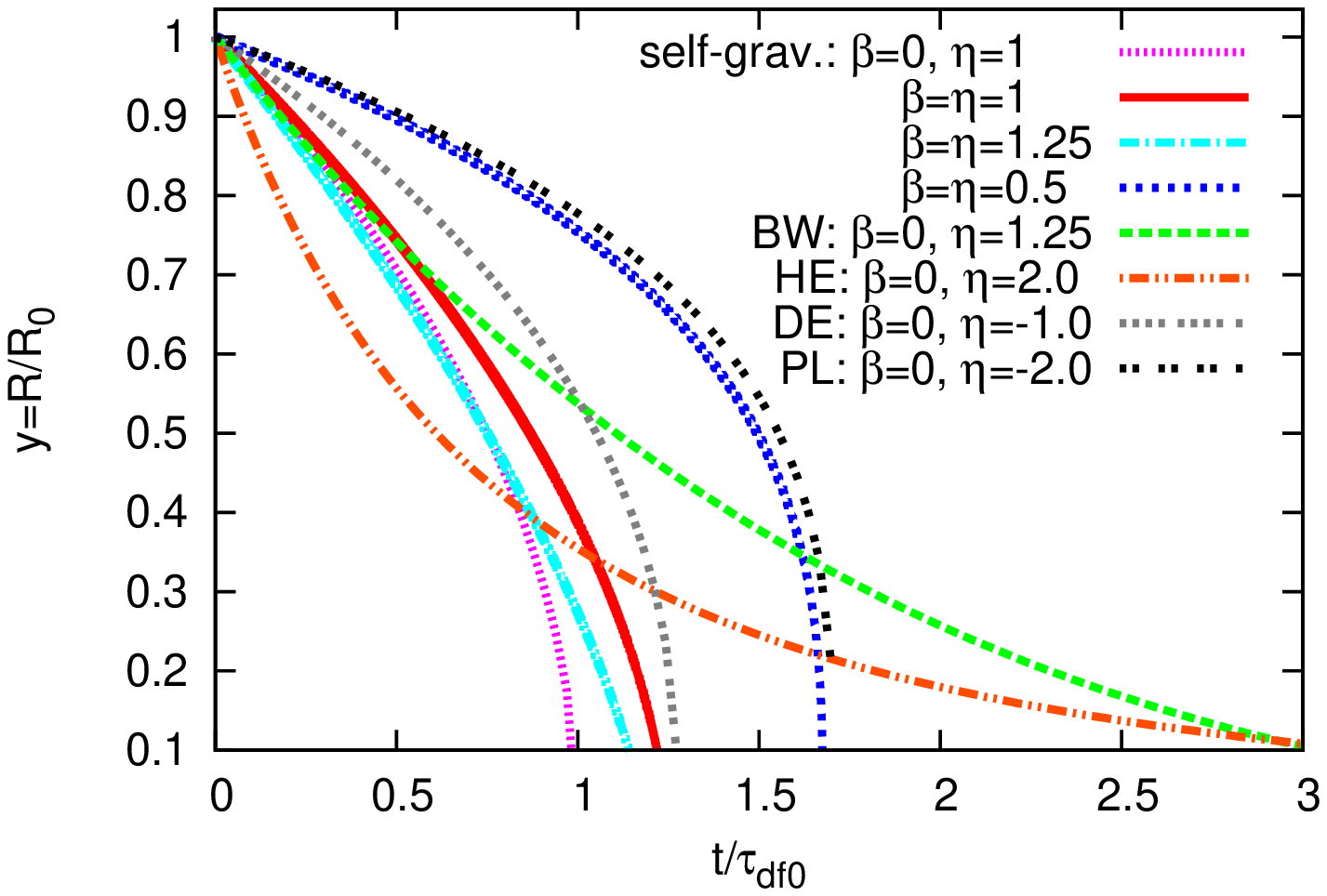}}
  }
\centerline{
  \resizebox{0.98\hsize}{!}{\includegraphics[angle=270]{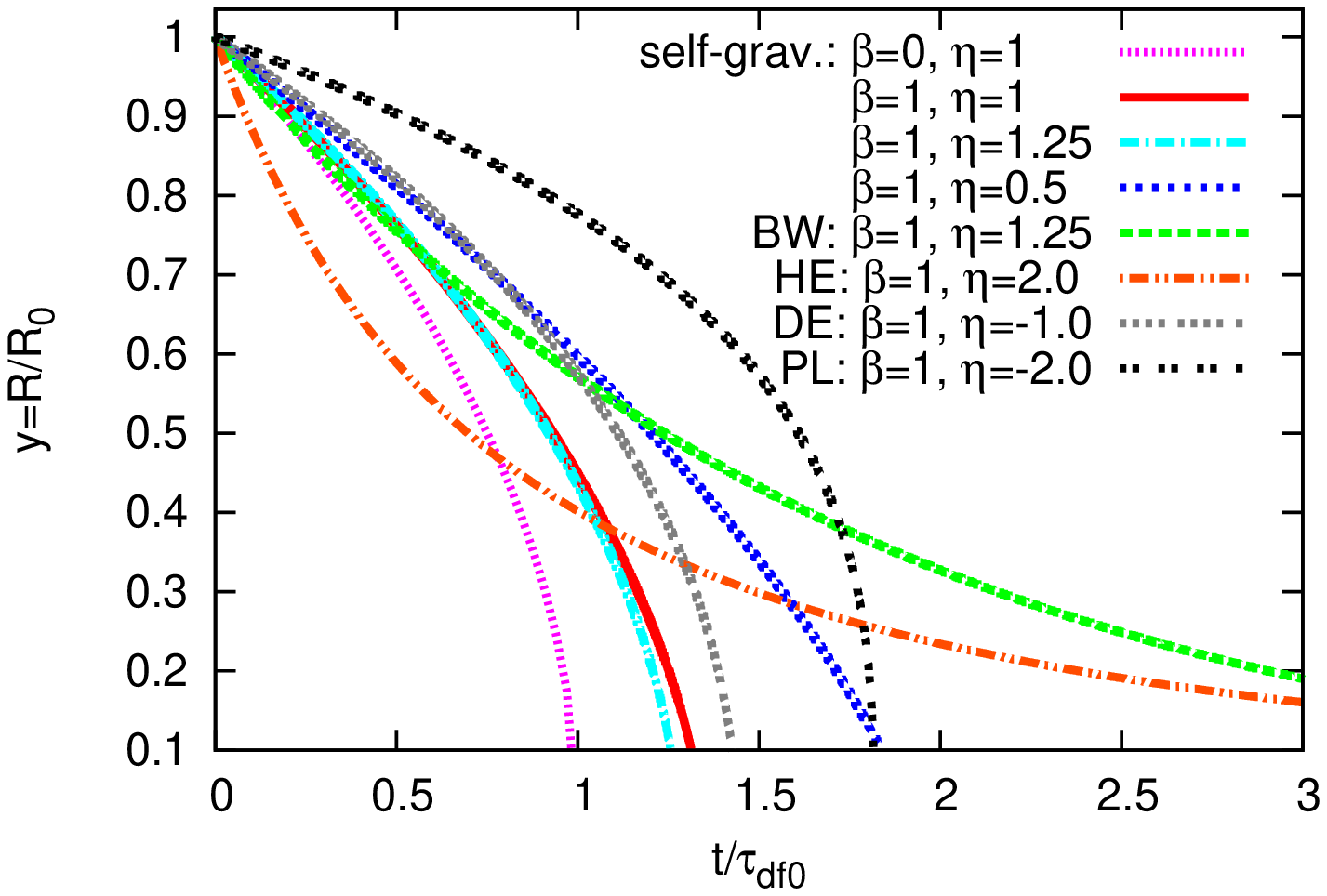}}
  }
\caption[]{
Radial evolution of circular orbits in cusps. We have used 
$M_{\rm bh}=10^{-2}M_0$ and $r_{\rm h}=6.7\times 10^{-3}R_0$ leading to the same 
$\ln\Lambda_0=5.0$ for the standard case with
$\beta=0, \eta=1$. For the four Kepler cases Bahcall-Wolf cusp (BW), Hernquist cusp (HE), Dehnen (DE) and Plummer (PL) outskirts with $\eta=1.25,2.0,-1.0,-2.0$ we have chosen the initial cusp masses $M_{\rm t}/M_{\rm c}=0.5,\,0.5,\, -0.225,\, -0.06$, respectively.
The top panel is for point-like objects and the lower panel for extended objects.
}\label{figyt}
\end{figure}
\begin{figure}
\centerline{
  \resizebox{0.98\hsize}{!}{\includegraphics[angle=270]{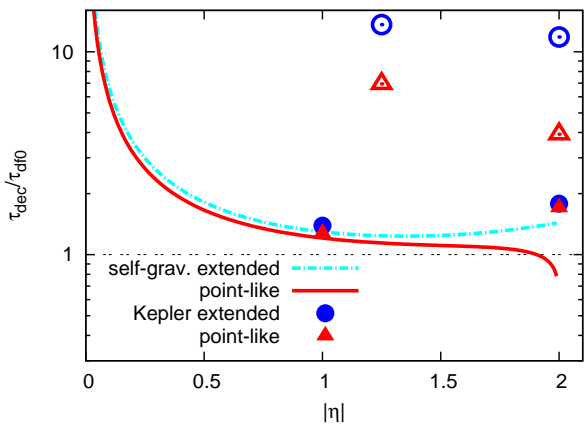}}
  }
\caption[]{
Relative variation of total decay times (eq. \ref{taudec}) normalised to the standard case of constant Coulomb logarithm. We used the same parameters as in Fig. \ref{figyt}. Open symbols are for positive $\eta$ and full symbols for negative $\eta$.
}\label{figtdec}
\end{figure}

In Fig. \ref{figyt} the orbital decay  $y(t)$ of circular orbits is presented.
The differences  in the orbital evolution are caused by the combination of using
self-consistent density profiles and distribution functions and by the position
dependence of the Coulomb logarithm.
The standard case with $\ln\Lambda=\mathrm{const.}$ ($\beta=0$, $\eta=1$) is given by
the black broken-dotted line. For point-like objects $y(t)$  is given in the top panel.
Orbits for different power law indices
$\eta$ with varying $\ln\Lambda$ are plotted. The full red line shows in an
 isothermal core the delay
due to the position dependence of $\ln\Lambda$ and the slightly smaller initial
value of $\ln\Lambda_0=4.6$ instead of 5.0. In the bottom panel the
evolution for the same values of $\eta$ are shown for extended bodies. The
parameters are chosen to give the same Coulomb logarithm
 $\ln\Lambda_0=5.0$ in the standard case.

In Fig. \ref{figtdec} the variation of the effective
decay time as a function of $\eta$ is shown for point-like and
extended bodies. Some care should be taken to use these numbers, because the
innermost radius reached at time $\tau_{\rm dec}$ depends on $\eta$. But the
general parameter dependence of $\tau_{\rm dec}$ gives some insight in the
physics of the orbital decay in cusps.
The effective decay time of circular orbits in self-gravitating
cusps is affected by the following aspects:
\begin{itemize}
\item {\bf Density profile} With decreasing $\eta$ the mass $M_0$ is more and more
concentrated to the centre leading to a smaller density in the outer regions.
This results in a smaller dynamical friction force and prolonged
$\tau_{\rm dec}$.
\item {\bf Distribution function} With decreasing $\eta$ the fraction of slow
particles $\chi_0$ increases considerably from 0.2 to 0.7, 
reducing the effect of the smaller density in the outer parts.
\item {\bf Coulomb logarithm} The position dependence of $\ln\Lambda$ leads to a
moderate delay in orbital decay in the later phase. The effect is strongest for
large values of $\eta$.
\item {\bf Extended  bodies} For extended bodies like star clusters $\ln\Lambda$ is
generally smaller compared to point-like bodies, because the minimum impact
parameter is larger. Compared to the standard case, the prolongation factor is
only weakly dependent on $\eta$.
\end{itemize}

In the case of eccentric orbits there is an additional effect of the variation
of $\ln\Lambda$, because along these orbits the relative strength of the
friction force at apo- and peri-centre is changed.

\subsection{Kepler potential}\label{subsec-kep}

In case of a Kepler potential the cusp mass distribution is decoupled from the potential and the enclosed cusp mass $M_{\rm t}$ is an additional free parameter.
We discuss the explicit solutions for four cases. Well inside the influence radius of a central SMBH the stellar distribution can be described by a cusp in a Kepler potential. We present the orbital decay in the Bahcall-Wolf cusp and the shallow cusp of a Hernquist profile. In the outskirts of self-gravitating systems the density distribution may be approximated by a power law and the potential by a point-mass potential, if the density profile is steep enough. We investigate two cases with steep power law distributions to test the maximum impact parameter dependence of the Coulomb logarithm. The Plummer sphere with an outer density slope of $-5$ and the Dehnen models with a slope of $-4$ are the ideal cases.

\subsubsection{Bahcall-Wolf cusp}\label{ssubsec-bwc}

Here we look to the orbital decay inside the gravitational influence
radius, where the enclosed mass of the stellar component $M_{\rm t}$ is smaller than
the mass $M_{\rm c}$ of the central SMBH. We neglect in this region the
contribution of the stellar component to the mean gravitational field.

The general equations are already given in the previous sections, but we evaluate the terms explicitly for the Bahcall-Wolf cusp with
\bq
\eta=5/4 \quad \rho\propto y^{-7/4}\,,
\eq
leading to
\bqn
\kappa=\frac{1}{4} &&
X_{\rm c}^2=\frac{11}{8}\,. \label{Xckep}
\eqn
For circular orbits the minimum impact parameter for a point-like object becomes
\bq
a_{\rm 90} = \frac{11}{19}\frac{M_{\rm bh}}{M_{\rm c}}R\,,\label{a90kep}
\eq
leading to the initial Coulomb logarithm
 \bq
\Lambda_0=
\left\{\begin{array}{lll}
\frac{76}{77}\frac{M_{\rm c}}{M_{\rm bh}}&\beta=0& \mbox{point-like}
        \\ \\
\frac{4}{7}\frac{R_0}{b_{\rm min}}&\beta=1& \mbox{unres. or ext.}
\end{array}\right.\,.
        \label{lam0kepe}
\eq
The distribution function $F(u)$ and the corresponding $\chi(U)$ are shown in Figs. \ref{figfe} - \ref{figchiR}. 
The decay time-scale 
$\tau_{\rm df}$ from Eq. \ref{taudf} reads
\bqn
\tau_{\rm df} &=& \frac{55}{\ln\Lambda_0} 
        \left[\frac{R_0}{\pc}\frac{M_{\rm c}}{\msol}\right]^{3/2}
        \left[\frac{M_{\rm t}}{\msol}
	\frac{M_{\rm bh}}{\msol}\right]^{-1}\mathrm{Myr} \,,
        \label{taukep1}
\eqn
For point-like objects the total decay time $\tau_{\rm dec}$ equals 
$\tau_{\rm df}$ and for extended bodies the corresponding equation is (using Eq.
\ref{taudec})
\bq
\tau_{\rm dec}=\frac{\ln\Lambda_0+4}{(\ln\Lambda_0)}
	\tau_{\rm df}\qquad\mbox{unres. or ext.}
        \label{tauphyse}
\eq
The orbital decay of circular orbits is given by Eqs. \ref{yt} and \ref{ty1} for point-like and extended bodies, respectively,
\bq
y=\left[1-\frac{t}{\tau_{\rm df}}\right]^4 \,,\label{ytkep}
\eq
which is very different to the standard case. 

In Fig. \ref{figyt} the orbit evolution in a Kepler
potential is plotted for $M_{\rm t}=0.5\,M_{\rm c}$. It shows the
strong slow-down in the inner part leading to long total decay times
$\tau_{\rm dec}$ (see Fig. \ref{figtdec}).

\subsubsection{Hernquist cusp}\label{ssubsec-HE}

\begin{figure}
\centerline{
  \resizebox{0.98\hsize}{!}{\includegraphics[angle=270]{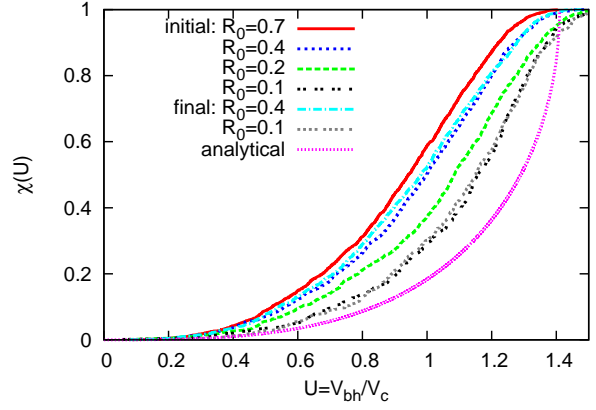}}
  }
\caption[]{
Initial and final cumulative distribution functions for the Hernquist cusp at different radii of runs E1 and E2. For comparison the analytic limit for $r\to\ 0$ is included. 
} \label{figchiE}
\end{figure}

For the Hernquist cusp the corresponding equations are
\bqn
\eta_0=2 && \rho\propto y^{-1}\\
\kappa=-\frac{1}{2} &&
X_{\rm c}^2=1\,. \label{Xche} \\
a_{\rm 90} &=&\frac{1}{2}\frac{M_{\rm bh}}{M_{\rm c}}R\,,\label{a90he}\\
\Lambda_0&=&
\left\{\begin{array}{lll}
2\frac{M_{\rm c}}{M_{\rm bh}}&\beta=0& \mbox{point-like}
        \\ \\
\frac{R_0}{b_{\rm min}}&\beta=1& \mbox{unres. or ext.}
\end{array}\right.\,.
        \label{lam0he}
\eqn
The decay timescales are
\bqn
-\tau_{\rm df} &=& \frac{15}{\ln\Lambda_0} 
        \left[\frac{R_0}{\pc}\frac{M_{\rm c}}{\msol}\right]^{3/2}
        \left[\frac{M_{\rm t}}{\msol}
	\frac{M_{\rm bh}}{\msol}\right]^{-1}\mathrm{Myr} \,,
        \label{taukephe}\\
\tau_{\rm dec}&=&-\frac{1}{2}\sqrt{\Lambda_0}\ln\Lambda_0\ln\left(\frac{\ln\Lambda}{2}\right)
	\tau_{\rm df}\qquad\mbox{unres. or ext.}\nonumber 
\eqn
and the orbital decay of circular orbits is for point-like objects
\bq
y=\left[1+\frac{t}{|\tau_{\rm df}|}\right]^{-2}\,.\label{ythe}
\eq
In Fig. \ref{figyt} we have chosen the same $y_{\rm h}$ as for the BW case. 
We find an initially faster decay than in the BW cusp and a stronger slow-down in the late phase leading to a 
comparable $\tau_{\rm dec}$ adopting a final Coulomb logarithm $\ln\Lambda=1.1$ (see figure  \ref{figtdec}). The comparison of the analytic and the numerically realised cumulative distribution function  $\chi(U)$ is shown in Fig. \ref{figchiE}. The figure demonstrates the influence of the outer boundary conditions deep into the cusp. 

\subsubsection{The outskirts of a Plummer sphere}\label{ssubsec-plum}

In the outskirts 
 of a Plummer sphere with total mass $M_{\rm c}$ the corresponding equations are
\bqn
\eta_0=-2 && \rho\propto y^{-5}\\
\kappa=\frac{7}{2} &&
X_{\rm c}^2=3\,. \label{Xcpl} \\
a_{\rm 90} &=&\frac{3}{4}\frac{M_{\rm bh}}{M_{\rm c}}R\,,\label{a90pl}\\
\Lambda_0&=&
\left\{\begin{array}{lll}
\frac{4}{15}\frac{M_{\rm c}}{M_{\rm bh}}&\beta=0& \mbox{point-like}
        \\ \\
\frac{1}{5}\frac{R_0}{b_{\rm min}}&\beta=1& \mbox{unres. or ext.}
\end{array}\right.\,.
        \label{lam0pl}
\eqn
\bqn
\tau_{\rm df} &=& \frac{0.78\,}{y_{\rm a}^2\ln\Lambda_0} 
        \left[\frac{R_0}{\pc}\right]^{3/2}
	\left[\frac{M_{\rm c}}{\msol}\right]^{1/2}
	\left[\frac{M_{\rm bh}}{\msol}\right]^{-1}\mathrm{Myr} \,,
        \label{tdfpl}\\
\tau_{\rm dec}&=&\frac{\ln\Lambda_0+2/7}{(\ln\Lambda_0)}
	\tau_{\rm df}\qquad\mbox{unres. or ext.}
        \label{tdecpl}
\eqn
We have used eq. \ref{mplumapp} for $M_{\rm t}$.
The orbital decay of circular orbits is for point-like objects
\bq
y=\left[1-\frac{t}{\tau_{\rm df}}\right]^{2/7}\,.\label{ytpl}
\eq
and the distribution function $F(u)$ and the corresponding $\chi(U)$ are shown in Figs. \ref{figfe} - \ref{figchiR}. 
In Figs. \ref{figyt} and \ref{figtdec} we have chosen $y_{\rm a}=0.2$ for the Plummer radius in units of $R_0$.

\subsubsection{The outskirts of Dehnen models}\label{ssubsec-dehn}

In the outskirts 
 of Dehnen models with total mass $M_{\rm c}$ and inner cusp slope $\eta$ the corresponding equations are
\bqn
\eta_0=-1 && \rho\propto y^{-4}\\
\kappa=\frac{5}{2} &&
X_{\rm c}^2=\frac{5}{2}\,. \label{Xcde} \\
a_{\rm 90} &=&\frac{5}{7}\frac{M_{\rm bh}}{M_{\rm c}}R\,,\label{a90de}\\
\Lambda_0&=&
\left\{\begin{array}{lll}
\frac{7}{20}\frac{M_{\rm c}}{M_{\rm bh}}&\beta=0& \mbox{point-like}
        \\ \\
\frac{1}{4}\frac{R_0}{b_{\rm min}}&\beta=1& \mbox{unres. or ext.}
\end{array}\right.\,.
        \label{lam0de}
\eqn
\bqn
\tau_{\rm df} &=& \frac{2.36\,}{y_{\rm a}\ln\Lambda_0} 
        \left[\frac{R_0}{\pc}\right]^{3/2}
	\left[\frac{M_{\rm c}}{\msol}\right]^{1/2}
	\left[\frac{M_{\rm bh}}{\msol}\right]^{-1}\mathrm{Myr} \,,
        \label{tdfde}\\
\tau_{\rm dec}&=&\frac{\ln\Lambda_0+0.4}{(\ln\Lambda_0)}
	\tau_{\rm df}\qquad\mbox{unres. or ext.}
        \label{tdecde}
\eqn
We have used eq. \ref{mdehnapp} for $M_{\rm t}$ with  $\eta=1.5$.
The orbital decay of circular orbits is for point-like objects
\bq
y=\left[1-\frac{t}{\tau_{\rm df}}\right]^{2/5}\,.\label{ytde}
\eq
and the distribution function $F(u)$ and the corresponding $\chi(U)$ are shown in Figs. \ref{figfe} - \ref{figchiR}. 
In Figs. \ref{figyt} and \ref{figtdec} we have chosen $y_{\rm a}=0.15$ for the scale radius in units of $R_0$.

\section[]{Code description}\label{sec-code}

We are using two different numerical codes to evolve our initial models. In order to study the orbital decay of the massive BH in self-gravitating cusps and in the outskirts of the Plummer sphere and Dehnen models, we used the PM code {\sc Superbox}. In order to study the decay of a secondary massive BH in a Bahcall-Wolf cusp around a SMBH, we used the PP code $\phi GRAPE$. For comparison of the numerical results with the analytic predictions for eccentric runs, the semi-analytic code  {\sc intgc} has been developed. A brief description of these codes is given below.

\subsection[]{SUPERBOX}\label{sup}

The PM code {\sc Superbox} \citep{fell00} is a highly efficient code with fixed time step for galaxy
dynamics, where more than 10 million particles per galaxy in co-moving nested
grids for high spatial resolution at the galaxy centres can be simulated. The
code is intrinsically collision-less, which is necessary for long-term simulations of
galaxies. Another advantage of a PM code is the large particle number which
can be simulated in a reasonable computing time (up to a few days or a week per
simulation).
Three grid levels with different resolutions are used which resolve the core of each component/galaxy, the major part of the component/galaxy and the whole simulation area. The spatial resolution is determined by the number of grid cells per dimension $N_{c} = 2^{m}$ and the size of the grids. The SMBH is included as a moving particle with own sub-grids for high spatial resolution in its vicinity. The grid sizes are chosen such that the full orbit of the secondary BH falls into the middle grid.  

All {\sc Superbox} runs were performed using the Astronomisches Rechen-Institut (ARI) fast computer facilities with no special hardware.

\subsection[]{$\phi$GRAPE}\label{phi}

This programme is a direct $N$-body code, which calculates the pairwise forces
between all particles. In order to combine high numerical accuracy and 
fast calculation of the gravitational interactions a very sophisticated 
numerical scheme and a special hardware is used. This PP code uses fourth order Hermite Integrator with individual block 
time steps. The acceleration and its time derivative are
calculated with parallel use of GRAPE6A cards. In order to minimise 
communication among different nodes MPI parallelisation strategy is 
employed. For the simulations in a dense cusp around a massive central SMBH, we are using a 
specially developed $\phi$GRAPE code. We include the central SMBH 
as an external potential in order to avoid the relatively large random motion of a live SMBH due to the small particle number. Secondly the time step criterion is modified. We add a reduction factor for the BH time step, which compensates the effect of the relatively small accelerations compared to the field particles. The code and special GRAPE hardware are described 
in \citet{harfst}. For our calculations we are using the parallel 
version of the programme on 32 node cluster 
$Titan$\footnote{http://www.ari.uni-heidelberg.de/grace/}, built
at the Astronomisches Rechen-Institut in Heidelberg.

Part of the calculations were done on the special 85 node Tesla
C1060 GPU cluster installed on the National Astronomical
Observatories of China, Chinese Academy of Sciences (NAOC/CAS).
For these runs we used the modified version of our $\phi$GRAPE code
including the SAPPORO library to work on the GPU cards \citep{gab09}.

\subsection[]{Semi-analytic code - INTGC}\label{intgc}

The program {\sc intgc} is an integrator for orbits in an analytic
background potential of a galactic centre including the Chandrasekhar
formula for dynamical friction. Different analytic models with
their $\chi$ functions and a variable Coulomb logarithm
\citep{jus05} are implemented. An 8th-order composition
scheme is used for the orbit integration \citep{yos};
for the coefficients see \citet{mcl}).
Since the symplectic composition schemes are by construction suited
for Hamiltonian systems, the dissipative friction
force requires special consideration. It is implemented in {\sc intgc}
with an implicit midpoint method \citep{mik02}. Four iterations turned out to guarantee
an excellent accuracy of the scheme. Gravitational potential and density are given analytically.

\section[]{Numerical models and results} \label{sec-model}

The contributions to the dynamical friction force covers a large range of parameters for the 2-body encounters, which must be fully covered by the numerical simulations in order to reach a quantitative measure of the Coulomb logarithm. The numerical representation is mainly restricted by the resolution of small impact parameters determined by the spatial resolution, the time resolution and the number statistics. For PP codes the number statistics in the main limitation. Therefore we can use the $\phi$GRAPE code for the BW cusp only with a very high local density in the inner cusp. We reach a few encounters per decay timescale $\tau_0$ with impact parameters below twice the minimum value. For the {\sc Superbox} runs the spatial resolution is limited, which we can take into account by a correct choice of the minimum impact parameter. But even in that case it turns out that due to the fixed time step the time resolution is  still a bottleneck, which limits the total time of some simulations. In all models we use the angular momentum to measure the orbital decay (eq. \ref{Lc}).

\subsection[]{Bahcall-Wolf cusp} \label{bwcn}

We used the extended $\eta$-model of \citet{mm07} to generate the initial particle distributions in phase space.
In all our runs we are using $N=64,000$ particles. The particles positions are generated so that their spatial distribution satisfy Eq. (\ref{density}) and the velocities were assigned to these particles according to Eq. (\ref{edis}). In all our simulations we are using the normalisation $G=M_{\rm c}=a = 1$ leading to $y_{\rm a}=R_0^{-1}$ in N-body units. The mass of the central black hole is $M_{\rm c}$ and the total cusp mass $M_{t}=0.1$. For the setup we used a radius range of $10^{-4}-20$ in all our runs. Inside one unit length our cusp follows density profile $-7/4$ and then turns over to a Plummer density profile with slope $-5$ for far out distances. Table \ref{table:BW} shows the list of parameters for the series of runs. 

\begin{table}
\caption{Parameters of the runs for the BW cusp.}
\begin{tabular}{c c c c c c c c c c c c}
\hline
Run & $N/10^{3}$ & $\epsilon/10^{-4}$ & $M_{\rm bh}$ & $R_{0}$ &
 $V_\mathrm{c,0}$ &$V_0/V_\mathrm{c,0}$ & $a_{90}/10^{-4}$ & $\ln\Lambda_0$\\
\hline
A0& 64 &	$1.0$	& --   &  --  & --     & --  & --     &  --  \\
A1& 64 &	$0.1$	& .005 & 0.2  & $2.25$ & 1.0 & $5.79$ & 5.28 \\
A2& 64 &	$0.1$	& .005 & 0.1  & $3.17$ & 1.0 & $2.89$ & 5.29 \\
A3& 64 &	$0.1$	& .01  & 0.2  & $2.25$ & 1.0 & $11.6$ & 4.59 \\
B1& 64 &	$0.1$	& .005 & 0.2  & $2.25$ & 0.7 & $7.01$ & 5.09 \\
B2& 64 &	$0.1$	& .005 & 0.1  & $3.17$ & 0.5 & $4.07$ & 4.94 \\
C1& 64 &	$1.0$	& .005 & 0.2  & $2.25$ & 1.0 & $5.79$ & 5.25 \\
C2& 64 &	$5.0$	& .005 & 0.2  & $2.25$ & 1.0 & $5.79$ & 4.79 \\
C3& 64 &	$10.0$	& .005 & 0.2  & $2.25$ & 1.0 & $5.79$ & 4.26 \\
E1& 64 &	$0.1$	& .005 & 0.7  & $1.20$ & 1.0 & $17.5$ & 5.33 \\ 
E2& 128 &	$0.1$	& .002 & 0.2  & $2.27$ & 1.0 & $2.0 $ & 6.91 \\ 
\hline
\end{tabular}

{\it Note.} For all runs the gravitational constant $G$, the SMBH mass
$M_\mathrm{c}$ and the scale radius $a$ are normalised to unity.
$N$ is the total number of particles in the cusp, $\epsilon$ the
softening length of the particles, $M_{\rm bh}$ the mass of the
secondary black hole with  initial distance $R_{0}$ in units of $a$, the circular velocity $V_{\rm c,0}$ at $R_{0}$ in units of $\sqrt{GM_\mathrm{c}/a}$ and the initial velocity  $V_0$  in units of the circular velocity,
the initial value of $a_{90}$ from Eq. \ref{a90} is in units of $a$.
\label{table:BW}
\end{table}

\subsubsection{Cusp stability analysis} \label{ stability}
\begin{figure}
\centerline{
  \resizebox{0.98\hsize}{!}{\includegraphics[angle=270]{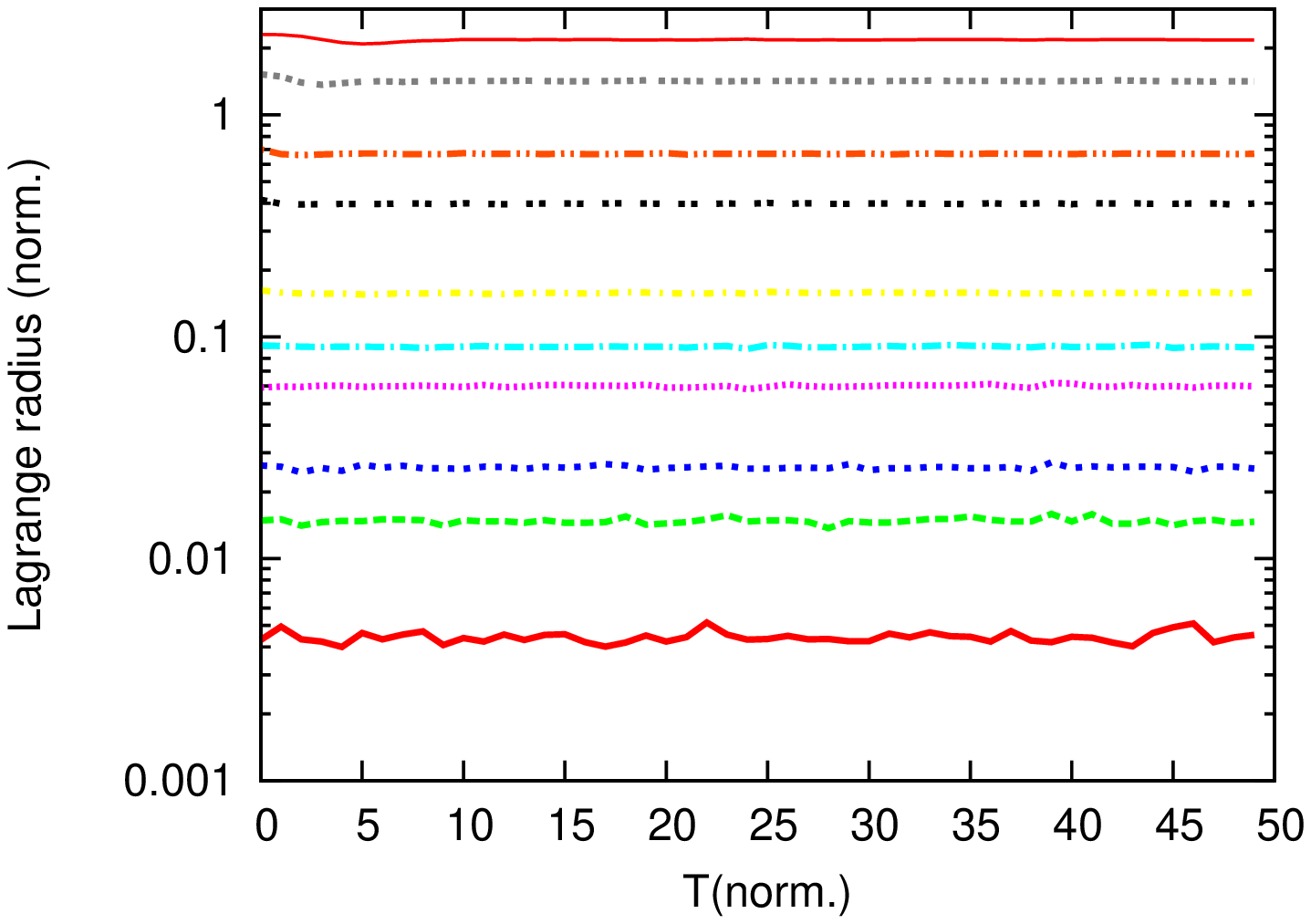}}
  }
\centerline{
  \resizebox{0.98\hsize}{!}{\includegraphics[angle=270]{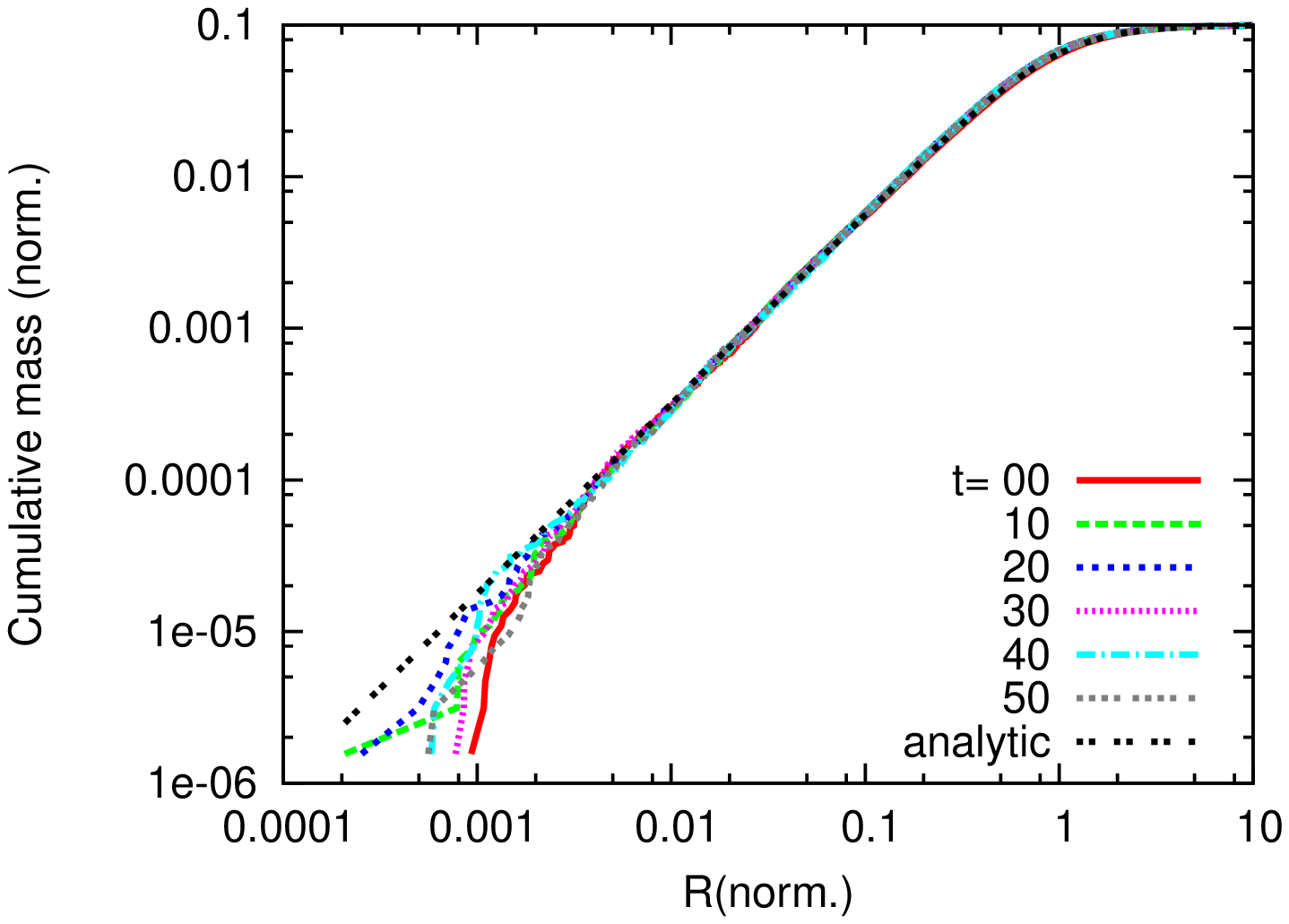}}
  }
\caption[]{
{\bf Top:} Figure shows the evolution of Lagrange radii of 0.1, 0.5, 1, 3, 5, 10, 30, 50, 80, 90$\%$ enclosed mass (from bottom to top) for the BW cusp. The Lagrange radii do not show any systematic evolution with time. 

{\bf Bottom:} Cumulative mass profile at various time steps. The cumulative mass profile is practically indistinguishable from the theoretical one except the inner few dozen particles, where deviations due to the inner cutoff and noise are expected.
} \label{stab}
\end{figure}

Since we are using an approximate DF, we first performed a run (run A0 in the table) without a secondary BH to test whether or not the cusp is stationary around the central SMBH. We run this model up till 50 time units. Figure \ref{stab} shows the evolution of Lagrange radii and also the cumulative mass profile at various time steps. We can see that the cusp is very stable. So the problem how to get a stationary cusp in a Kepler potential
was solved by this kind of a compromise distribution between Bahcall-Wolf cusp 
in the inner (high binding energy) regime and a Plummer distribution in the outskirts.

\subsubsection{Circular Runs}

\begin{figure}
\centerline{
  \resizebox{0.98\hsize}{!}{\includegraphics[angle=270]{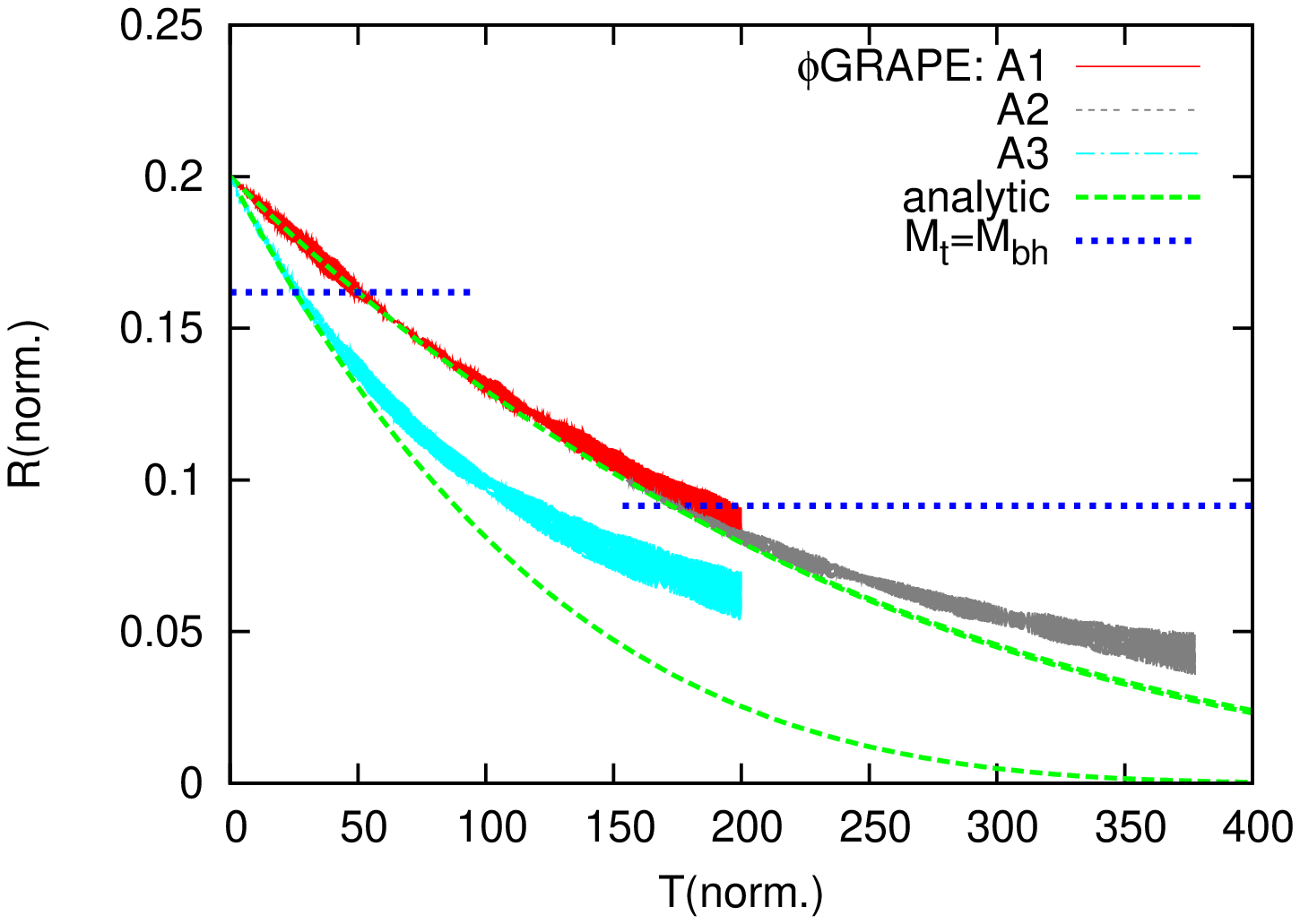}}
  }
\centerline{
  \resizebox{0.98\hsize}{!}{\includegraphics[angle=270]{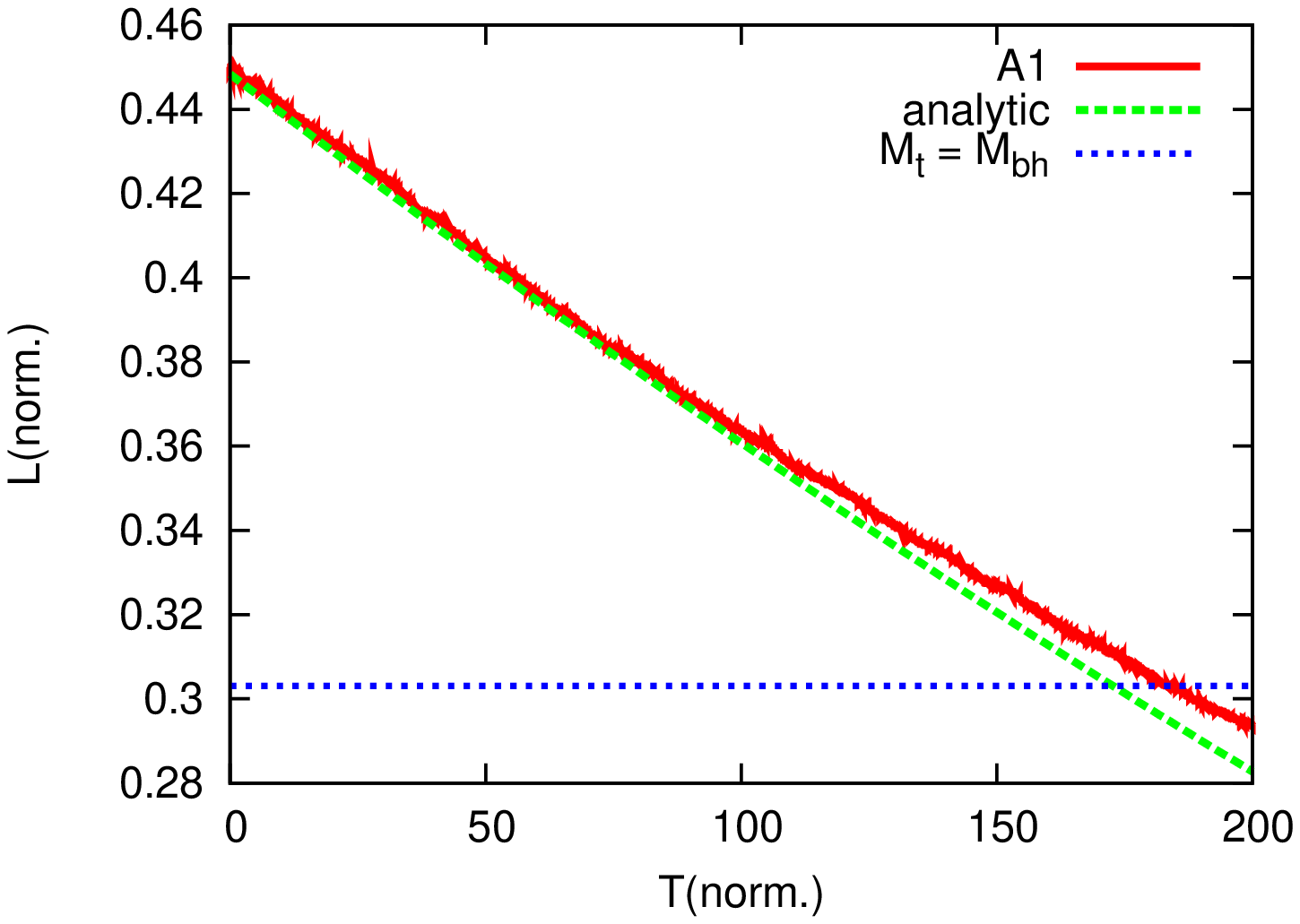}}
  }
\caption[]{
Top panel: Comparison of the orbit evolution of the $\phi$GRAPE data and the analytic estimates (Eq. \ref{ytkep}) for the circular orbits A1, A2, A3. The run A2 is shifted by $T_0=155$ in order to continue the theoretical line of A1. The horizontal lines shows the radii, where the enclosed cusp mass equals $M_{\rm bh}$ (for A1, A2 at $T>150$ and for A3 at $T<100$).
Bottom panel: Comparison of the angular momentum evolution of the $\phi$GRAPE data and the analytic estimates (Eq. \ref{Lc}) for the circular orbit A1. The horizontal line shows $L_{\rm c}$ at the radius, where the enclosed cusp mass equals $M_{\rm bh}$.
} \label{A1}
\end{figure}

\begin{figure}
\centerline{
  \resizebox{0.98\hsize}{!}{\includegraphics[angle=270]{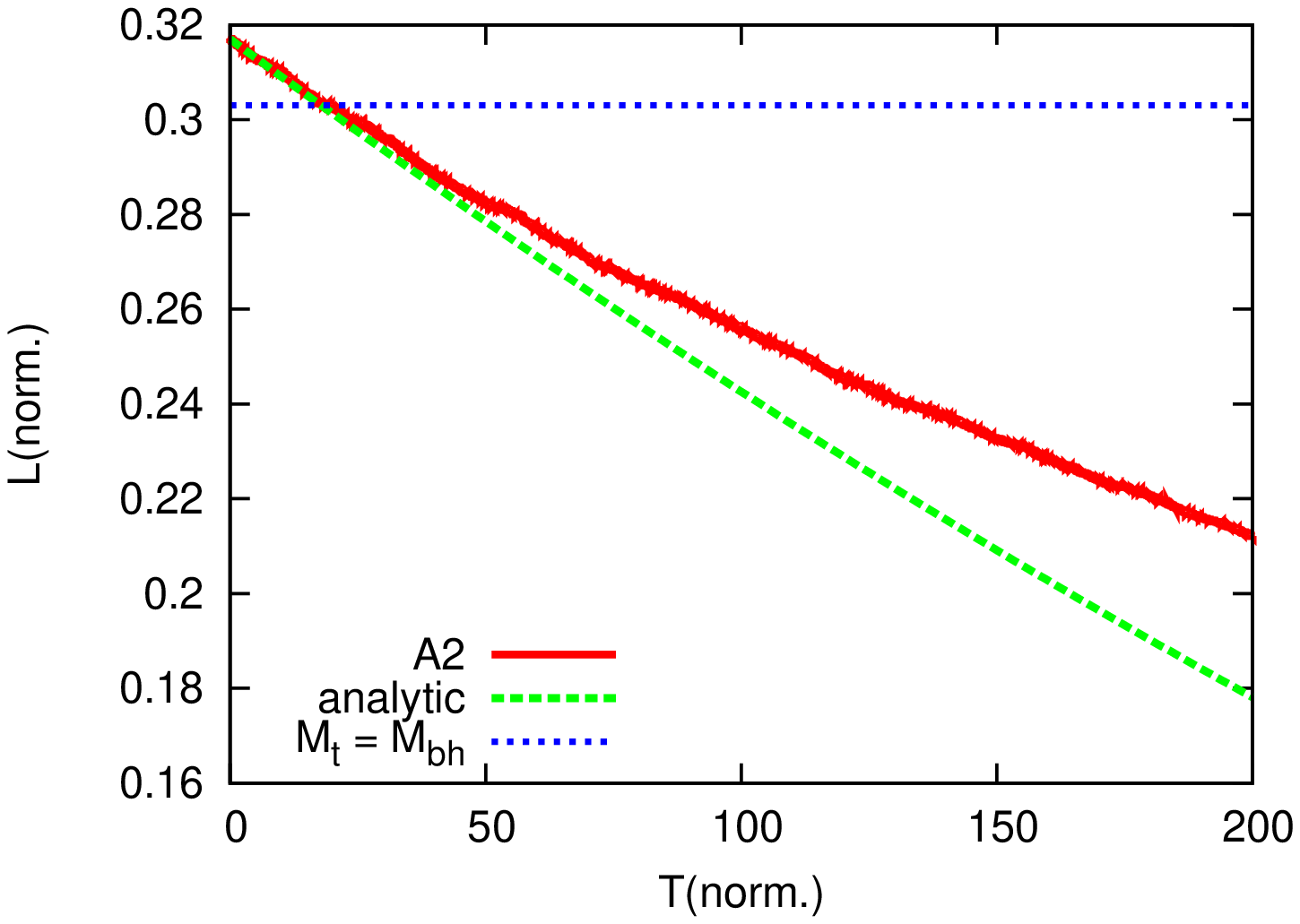}}
  }
\centerline{
  \resizebox{0.98\hsize}{!}{\includegraphics[angle=270]{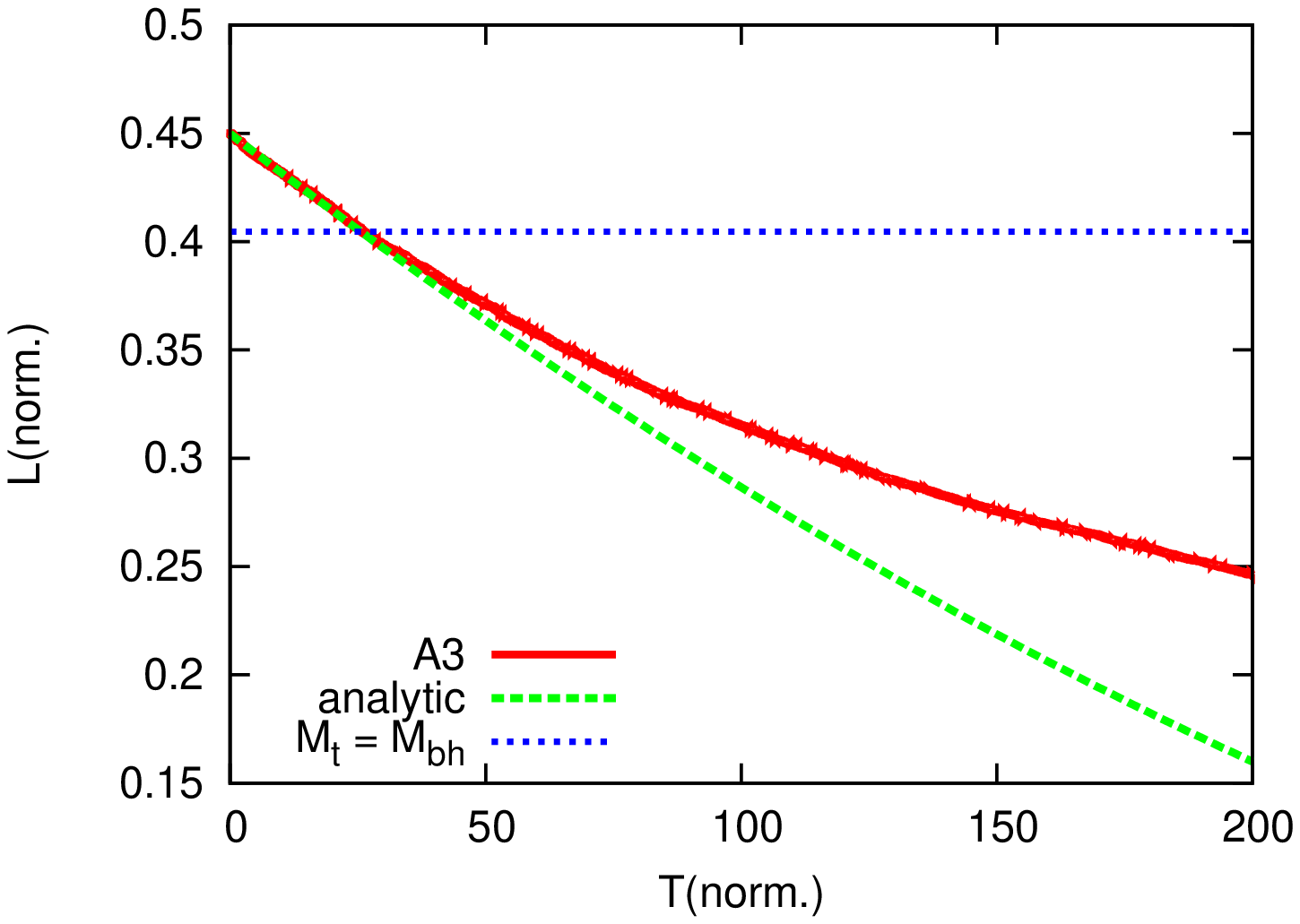}}
  }
\caption[]{
Same as in the bottom panel of figure \ref{A1} for runs A2 and A3.
} \label{delay}
\end{figure}

\begin{figure}
\centerline{
  \resizebox{0.98\hsize}{!}{\includegraphics[angle=270]{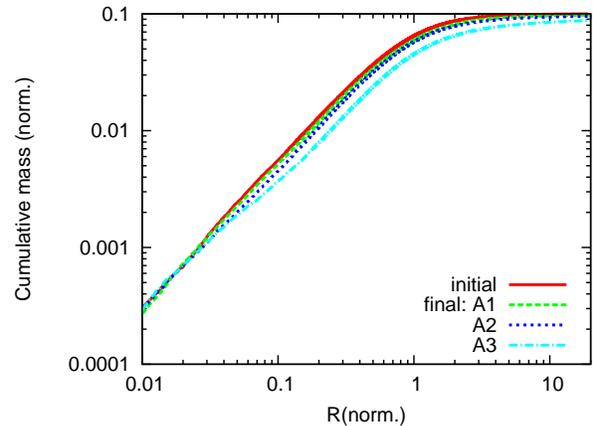}}
  }
\caption[]{
Cumulative mass profiles for the runs A1--A3 demonstrating the reduced local density in runs A2 and A3 at the position of the BH.
} \label{massA123}
\end{figure}

We performed two series of circular runs (see table \ref{table:BW}).
The runs A1--A3 with different BH masses and initial radii are all resolved, i.e. the softening parameter $\epsilon = 10^{-5}$  is much smaller than the initial value of $a_{90}$. Since $a_{90}$ decreases linearly in $y$ (Eq. \ref{a90kep}), the minimum impact parameter is fully resolved to very small radii and we can use Eq. \ref{ytkep} for analytic estimates. 
In figures \ref{A1} and \ref{delay} we show the distance and the angular momentum evolution, because already small deviations from circularity smear out the appearance of the orbits due to the very short orbital time compared to the decay time. The top panel of figure \ref{A1} shows the distance evolution for runs A1, A2, and A3. Run A2 corresponds to a restart of A1 after time $T=155$ but using the initial particle distribution of the cusp. We see that there is no significant long-term evolution of the cusp, which influences the orbital decay, until the end of run A1. The dashed (green) lines show the analytic predictions of the orbital evolution from Eq. \ref{ytkep}. There is an excellent agreement in the first phase with a small delay in the later phase of run A1, which occurs much earlier in A2. The reason for the reduced friction is investigated in run A3 further.
In run A3 we increased the mass of the BH and put it back at $R_0=0.2$ such that in case A3 the radius, where the enclosed cusp mass $M_{\rm t}$ equals $M_{\rm bh}$ (horizontal lines in figures \ref{A1} and \ref{delay}), is twice that of run A2. In run A3 the delay starts also very early. 
In the bottom panel of figure \ref{A1} and in figure \ref{delay} the same evolution is shown much clearer in angular momentum $L$ using Eq. \ref{Lc} for the analytic predictions. The horizontal lines show the distance, where $M_{\rm bh}=M_{\rm t}$ in all figures. This radius coincides with the distance, where the BH mass exceeds the mass in a shell centred at the orbit. An inspection of the cumulative mass profiles shows that the back-reaction of the scattering events to the cusp distribution becomes significant (see figure \ref{massA123}). The cumulative mass profile becomes shallower, which means that the local density is reduced leading to a smaller dynamical friction force.

For a circular orbit in a Bahcall-Wolf cusp the new friction formula is very close to the standard formula: the Coulomb logarithm is also constant, the value deviates only by $\Delta\ln\Lambda$=-0.1 from $\ln\Lambda_\mathrm{s}$, the $\chi$ value is 0.430 instead of $\chi_\mathrm{s}$=0.428. This leads to an indistinguishably faster decay when applying the standard formula. The picture changes slightly for the eccentric orbits (see below).

In all simulations the eccentricity of the orbits vary slightly. The increasing eccentricities in the later phases of the runs may correlate to the decreasing mean density, i.e. may be connected to the feedback of the BH on the cusp. The eccentricity evolution will be investigated in more detail in future work.

\begin{figure}
\centerline{
  \resizebox{0.98\hsize}{!}{\includegraphics[angle=270]{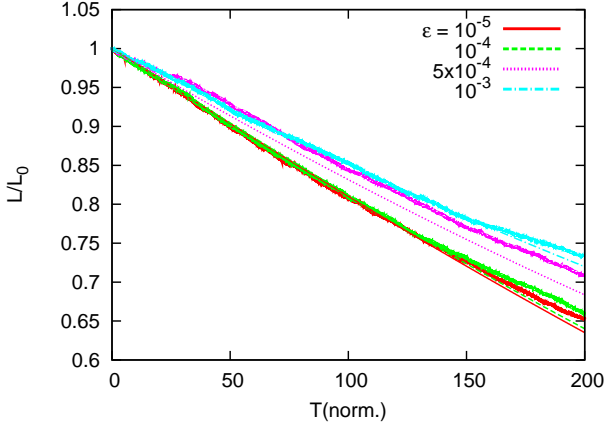}}
  }
\caption[]{
Evolution of a circular orbit in Bahcall-Wolf cusp for with different softening lengths $\epsilon$ (runs A1, C1--C3). Thin lines show the semi-analytic calculations with {\sc Intgc} with eq. \ref{loglam} for the Coulomb logarithm with the best fit minimum impact parameter 
$b_{\rm min} = 1.5\epsilon $.
} \label{C1}
\end{figure}

In a second series of runs C1--C3, we study the impact of the minimum impact parameter by increasing the softening parameter until it is much larger than $a_{90}$. In figure \ref{C1} we can see that for the largest softening length $\epsilon$ (greater than $a_{90}$) the decay is slower as expected from the  smaller Coulomb logarithm (eq. \ref{lam0kepe}). We can use the variation to measure the effective resolution of the code. The best simultaneous fit of all curves lead to $b_{\rm min}=1.5\epsilon$ in Eq. \ref{loglam} for the PP code $\phi$GRAPE. The random variations of the orbital decay on time-scales of 10\dots 100 time units are expected from a rough estimate of the close encounter rate and the corresponding velocity changes.

\subsubsection{Eccentric runs}\label{sub-ecc}

\begin{figure}
\centerline{
  \resizebox{0.98\hsize}{!}{\includegraphics[angle=270]{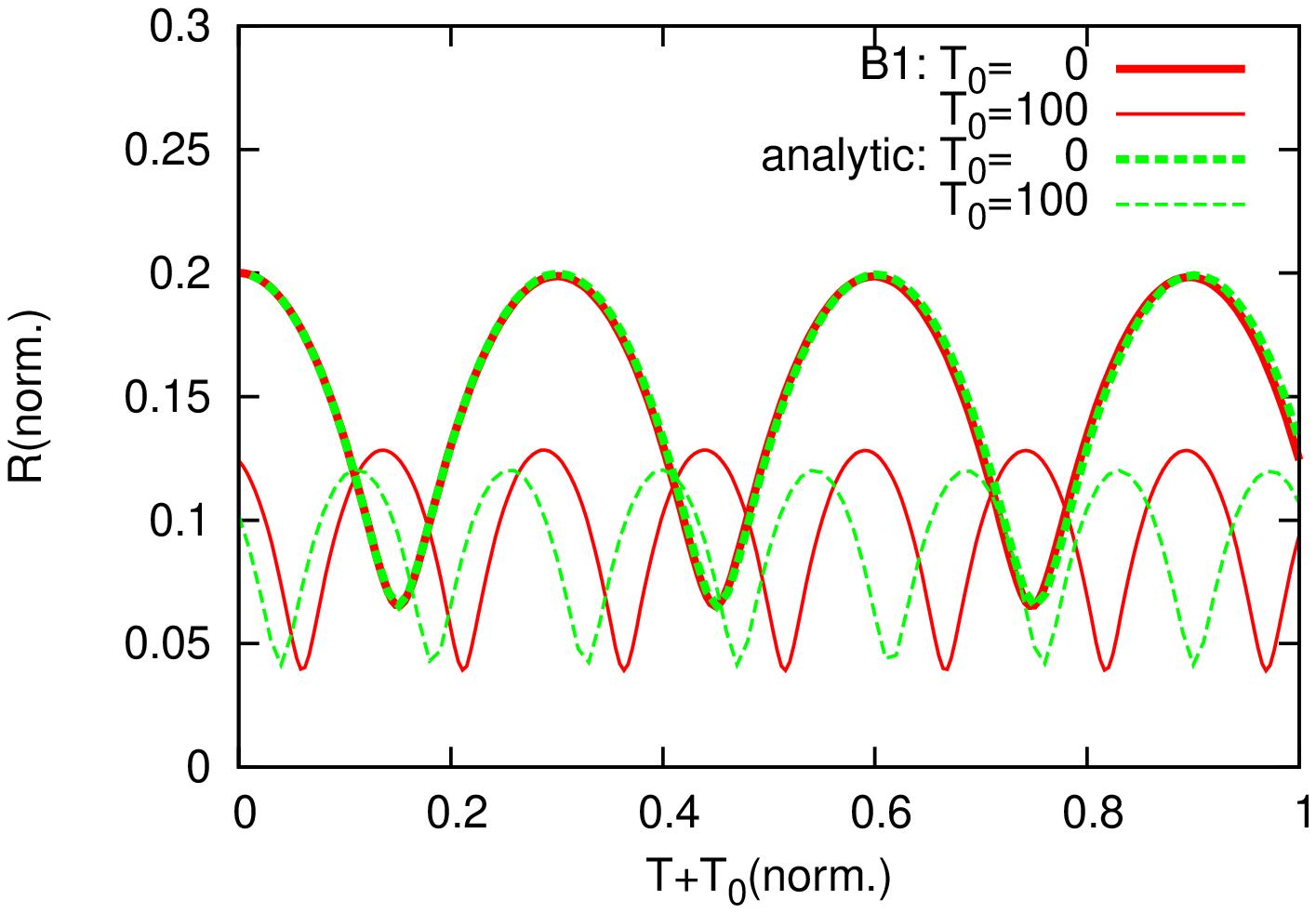}}
  }
\centerline{
  \resizebox{0.98\hsize}{!}{\includegraphics[angle=270]{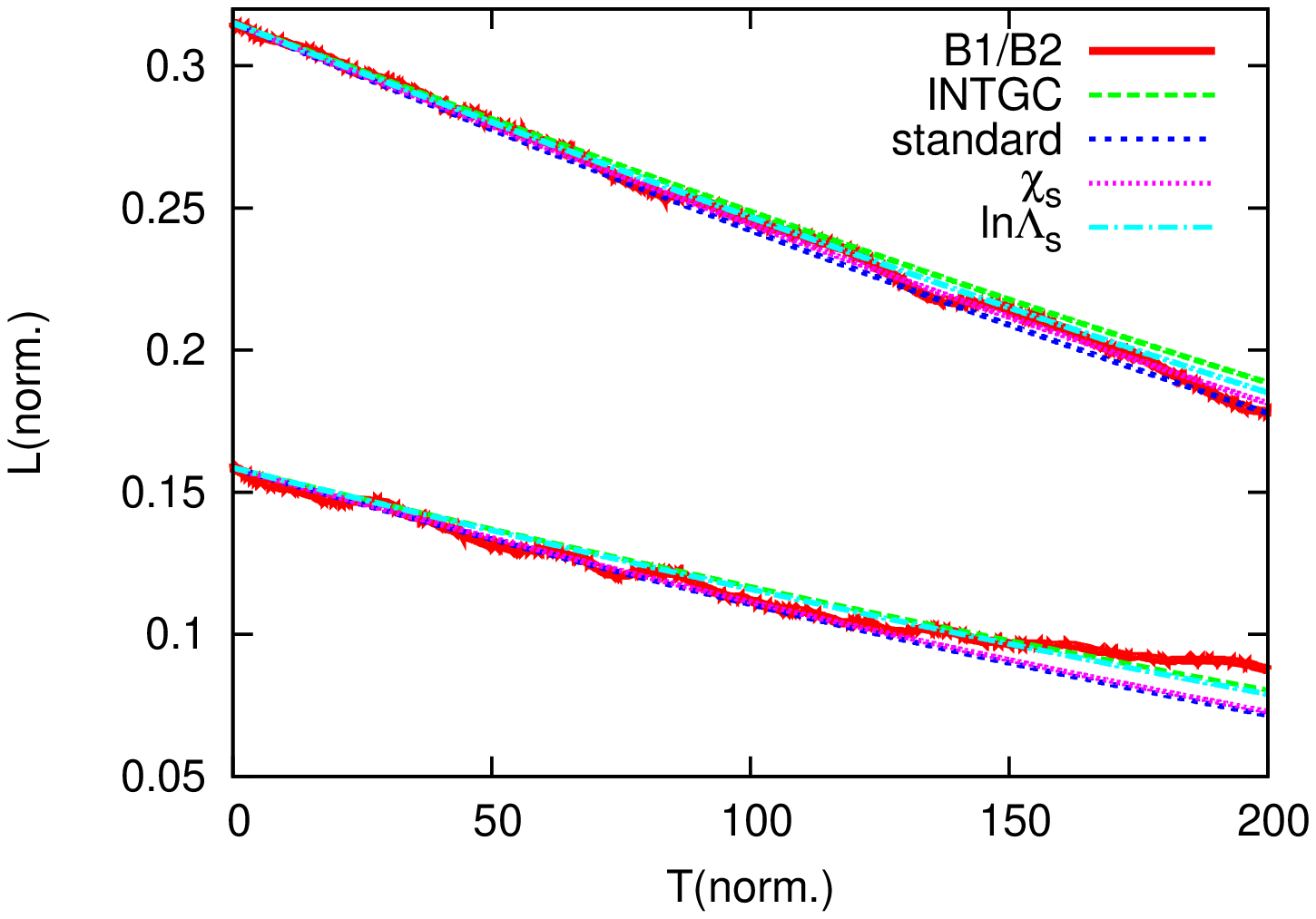}}
  }

\caption[]{Top Panel: Orbit evolution of run B1 in two time intervals (0,1) and and (100,101) compared to the semi-analytic predictions from {\sc intgc}.
Bottom panel: Angular momentum decay for runs B1 (top) and B2 (bottom) for the full simulated time compared to the semi-analytic predictions.
Label {\sc intgc} stands for the new $\ln\Lambda$ and $\chi$, 
'standard' for $\ln\Lambda_\mathrm{s}$ and $\chi_\mathrm{s}$, the other two for substituting only one parameter by the standard value. 
} \label{B12}
\end{figure}

In eccentric runs the additional effect of the  velocity dependence of $\chi(U)$ and also $a_{90}$ (eq. \ref{a90}) along the orbit occurs. We performed two runs B1 and B2 with different initial radii and velocities at apo-centre corresponding to eccentricities of $e=0.5,\,0.75$, respectively. The effective minimum impact parameter $a_{90}$ is well resolved along the orbits.
In the top panel of figure \ref{B12} the orbit of B1 is shown in two short time intervals of length $\Delta T=1$ in order to resolve individual revolutions. Since the orbital phase is very sensitive to the exact enclosed mass and apo- peri-centre positions, the cumulative phase shift after hundreds of orbits is expected. In the bottom panel the evolution in $L$ is shown for runs B1 and B2 for the full evolution time. The comparison with the semi-analytic results show very good agreement for run B1. The delay in the orbital evolution as in runs A1 --A3 does not occur. In run B2 there is a delay in the later phase. In contrast to run B1 the peri-centre passages of B2 suffer from low number statistics. The peri-centre distance decays from 0.015 at $T=0$ to 0.01 at $T=150$, where only 200 cusp particles are enclosed inside the orbit.

In the lower panel of figure \ref{B12} we show also the orbital evolution using the standard parameters. In the standard case with constant $\ln\Lambda_\mathrm{s}$ (equation \ref{loglams}) and $\chi_\mathrm{s}$ from a Gaussian velocity distribution are applied. The decay is slightly faster. In order to separate the effect of the new Coulomb logarithm and the correct distribution function we show also the orbital decay substituting only $\chi$ by $\chi_\mathrm{s}$ (dotted blue line) and only $\ln\Lambda$ by $\ln\Lambda_\mathrm{s}$ (dashed-dotted cyan line), respectively. The effect of the new Coulomb logarithm is larger, but still not significant in the Bahcall-Wolf cusp. 

\subsection{Hernquist cusp}\label{HE}

\begin{figure}
\centerline{
  \resizebox{0.98\hsize}{!}{\includegraphics[angle=270]{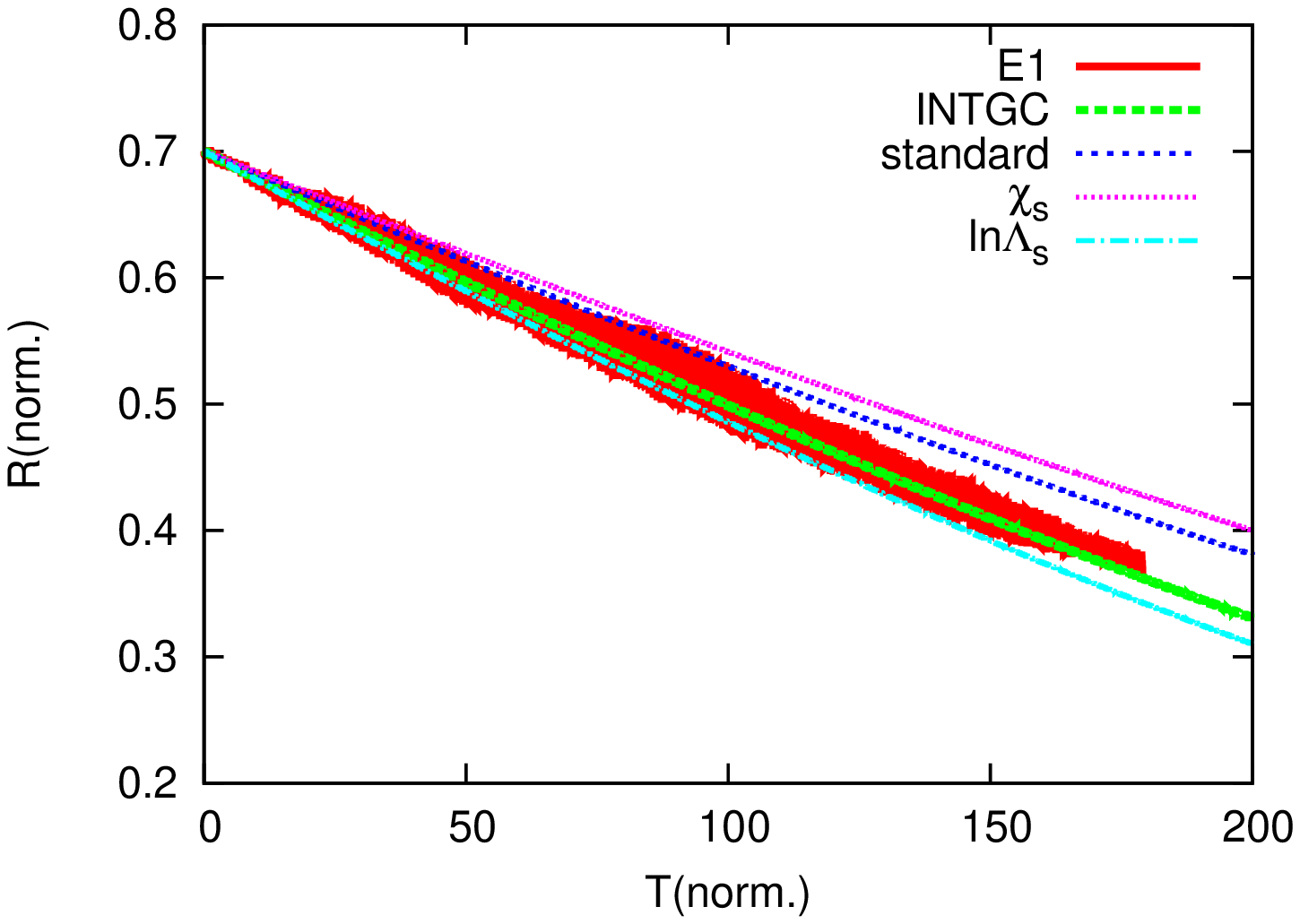}}
  }
\centerline{
  \resizebox{0.98\hsize}{!}{\includegraphics[angle=270]{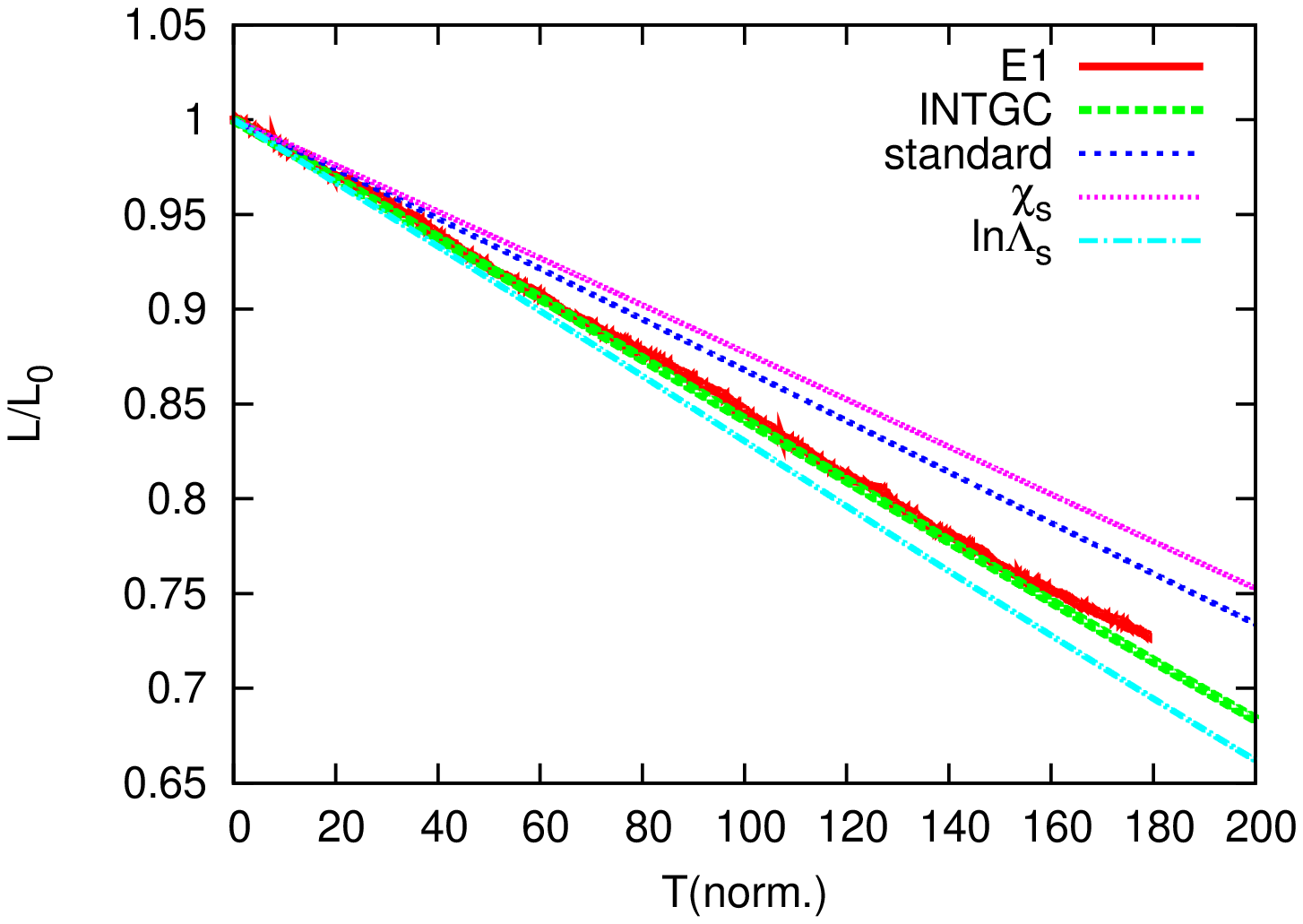}}
  }
\caption[]{
Orbital decay of a BH in the Hernquist cusp of the circular orbit E1 in 
$r(t)$ (top panel) and $L(t)/L_0$ (bottom panel). The $\phi$GRAPE simulation is compared to the semi-analytic results with {\sc intgc}.
Same notation as in figure \ref{B12}.
} \label{E1}
\end{figure}

\begin{figure}
\centerline{
  \resizebox{0.98\hsize}{!}{\includegraphics[angle=270]{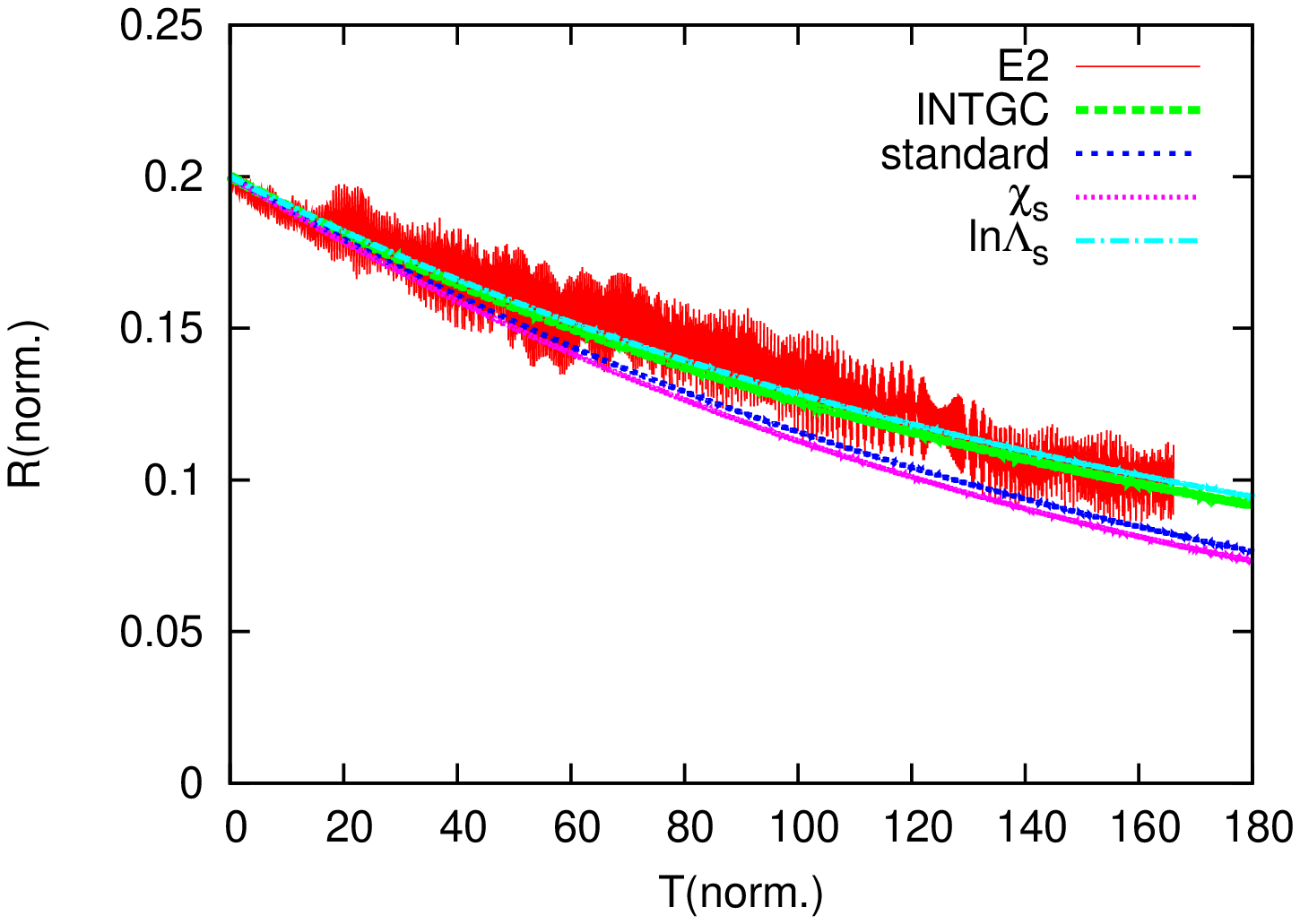}}
  }
\centerline{
  \resizebox{0.98\hsize}{!}{\includegraphics[angle=270]{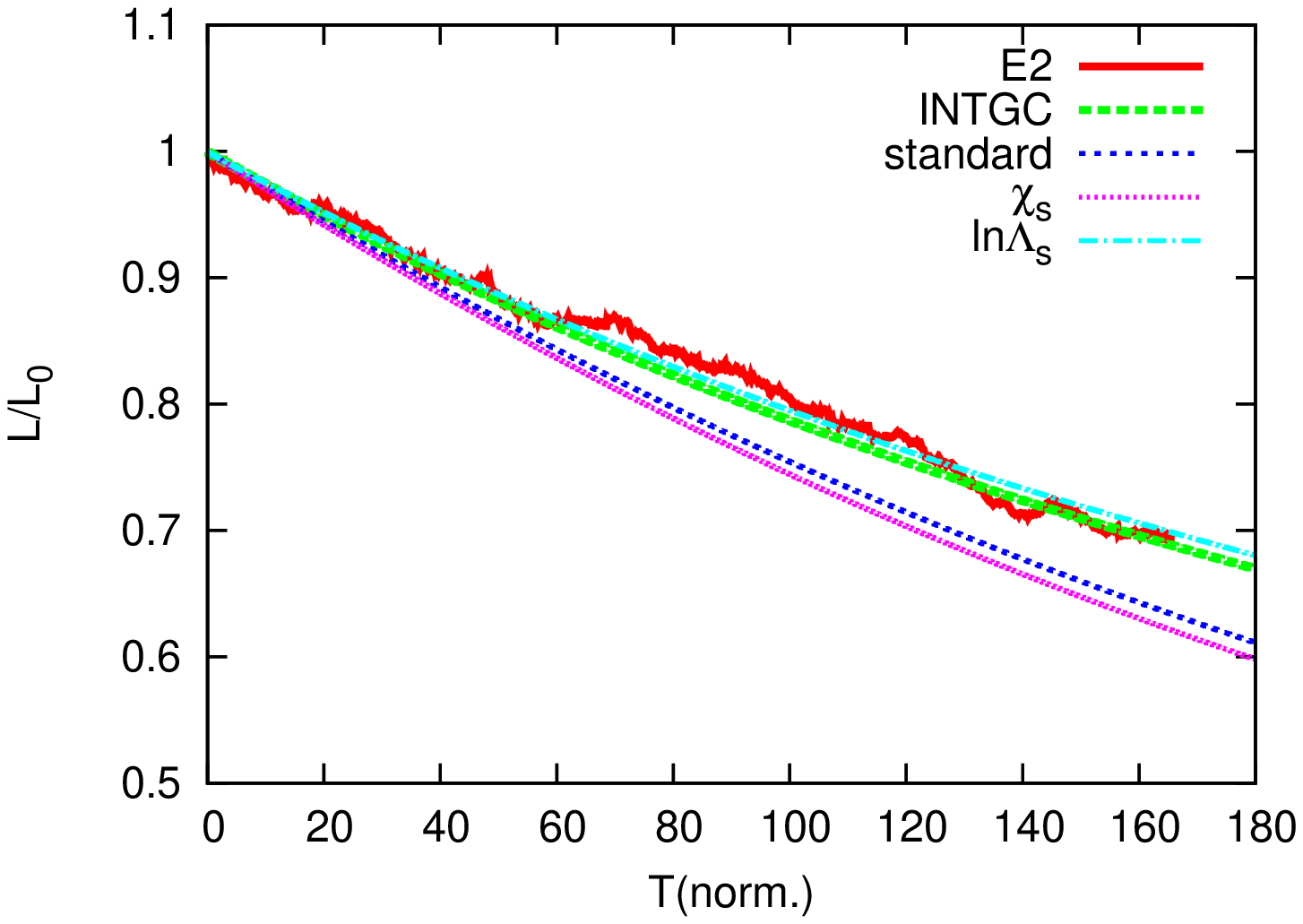}}
  }
\caption[]{
Same  as in figure \ref{E1} but for run E2 with higher resolution and closer in.
} \label{E2}
\end{figure}

We performed two circular runs E1 and E2 in the shallow Hernquist cusp to test the limit of our friction formula using the $\phi$GRAPE code. Run E1 is performed on the GRAPE cluster at ARI and E2 with larger $N$ on the GPU cluster of NAOC/CAS. The orbital decay is shown in figures \ref{E1} and  \ref{E2}. The velocity distribution function deviates significantly from the analytic limiting case of an idealised cusp (see figure \ref{figchiE}). The $\chi$ function is stable over the simulation time and depends only weakly on the distance to the central SMBH. Therefore we used for the semi-analytic simulations with {\sc Intgc} constant mean values for the circular orbits. The local scale length of run E1 entering the Coulomb logarithm is smaller than the distance to the centre, which is the limiting value, because $R_0=0.7$ is close to the scale radius. The orbital decays are very well reproduced in $R$ and in $L$. A comparison with the standard values $\ln\Lambda_\mathrm{s}$ and $\chi_\mathrm{s}$ show a significant deviation mainly due to the different $\chi$ function.


\subsection{Outskirts of the Plummer sphere}\label{skirts}

The background distribution is a self-gravitating Plummer sphere and the outskirts are described by eq. \ref{mplumapp}. We place the orbits of the circular run P1 and eccentric run P2 far outside the Plummer radius in order to be very close to the power law density with $\eta_0-3=-5$ and a Kepler potential according to eqs. \ref{mplumapp}. Since the number density is very small, we cannot use a PP code. Instead we use {\sc Superbox} with ten million particles. The parameters of all {\sc Superbox} runs are listed in table \ref{table:PM}. 

Due to the very long orbital decay time the density distribution evolves slowly.  In steep cusps the representation of the local density in the semi-analytic calculations is the most critical parameter. This affects the decay time scale via the local density and also the enclosed mass (eq. \ref{tau0}). In the Kepler limit the enclosed mass is given by the constant central mass $M_{\rm c}$. In lowest order the density evolution can be modelled by a linear expansion of the Plummer radius. The analytic solution of circular orbits is still given by eqs. \ref{ty1} and \ref{yt} with the substitution $t\to s(t)$ with $\nu=2$ (see eq.\ref{34t}).
By fitting the density profiles at different times we find $t_{\rm a}= 11.3$\, Gyr. 

\begin{figure}
\centerline{
  \resizebox{0.98\hsize}{!}{\includegraphics[angle=270]{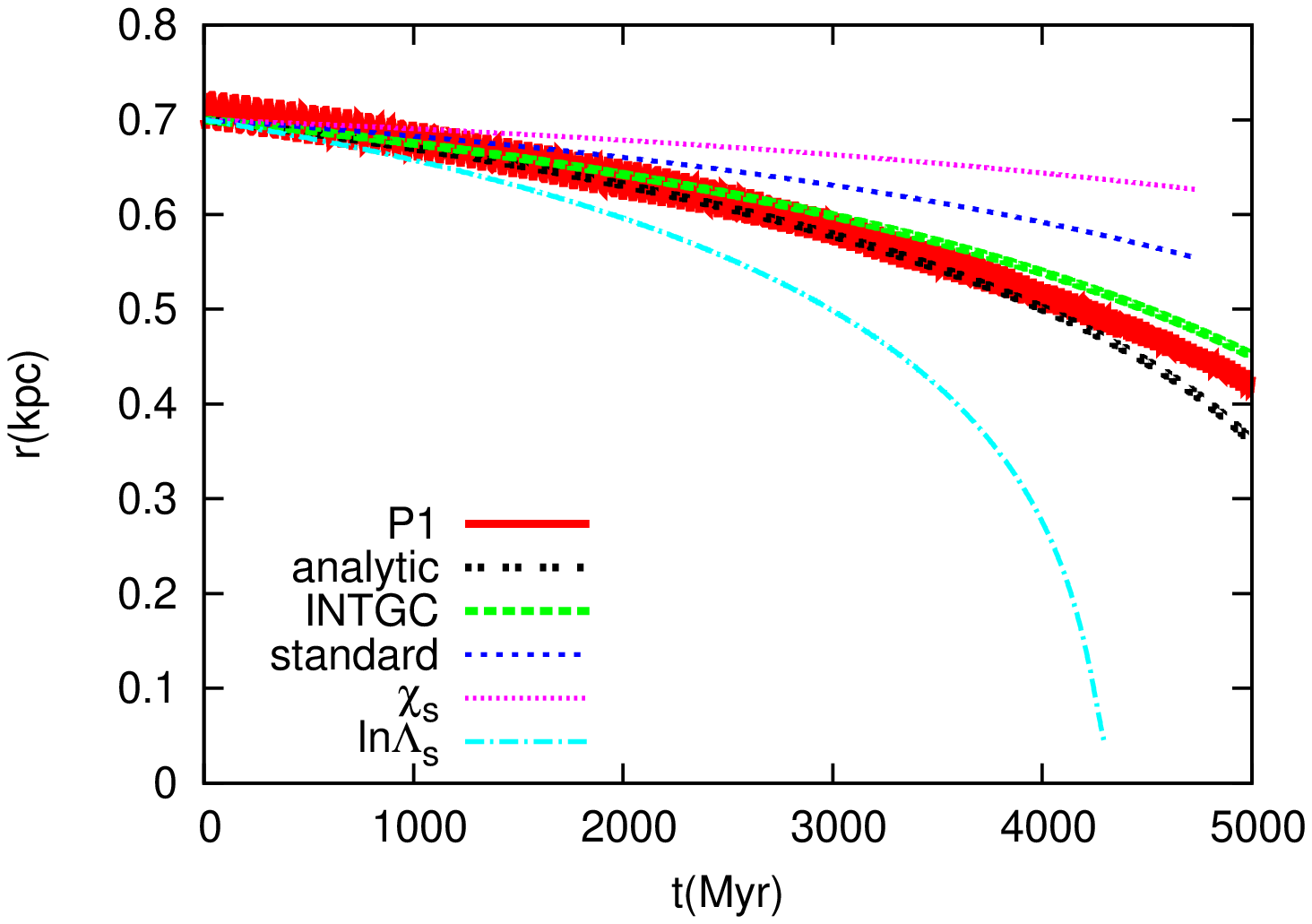}}
  }
\centerline{
  \resizebox{0.98\hsize}{!}{\includegraphics[angle=270]{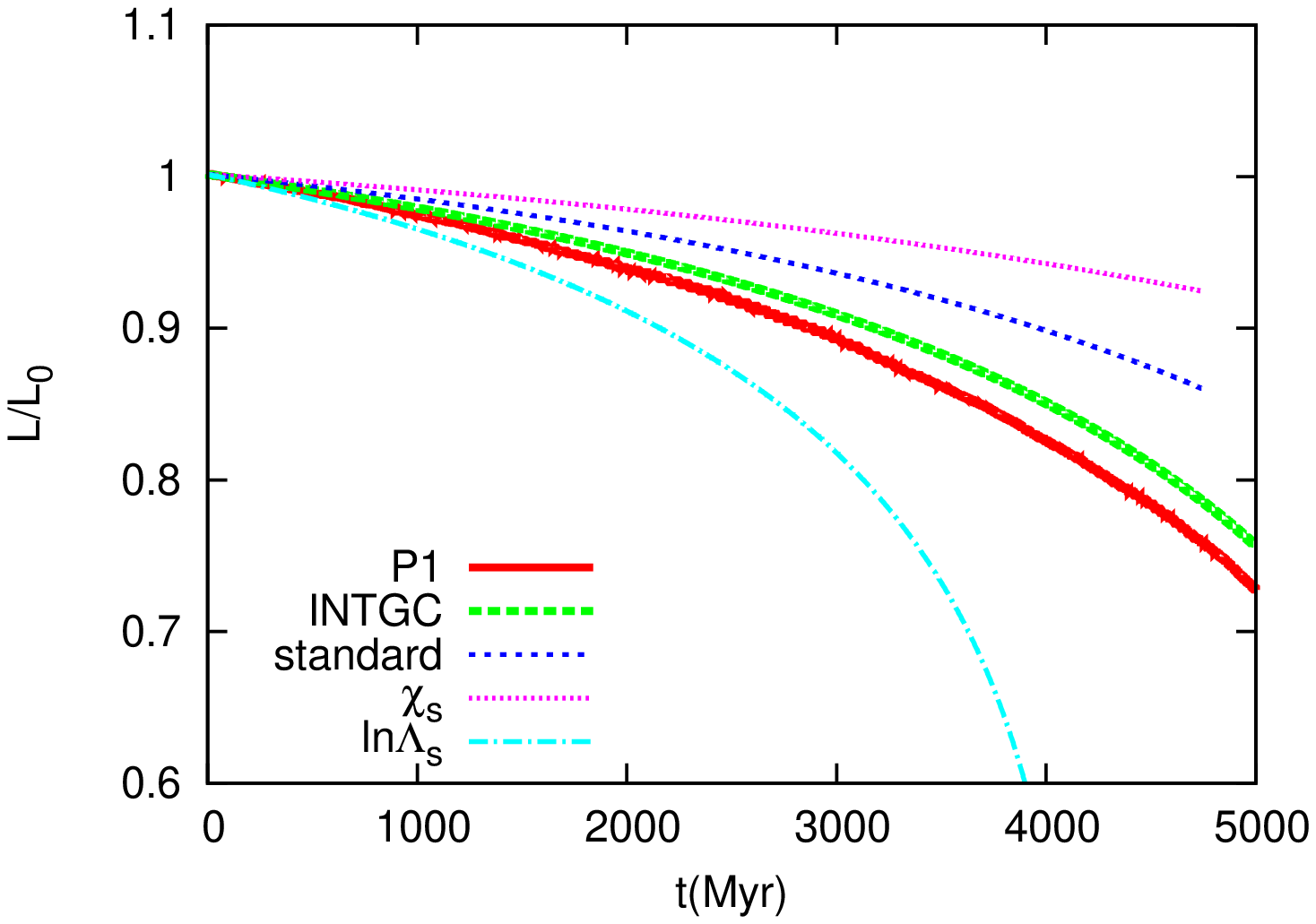}}
  }
\caption[]{
Orbital decay of a BH in the outskirts of a self-gravitating Plummer sphere for the {\sc Superbox} runs of the circular orbit P1
in $r(t)$ (top panel) and in $L(t)/L_0$ (bottom panel). The {\sc Superbox} run is compared to the semi-analytic results with {\sc intgc}. Same notation as in figure \ref{B12}. The top panel shows additionally the analytic approximation from equation \ref{ty1}.
} \label{Pck}
\end{figure}

\begin{figure}
\centerline{
  \resizebox{0.98\hsize}{!}{\includegraphics[angle=270]{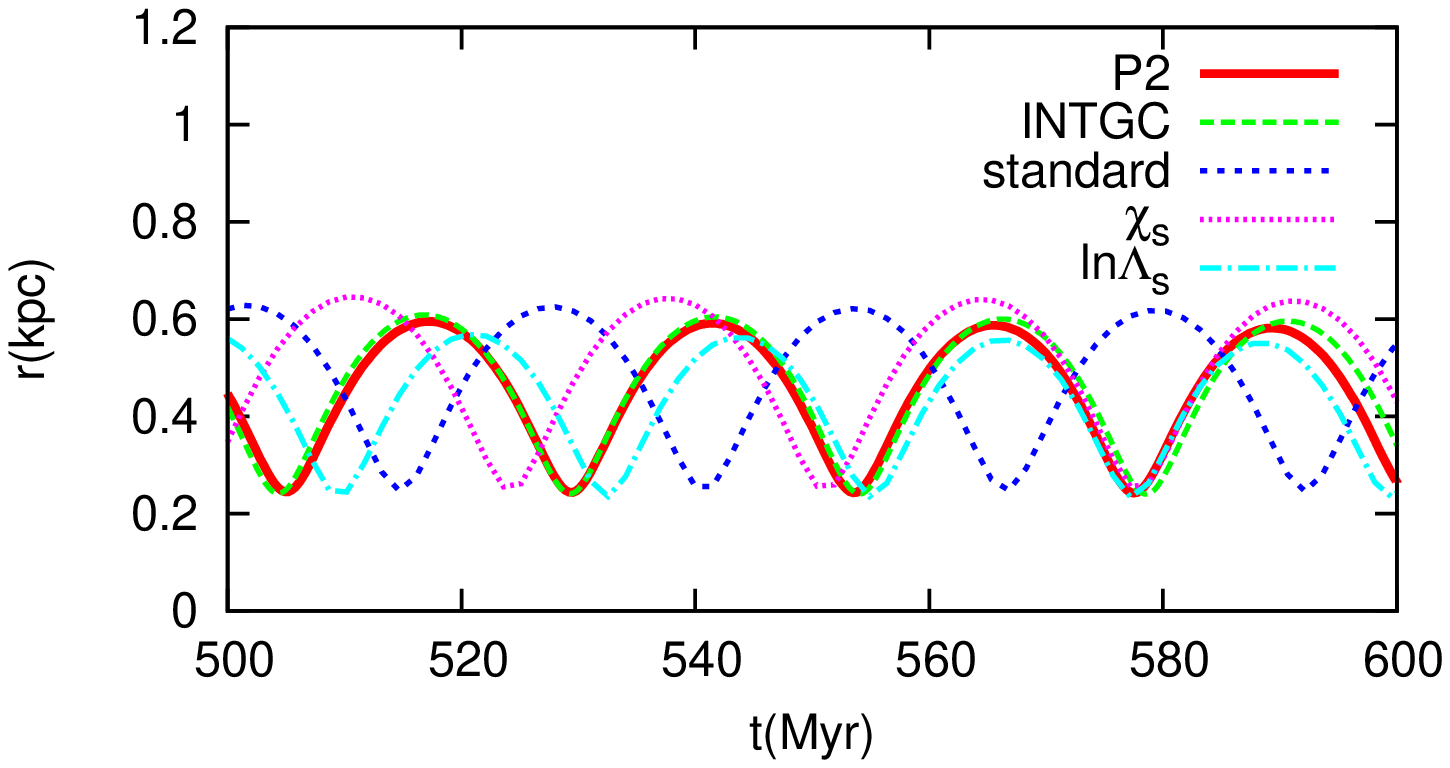}}
  }
\centerline{
  \resizebox{0.98\hsize}{!}{\includegraphics[angle=270]{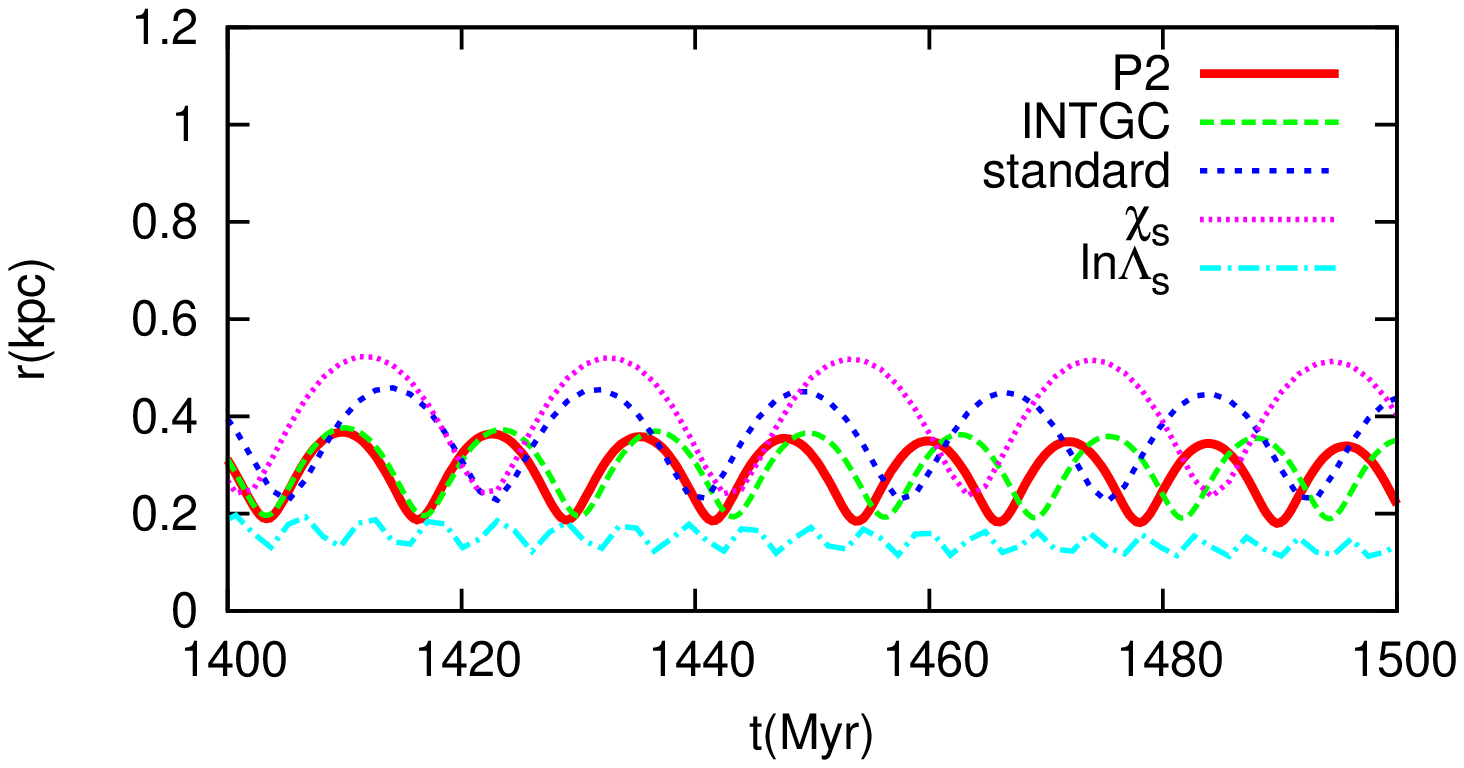}}
  }
\centerline{
  \resizebox{0.98\hsize}{!}{\includegraphics[angle=270]{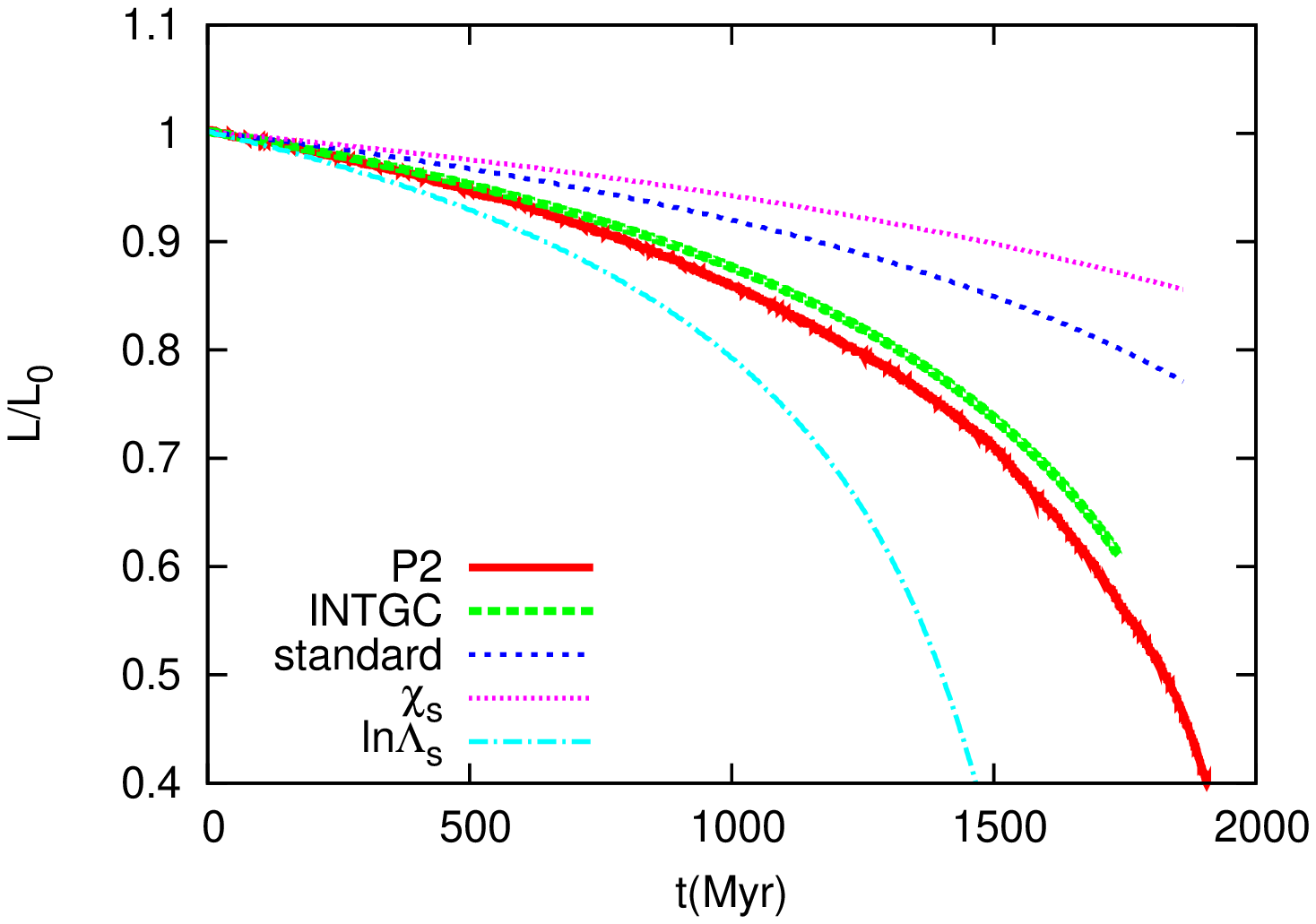}}
  }
\caption[]{
Same as in figure \ref{Pck} but for the eccentric orbit P2. The top panels show $r(t)$ for two time intervals with higher resolution and the bottom panel shows $L_z(t)$ for the full calculation. Same notation as in figure \ref{B12}.
} \label{Pek}
\end{figure}

The numerical results of the circular run P1 are compared in figure \ref{Pck} to the analytic solution of
eq. \ref{ty1} in the Kepler limit using an expanding Plummer profile. For the calculations with {\sc intgc} we correct additionally for the decreasing enclosed mass. Since in these simulations $a_{90}$ is not resolved by the grid cell size $d_{\rm c}$, we determined the best value for the minimum impact parameter to be $b_{\rm min}=d_{\rm c}/2$. The same factor 1/2 is used for all other {\sc Superbox} runs.  In the lower panel of figure \ref{Pck} the orbits with standard Coulomb logarithm $\ln\Lambda_\mathrm{s}$
and standard $\chi_\mathrm{s}$ as in figure \ref{B12} are shown for comparison.
In figure \ref{Pek} the orbital evolution is shown for the eccentric run P2.

In both cases we find a satisfying agreement of the numerical results and our analytic predictions. With the standard formula the decay is significantly delayed (figures \ref{Pck} and \ref{Pek}). Here the corrections due to the correct $\chi$ function and the new Coulomb logarithm have a different sign and cancel each other partly. In the circular run both corrections are equally important. In the eccentric run the correction by the $\chi$ function dominates, but the additional correction due to the new $\ln\Lambda$ is also significant.
\begin{table*}
\caption{Parameters of the SUPERBOX runs.}
\begin{tabular}{c c c c c c c c c c c c c c c c c}
\hline
Run & $M_{\rm tot}$ & $M_0$  & $a$  & $r_{\rm cut}$ & $N$ & $m$ & $dt$ & $d_{\rm c}$ & $M_{\rm bh}$ & $R_{0}$ & $V_\mathrm{c,0}$ & $V_0/V_\mathrm{c,0}$ &
 $a_{90}$ & $\ln\Lambda_0$\\
& $10^9\,\msol$ & $10^9\,\msol$ & kpc & kpc & $10^7$ && Myr & pc & $10^6\,\msol$ & kpc & km/s & & pc &\\
\hline
P1 & 1.0 & 0.970 & 0.1 & 10 & 1 & 7 & 0.3 & 16.1 & 1.0 & 0.7 & 77.2 & 1.0  
& 0.538 & 2.87 \\
P2 & 1.0 & 0.970 & 0.1 & 10 & 1 & 7 & 0.3 & 16.1 & 1.0 & 0.7 & 77.2 & 0.7  
& 0.869 & 2.87 \\
D1 & 1.0 & 0.826 & 0.1 & 1  & 1 & 6 & 0.1 & 16.7 & 1.0 & 0.4 & 94.2 & 1.0  
& 0.186 & 2.48 \\
D2 & 1.0 & 0.826 & 0.1 & 1  & 1 & 6 & 0.1 & 16.7 & 1.0 & 0.4 & 94.2 & 0.7  
& 0.207 & 2.48 \\
D3 & 1.0 & 0.128 & 1.0 & 10 & 1 & 8 & 0.1 & 2.78 & 1.0 & 0.3 & 42.8 & 1.0  
& 1.24 & 4.75 \\
D4 & 1.0 & 0.128 & 1.0 & 10 & 1 & 8 & 0.3 & 2.78 & 1.0 & 0.3 & 42.8 & 0.7  
& 1.70 & 4.44 \\
H1 & 1.0 & 0.197 & 1.0 & 10 & 1 & 7 & 1.0 & 16.1 & 1.0 & 0.7 & 34.8 & 1.0  
& 6.0 & 3.68\\
H2 & 1.0 & 0.197 & 1.0 & 10 & 1 & 7 & 0.3 & 16.1 & 1.0 & 0.7 & 34.8 & 0.56  
& 9.7 & 3.46\\
H3 & 1.0 & 0.052 & 1.0 & 10 & 1 & 8 & 0.1 & 4.76 & 1.0 & 0.3 & 30.3 & 1.0  
& 2.13 & 3.64\\
H4 & 1.0 & 0.052 & 1.0 & 10 & 1 & 8 & 0.1 & 4.76 & 1.0 & 0.3 & 30.3 & 0.9  
& 2.39 & 3.53\\
\hline
\end{tabular}

\flushleft{{\it Note.} All quantities are given in physical units. $M_{\rm tot}$ is the total mass inside the cutoff radius $r_{\rm cut}$, $M_0$ is the enclosed mass at the initial distance $R_0$,  $a$ is the scale radius, $N$ is the particle number, $2^m$ is the number of cells per dimension in each grid, and $d_{\rm c}$ is the cell size of the middle grid. The initial value of $a_{90}$ from Eq. \ref{a90} is calculated for the exact model.}
\label{table:PM}
\end{table*}

\subsection{Outskirts of Dehnen models}\label{Deskirts}

The background distribution is a self-gravitating Dehnen model with $\eta=1.5$ in the inner cusp. We place the orbits of the circular run D1 and the eccentric run D2 (table \ref{table:PM}) far outside the scale radius in order to be very close to the power law density with $\eta_0-3=-4$ and a Kepler potential according to eqs. \ref{mdehnapp}. 

\begin{figure}
\centerline{
  \resizebox{0.98\hsize}{!}{\includegraphics[angle=270]{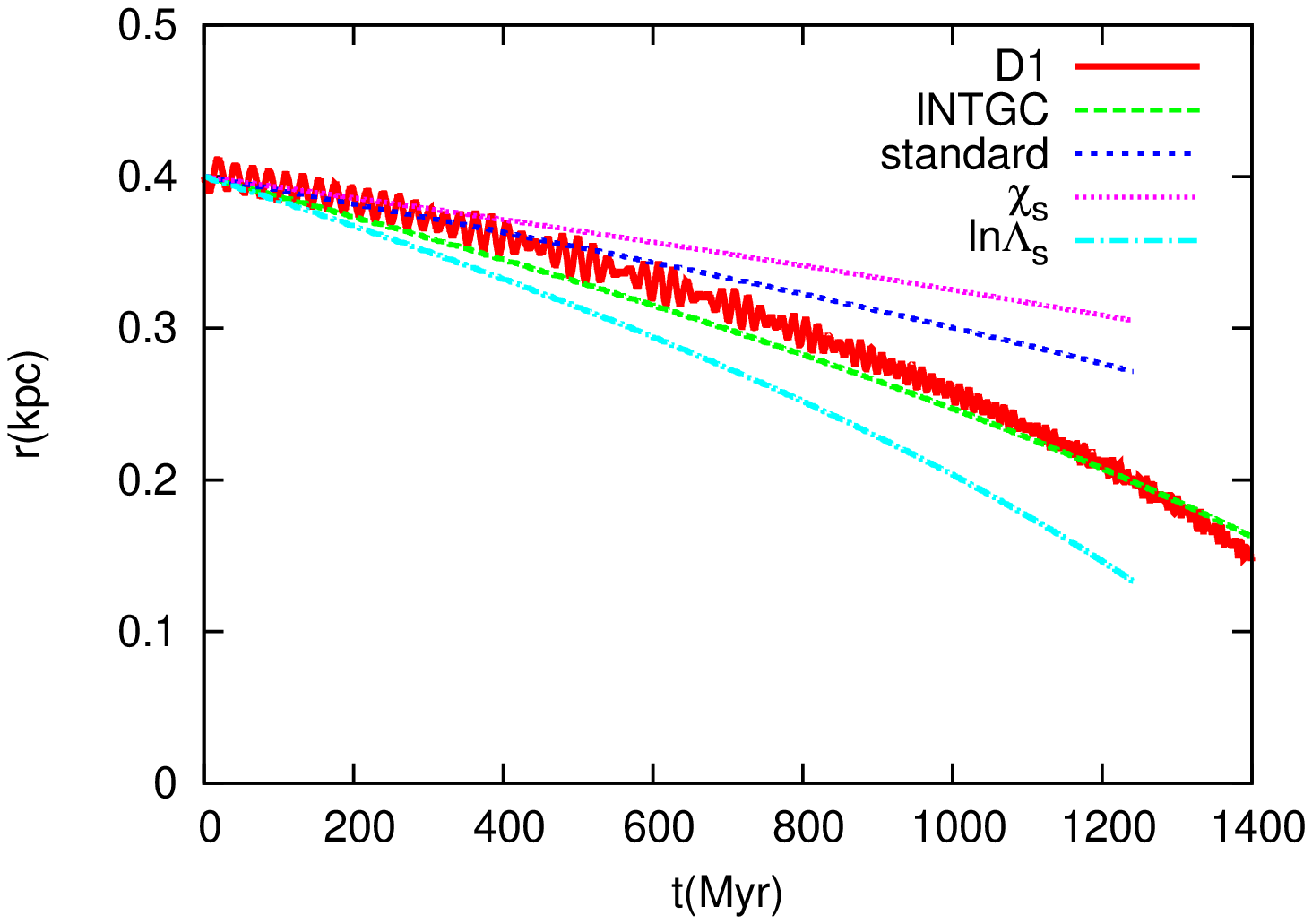}}
  }
\centerline{
  \resizebox{0.98\hsize}{!}{\includegraphics[angle=270]{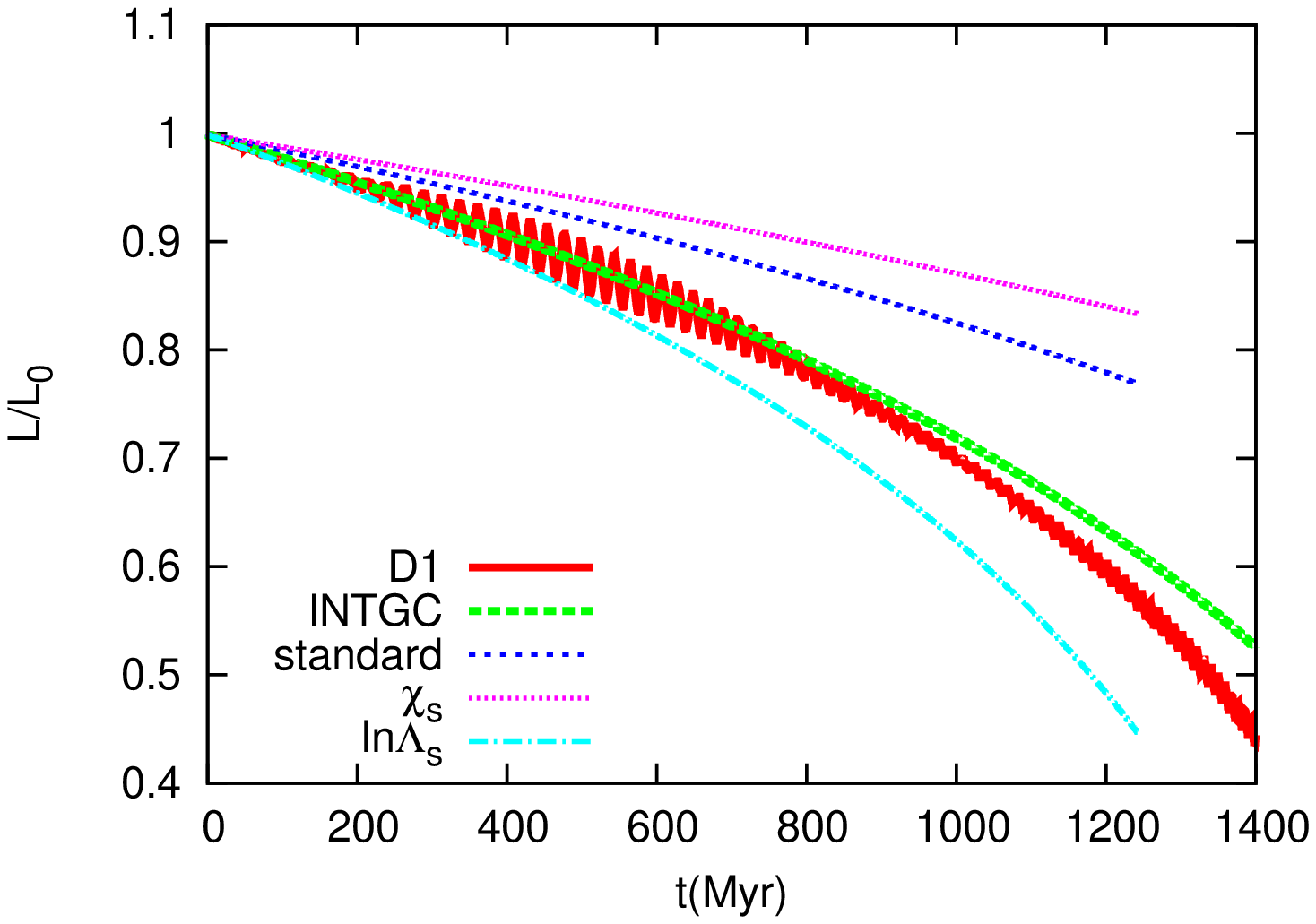}}
  }
\caption[]{
Same as in figure \ref{Pck} but for the orbit D1 in the outskirts of a Dehnen model. Same notation as in figure \ref{B12}.
} \label{Dck}
\end{figure}

\begin{figure}
\centerline{
  \resizebox{0.98\hsize}{!}{\includegraphics[angle=270]{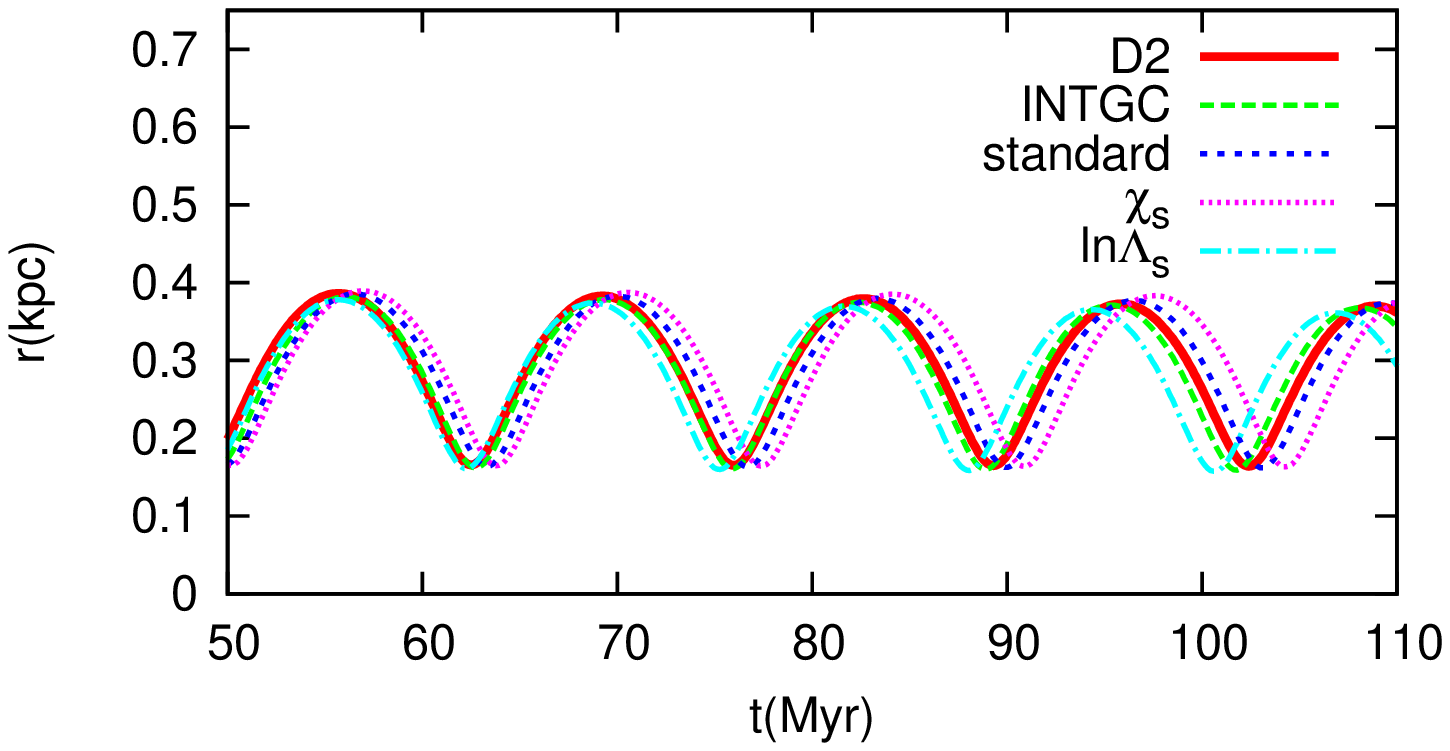}}
  }
\centerline{
  \resizebox{0.98\hsize}{!}{\includegraphics[angle=270]{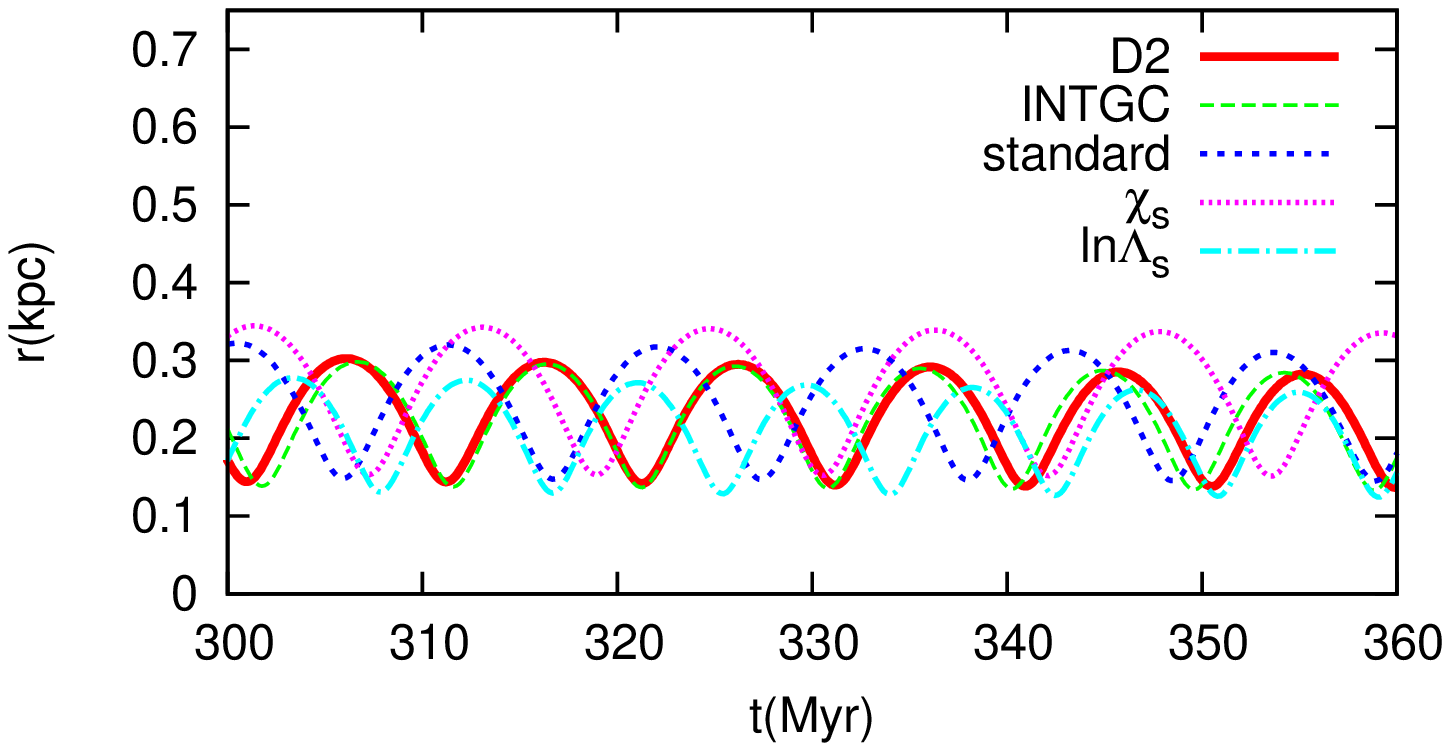}}
  }
\centerline{
  \resizebox{0.98\hsize}{!}{\includegraphics[angle=270]{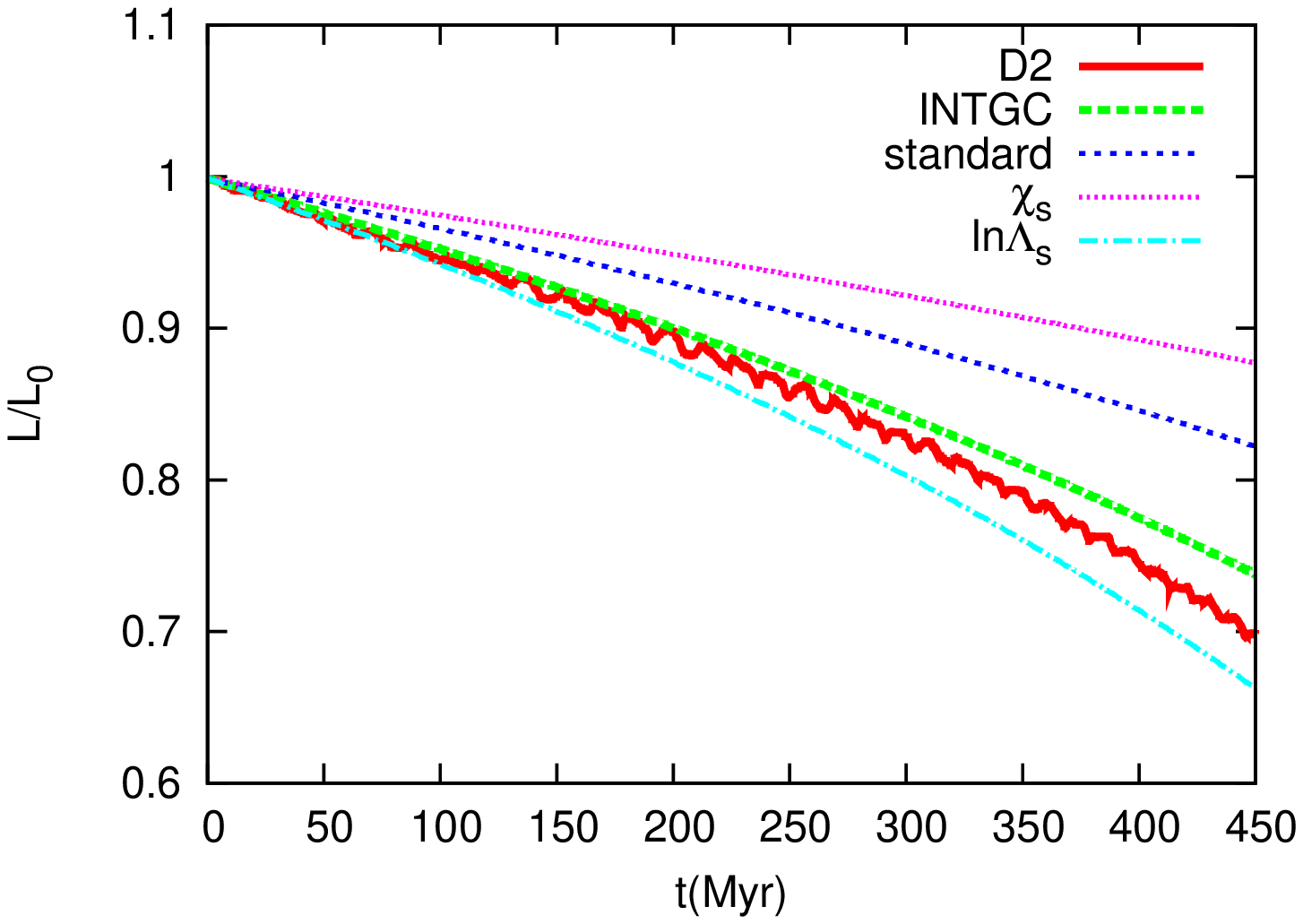}}
  }
\caption[]{
Same as in figure \ref{Dck} but for the eccentric orbit D2. Same notation as in figure \ref{B12}.
} \label{Dek}
\end{figure}

The numerical results are compared in figures \ref{Dck} and \ref{Dek} with the semi-analytic calculations using {\sc intgc}. In the numerical simulations the inner cusp becomes shallower and shrinks considerably leading to a decreasing enclosed mass and increasing density in the outer parts. We correct for that in {\sc intgc}. In these simulations $a_{90}$ is not resolved as in the Plummer case. The numerical results are in good agreement with the analytic predictions using the same minimum impact parameter $b_{\rm min}=d_{\rm c}/2$. The orbital resonances seen in figure \ref{Dck} are connected to a motion of the density centre of the cusp, which is used as the origin of the coordinate system. Therefore the angular momentum of the orbit oscillates due to the accelerated zero point.

\subsection{Self gravitating cusps}

 Here we study the orbital decay of  a BH in self-gravitating cusps of Dehnen models using {\sc Superbox} without a central SMBH. We performed two runs (one circular and one eccentric) for a Hernquist ($\eta=2$) and a Dehnen-1.5 ($\eta=1.5$) model each. For the semi-analytic calculations we use again the same minimum impact parameter $b_{\rm min}=d_{\rm c}/2$ in Eq. \ref{loglam}.

\subsubsection{Dehnen-1.5 model}

We studied the decay of a circular and an eccentric orbit in the inner cusp of a
Dehnen model with $\eta=1.5$ (D3 and D4 in table \ref{table:PM}). A comparison of $d_{\rm c}/2$ and $a_{90}$ shows that
the close encounters are marginally resolved. For this reason and because the transition of the inner and outer power law regimes is very wide, the analytic approximation of a self-gravitating cusp shows systematic deviations. 
With {\sc intgc} we find nevertheless a very good match to the {\sc Superbox} results of the circular run D3 (see Fig. \ref{Dck}) and for the eccentric run D4 (see Fig. \ref{Dek}). In the upper panel of Fig. \ref{Dck} we show also the circular run with a larger time-step in {\sc Superbox}, where for the close encounters with impact parameter comparable to the cell length $d_{\rm c}$ the motion of the perturber are not resolved in time. This leads to a larger effective minimum impact parameter and a slower orbital decay.
\begin{figure}
\centerline{
  \resizebox{0.98\hsize}{!}{\includegraphics[angle=270]{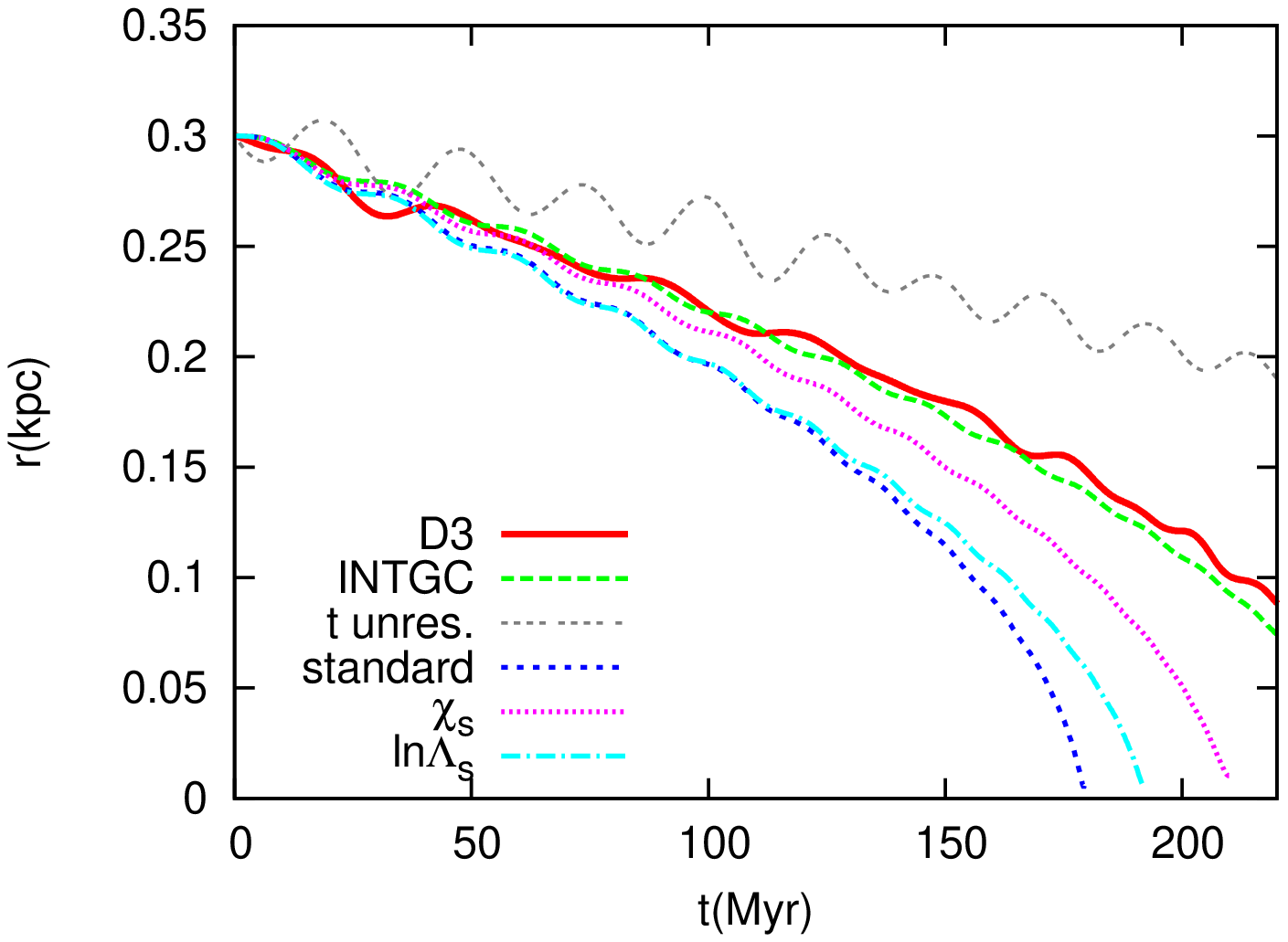}}
  }
\centerline{
  \resizebox{0.98\hsize}{!}{\includegraphics[angle=270]{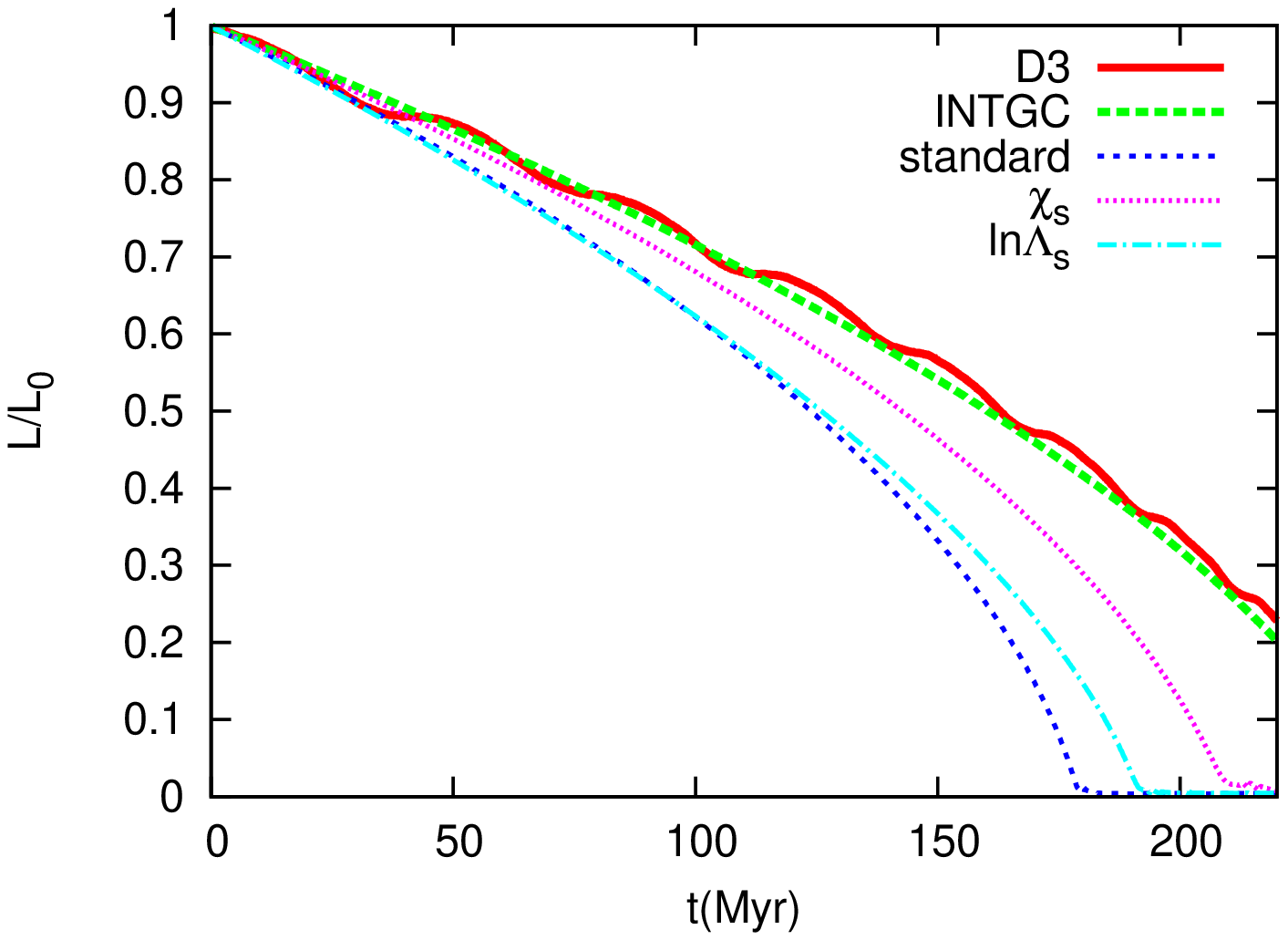}}
  }
\caption[]{
Orbital decay in distance (top panel) and angular momentum (bottom panel) for the circular orbit D3 in the self-gravitating inner cusp of the Dehnen-1.5 model. Same notation as in figure \ref{B12}. In the top panel there is additionally the orbit with insufficient time resolution shown.
} \label{Dcs}
\end{figure}

\begin{figure}
\centerline{
  \resizebox{0.98\hsize}{!}{\includegraphics[angle=270]{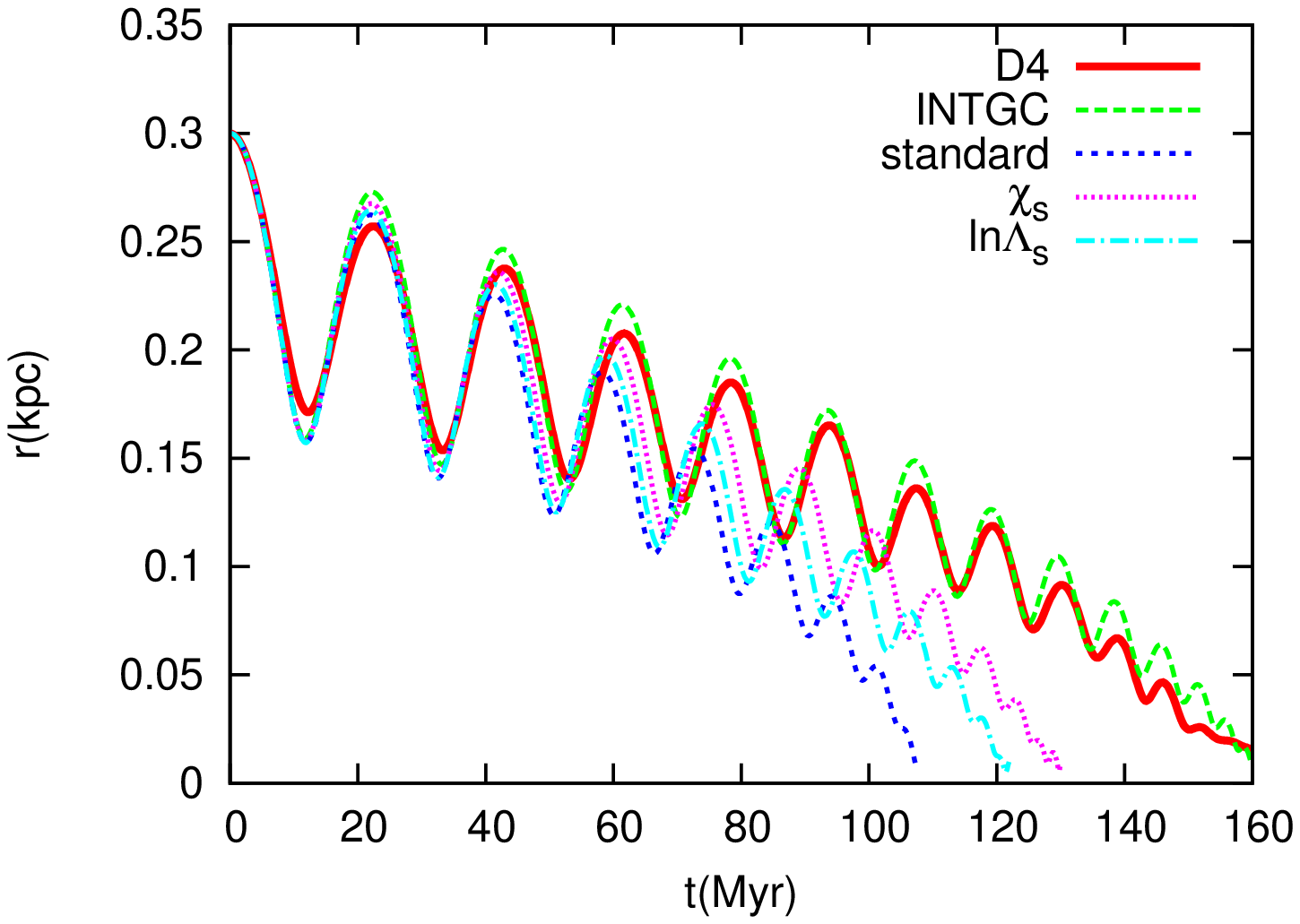}}
  }
\centerline{
  \resizebox{0.98\hsize}{!}{\includegraphics[angle=270]{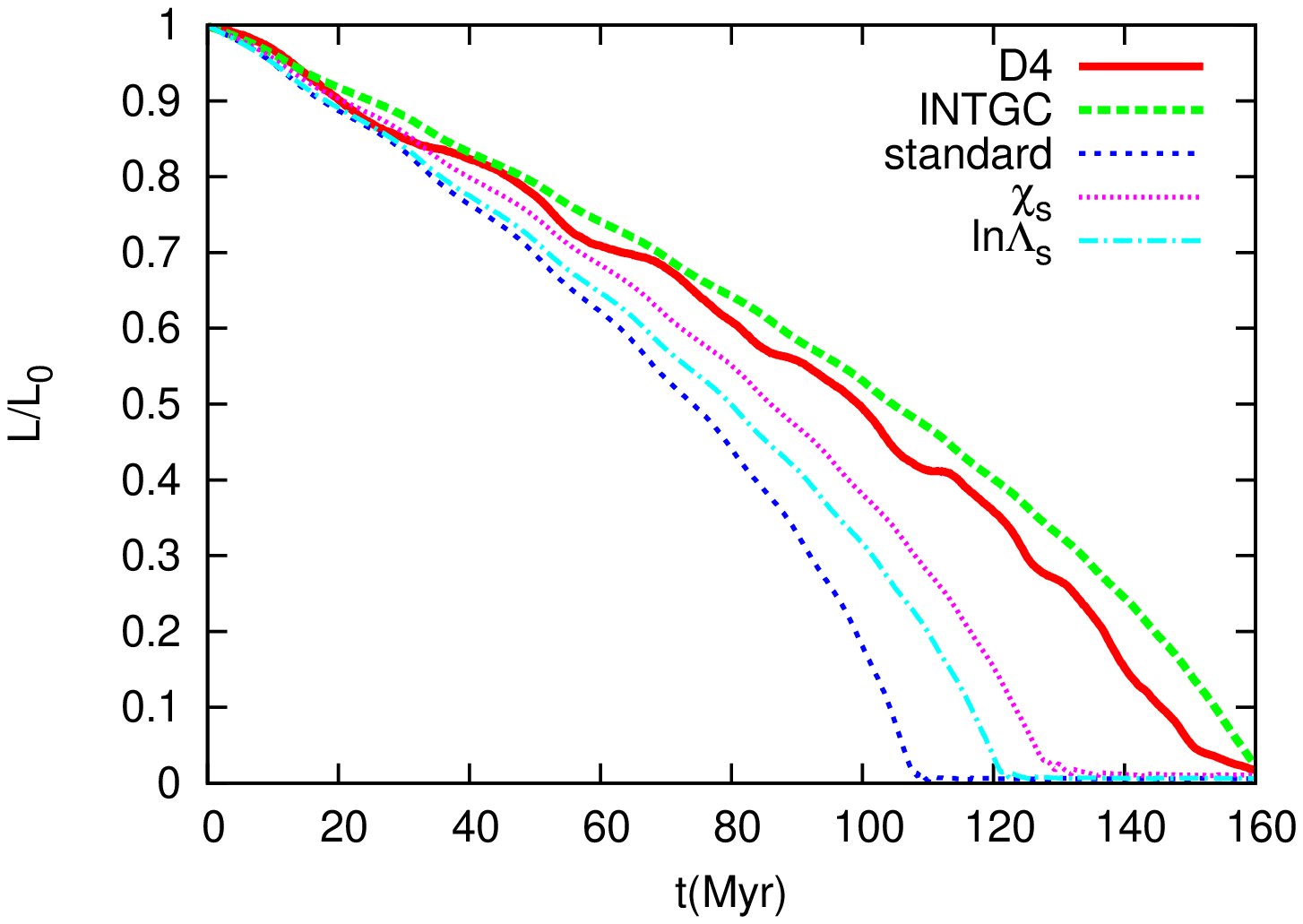}}
  }
\caption[]{
Same as in figure \ref{Dcs} but for the eccentric orbit D4.
} \label{Des}
\end{figure}
 
\subsubsection{Hernquist model}

\begin{figure}
\centerline{
  \resizebox{0.98\hsize}{!}{\includegraphics[angle=270]{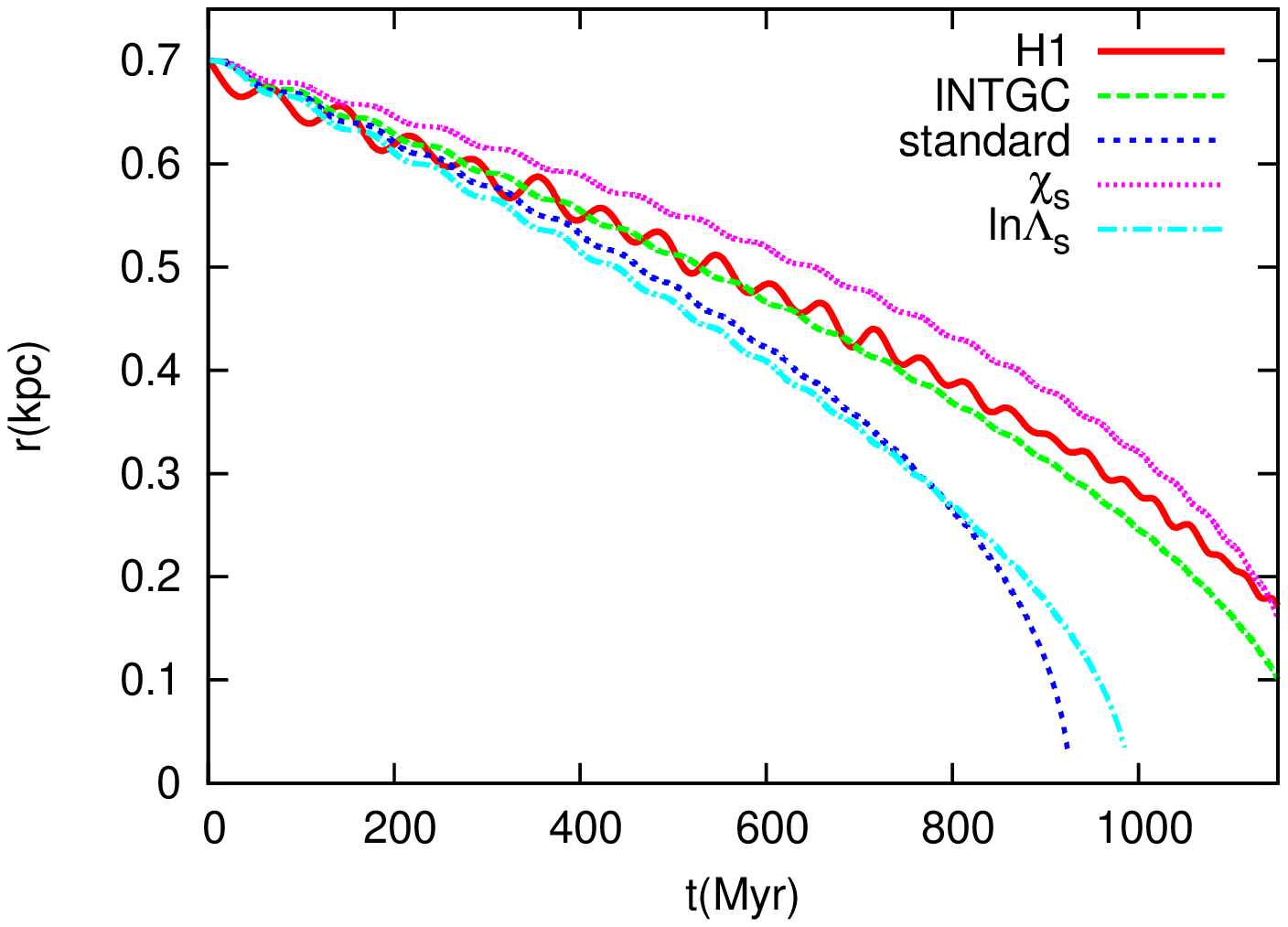}}
  }
\centerline{
  \resizebox{0.98\hsize}{!}{\includegraphics[angle=270]{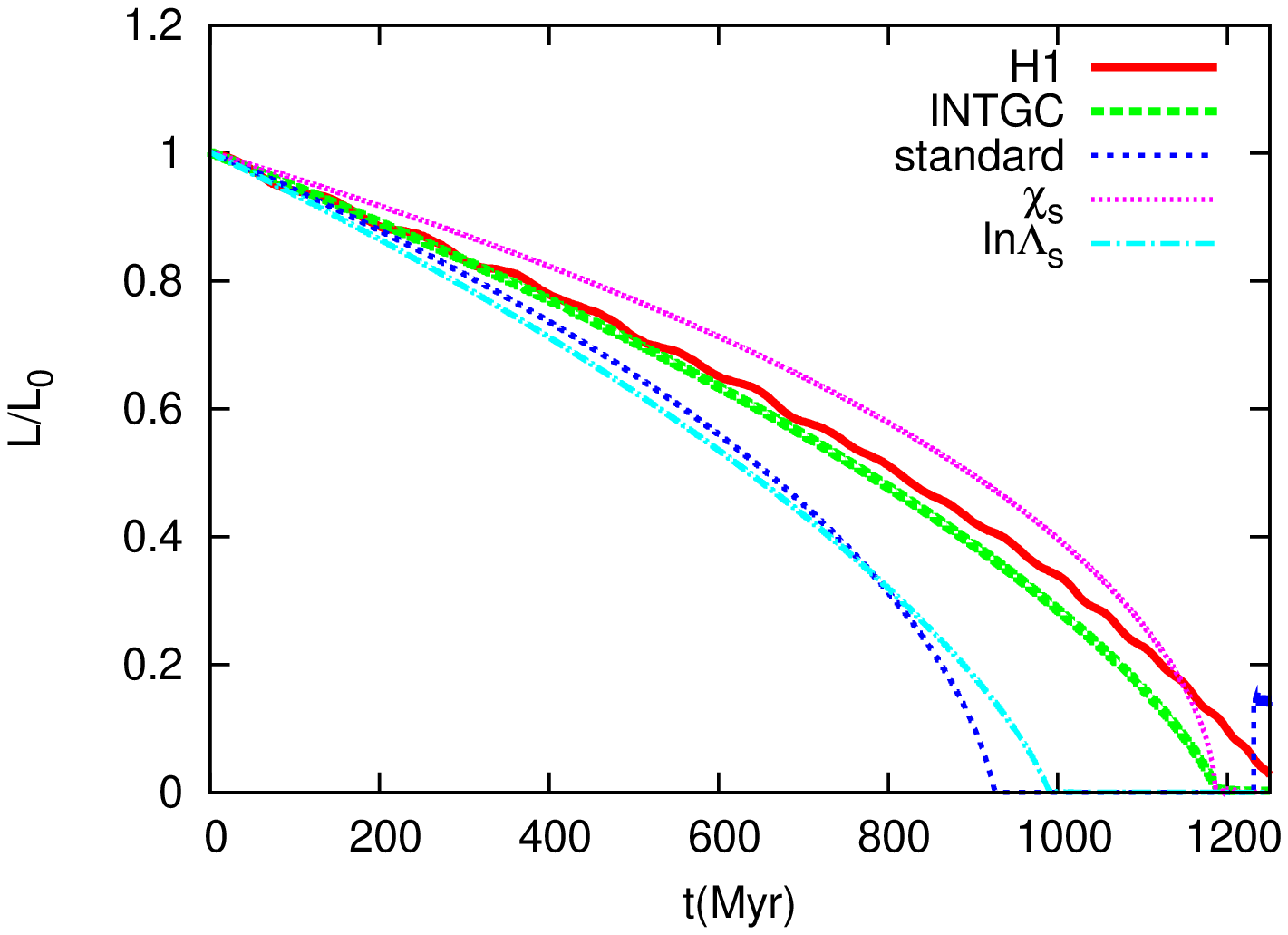}}
  }
\caption[]{
Same as in Fig. \ref{Dcs} for the circular orbit H1 in the self-gravitating inner cusp of the Hernquist model.
} \label{Hcs}
\end{figure}

\begin{figure}
\centerline{
  \resizebox{0.98\hsize}{!}{\includegraphics[angle=270]{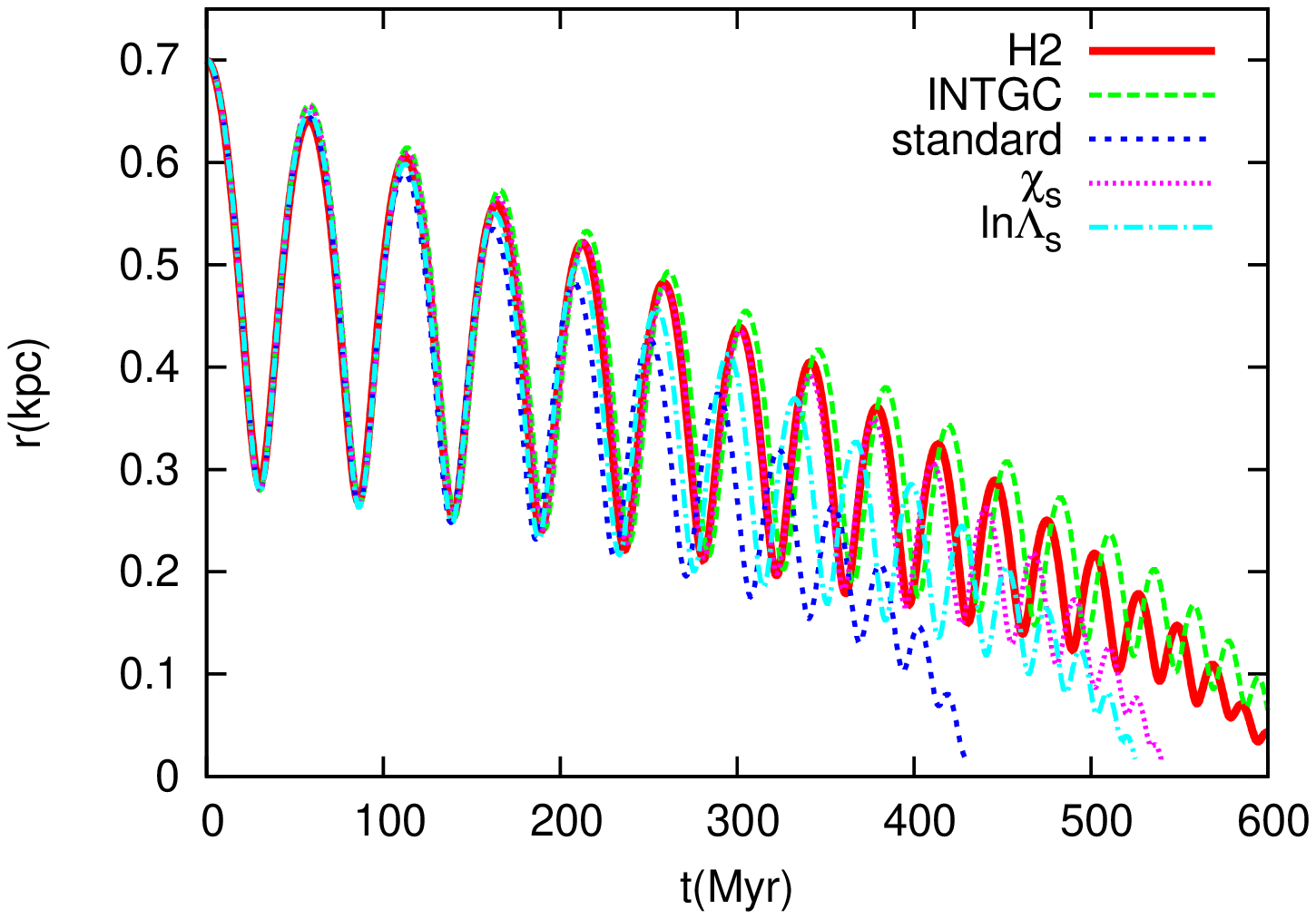}}
  }
\centerline{
  \resizebox{0.98\hsize}{!}{\includegraphics[angle=270]{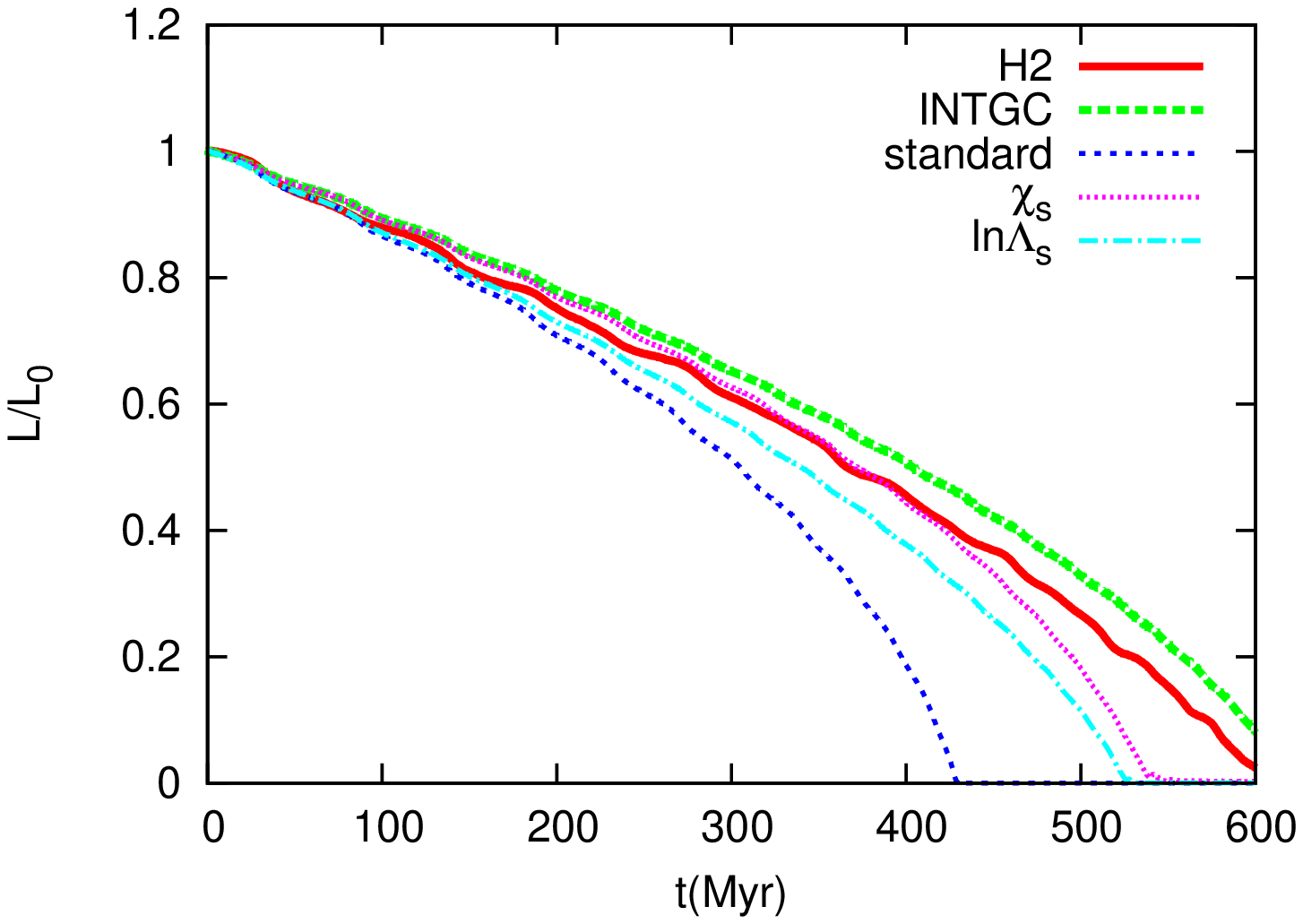}}
  }
\caption[]{
Same as in Fig. \ref{Hcs} for the eccentric orbit H2.
} \label{Hes}
\end{figure}

\begin{figure}
\centerline{
  \resizebox{0.98\hsize}{!}{\includegraphics[angle=270]{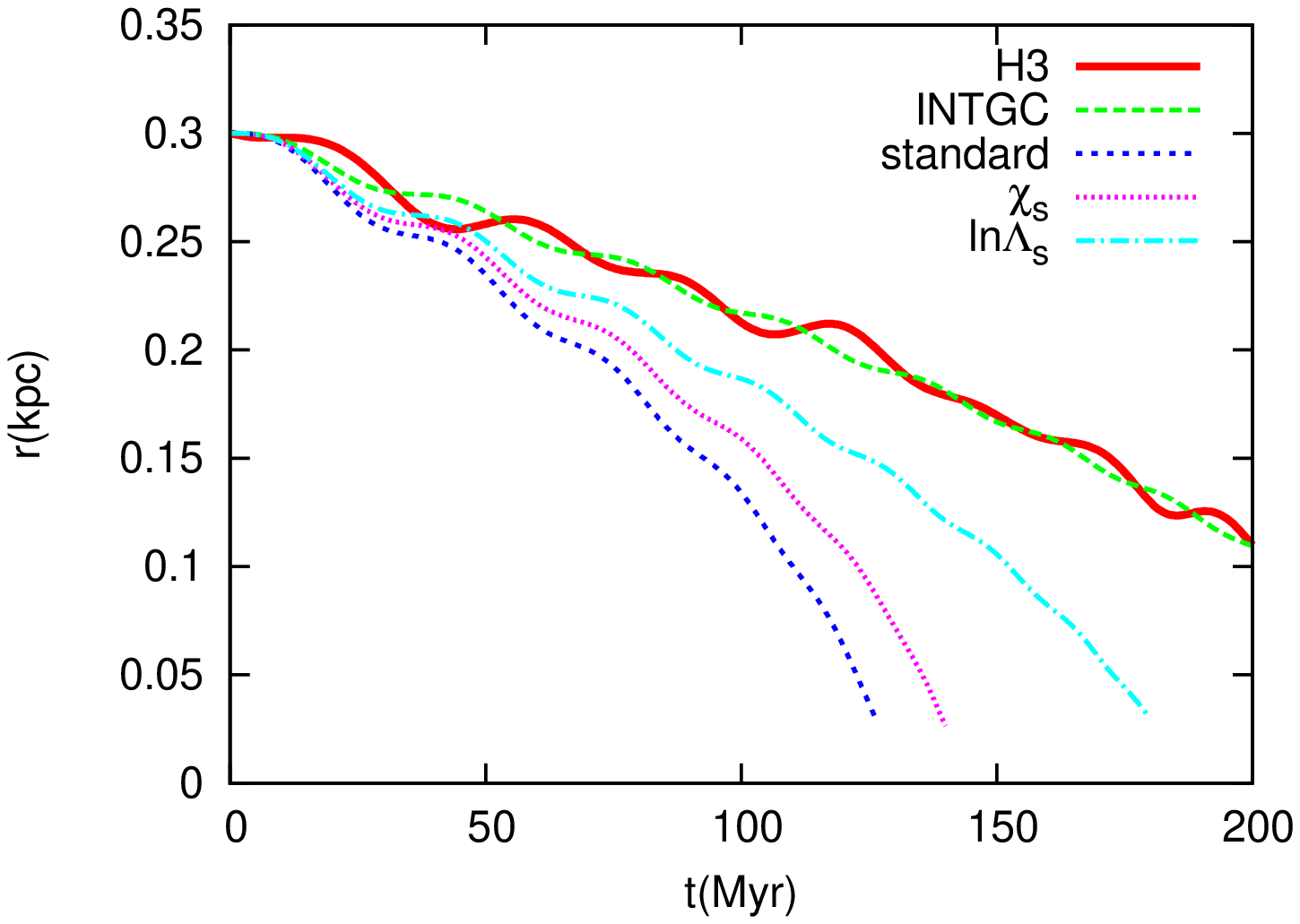}}
  }
\centerline{
  \resizebox{0.98\hsize}{!}{\includegraphics[angle=270]{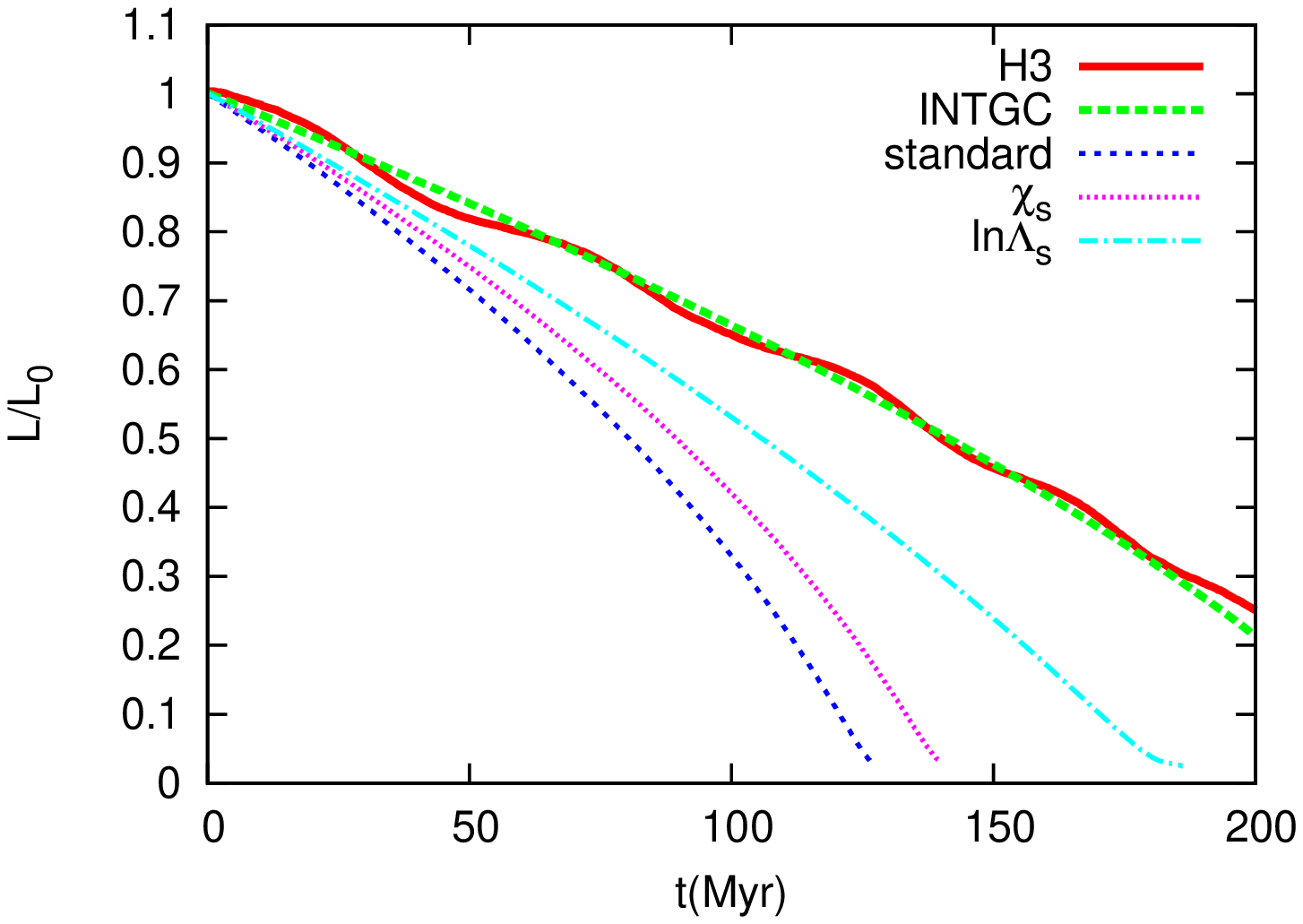}}
  }
\caption[]{
Same as in Fig. \ref{Hcs} for the circular orbit H3 with resolved $a_{90}$.
} \label{H3s}
\end{figure}

\begin{figure}
\centerline{
  \resizebox{0.98\hsize}{!}{\includegraphics[angle=270]{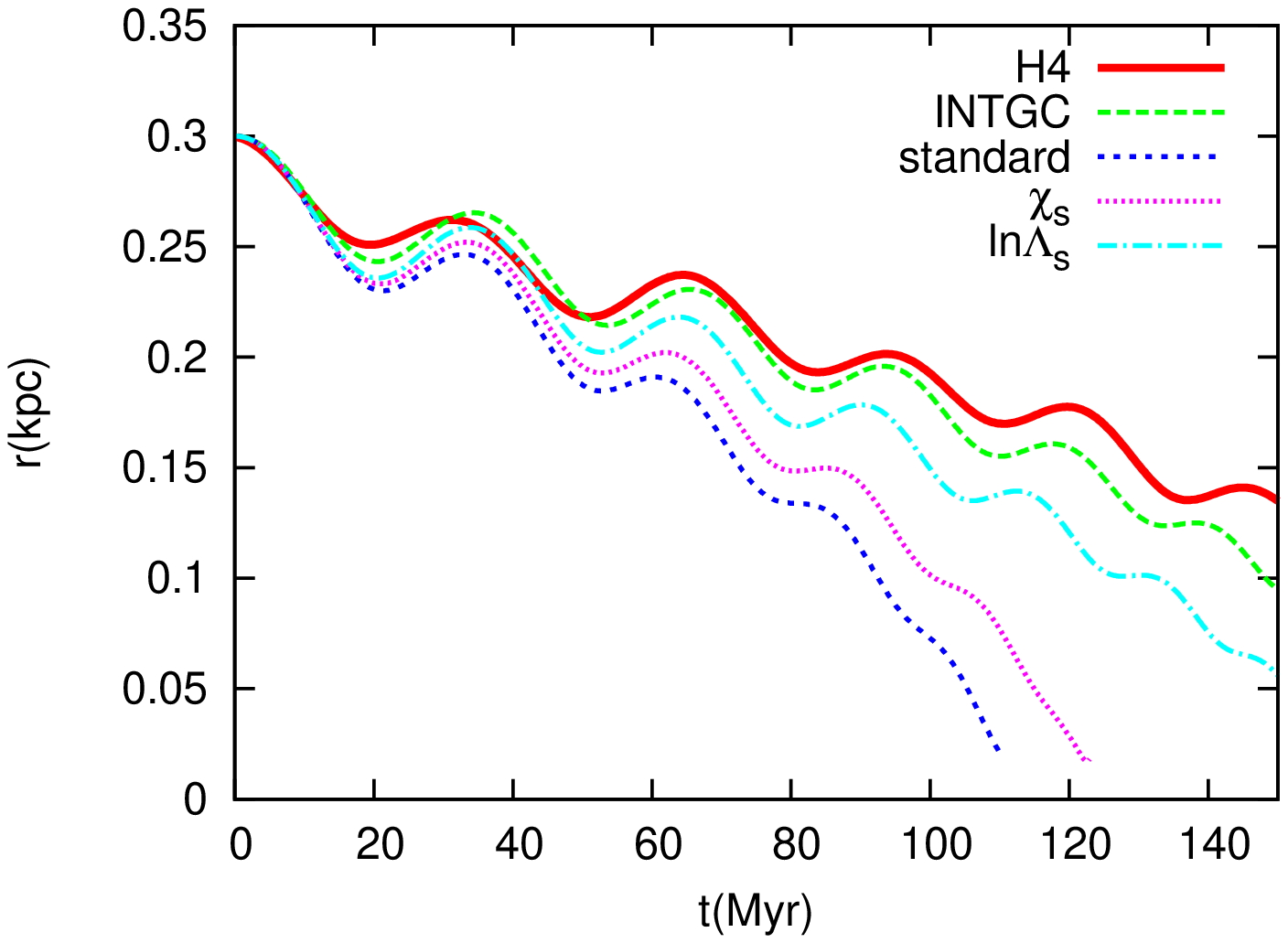}}
  }
\centerline{
  \resizebox{0.98\hsize}{!}{\includegraphics[angle=270]{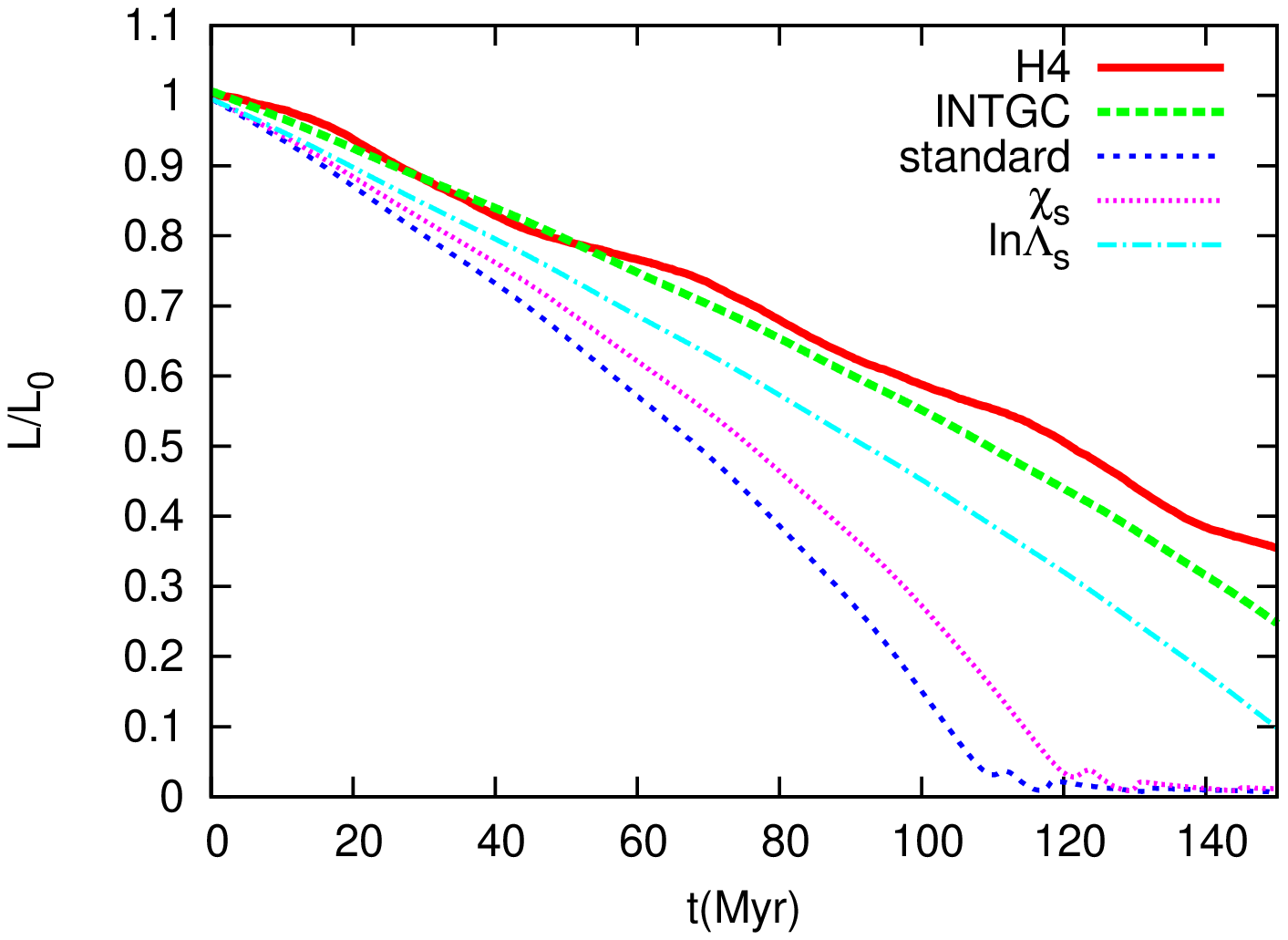}}
  }
\caption[]{
Same as in Fig. \ref{H3s} for the eccentric orbit H4.
} \label{H4s}
\end{figure}

The Hernquist model with shallower cusp ($\eta=2$) is even more complicated, because $X_{\rm c}^2$ is not constant and the $\chi$ function depends on the outer boundary conditions. For the two runs H1 (circular) and H2 (eccentric) the parameter range is similar to the Dehnen-1.5 case with marginally resolved $a_{90}$. The orbits of  H3 (circular) and H4 (eccentric) are further in and the grid resolution is higher leading to a fully resolved $a_{90}$. All orbits are reproduced reasonably well by  {\sc intgc} using the correct $\chi$ function and taking the correct velocity dispersion in $a_{90}$ \citep{tre94} into account (see figures \ref{Hcs} -- \ref{H4s}).

\subsection{Velocity distribution functions}

\begin{figure}
\centerline{
  \resizebox{0.98\hsize}{!}{\includegraphics[angle=270]{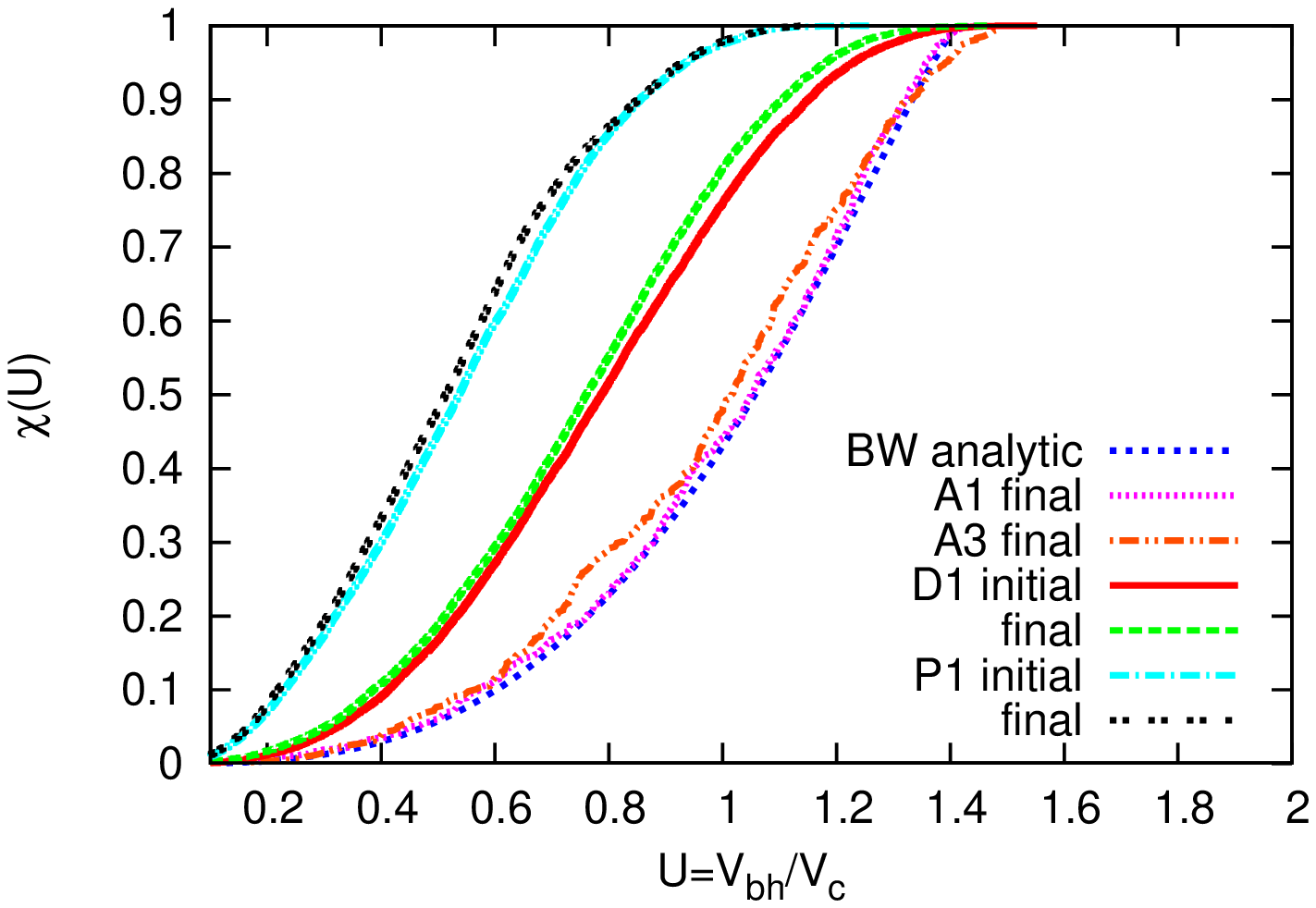}}
  }
\centerline{
  \resizebox{0.98\hsize}{!}{\includegraphics[angle=270]{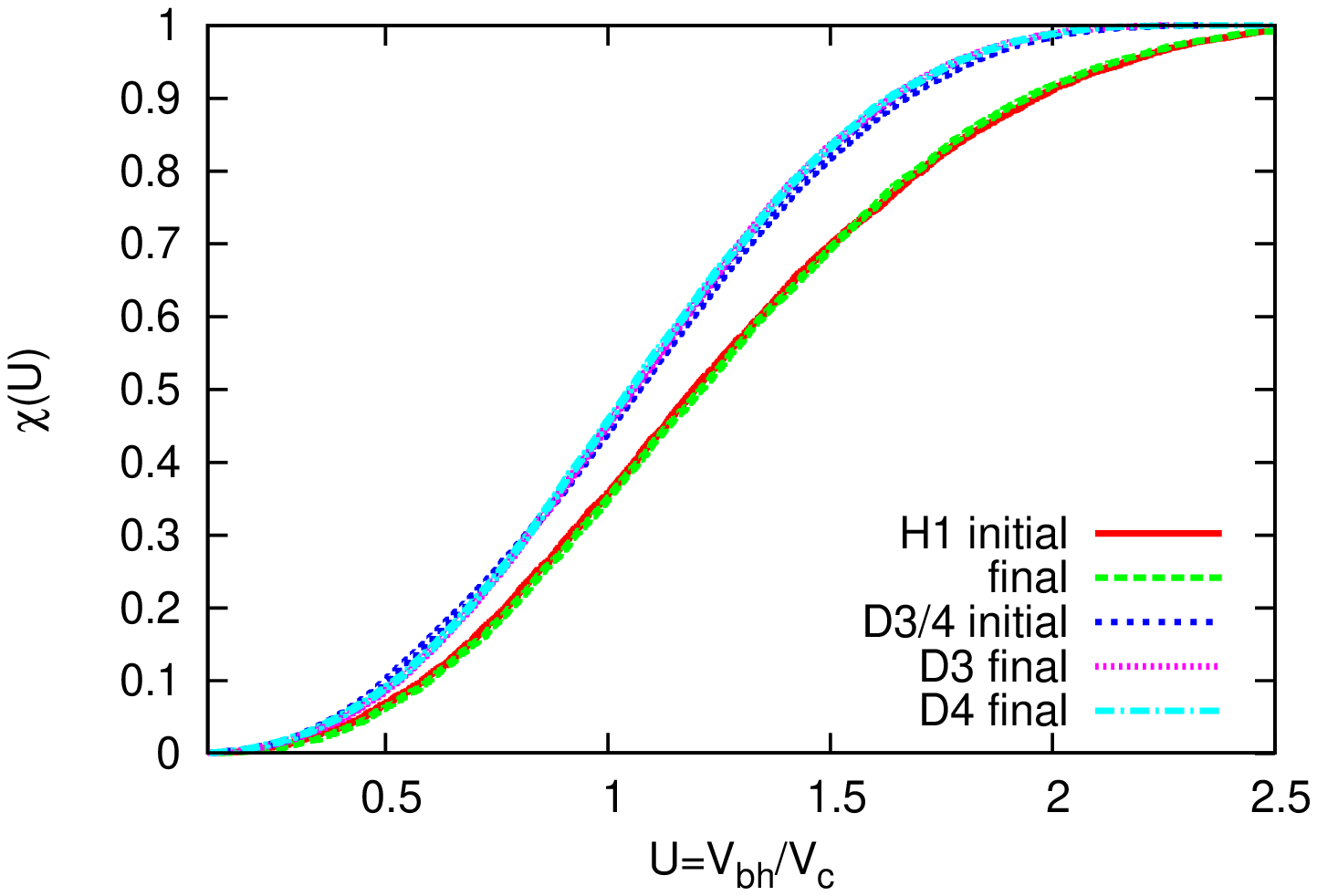}}
  }
\caption[]{
Top panel: Initial and final $\chi$ functions for the circular runs in the Bahcall-Wolf cusp (BW, A1, A3), the Dehnen case D1, and the Plummer case P1.
Bottom panel: Same for the self-gravitating cusps with Dehnen-1.5 (D3, D4) and Hernquist profiles (H1).
} \label{chi-kep}
\end{figure}

The local velocity distribution functions are crucial for the dynamical friction force. Therefore we check here the numerical realisation of the distribution functions. The best way to measure the distribution function is to determine the $\chi$ function in a spherical shell at the distance of the BH for different times. In figure \ref{chi-kep} the $\chi$ functions are shown for a few  examples. The top panel shows the Kepler cases and the bottom panel the self-gravitating cusps.

In the Kepler case we present the circular runs in the Bahcall-Wolf cusp (BW, A1, A3), the Plummer case P1, and the Dehnen case D1. In the Bahcall-Wolf cusp the final distribution function of A1 is very close to the theoretical line. The $\chi$ functions of the eccentric runs B1, B2 are very similar and therefore not plotted here. In the runs A2 and A3, where the density profile flattens slightly, a bump in the $\chi$ function at velocities below the circular velocity can be observed. It is more pronounced in run A3. The $\chi$ value at the circular speed is not influenced dramatically showing that the orbital delay is caused entirely by the reduced local density. The distribution functions in the outskirts of the Plummer and Dehnen cases are very stable and well represented by the theoretical expectations. 

The lower panel of figure \ref{chi-kep} shows the $\chi$ functions of the self-gravitating cusps for the circular runs H1 and D3. Here we added the case of the eccentric run D4 to demonstrate the stability of the velocity distribution functions independent of the shape of the orbit.

Overall the velocity distribution functions are very robust and well reproduced in the numerical simulations.

\section{Applications}\label{sec-appl}

\subsection{The Galactic Centre}\label{sub-galcen}

The central region of the Galactic Centre can  be modelled by a cusp with $\eta=1.2$ and enclosed mass $M_0=M(R_0)=1\cdot 10^9 \msol$ at $R_0=200\,\pc$, and a central BH with mass $M_{\rm c}=2.6\cdot 10^6 \msol$  \citep{gen87}. The gravitational influence radius is at $R=1.4\,\pc$, where we may assume a slightly shallower cusp with $\eta=1.25$. If the central BH would have entered the cusp on a circular orbit at some later time, the decay time from $R_0=200\,\pc$ to the centre would be 110\,Myr (Eq. \ref{taudec} with $\beta=\eta$). An intermediate mass BH with $M_{\rm bh}=1\cdot 10^4 \msol$ can reach the centre from $R_0=60\,\pc$ in $2.5$\,Gyr. For the final decay inside the influence radius of $R=1.4\,\pc$ the decay time is $15\,\Myr$.

In the central cusp of the Galaxy there are young star clusters like the Arches and the Quintuplet cluster with a projected distance from the Galactic centre of about 30 pc and an age of a few $\Myr$. The stellar mass  is $\approx 1\cdot 10^4 \msol$ with a half-mass radius of $r_{\rm h}\approx 0.2\,\pc$ \citep{cot96,fig99}. Assuming that the initial distance is $R_0=30\pc$ we find with Eq. \ref{lam0} for the initial Coulomb logarithm $\ln\Lambda_0=4.4$ leading to the decay time $\tau_{\rm dec}=720\,\Myr$. If we assume that the star clusters are still embedded in their parent molecular cloud with mass $1\cdot 10^6 \msol$ and initial half-mass radius $r_{\rm h0}\approx 3\,\pc$, the decay  time becomes considerably smaller. Adopting cluster mass loss linear in time we get $\tau_{\rm dec}\approx 30\,\Myr$, which is still large compared to the actual age of the clusters. This is in contrast to \citet{ger01}, who investigated the infall of massive clusters, also embedded in giant molecular clouds, to the Galactic centre. He found much shorter time-scales, because he used unrealistically high values for the Coulomb logarithm $\ln\Lambda\approx10-20$.

\subsection{Minor merger}\label{sub-merger}

One important application is the orbital decay of the SMBHs after the merger process of two galaxies. Lets assume a 10:1 merger with primary BH of mass $M_{\rm c}=1\cdot 10^8 \msol$ and the secondary BH with $M_{\rm bh}=1\cdot 10^7 \msol$. After violent relaxation of the stellar components and settling of the primary BH to the centre we adopt a shallow new central cusp with $\eta=1.75$ at radii large compared to the gravitational influence radius of the central SMBH. With an enclosed mass $M_0=M(R_0)=1\cdot 10^{10} \msol$ at $R_0=1\,\kpc$ the circular velocity is $V_{\rm c0}=207\,\kms$ corresponding to a velocity dispersion of $\sigma_0=293\,\kms$ (eq. \ref{Xc2}). For the secondary BH we find with Eq. \ref{taudec} a decay time of $830\,\Myr$. For the inner $500\,\pc$ with enclosed mass of $M(500\pc)=3\cdot 10^9 \msol$ the BH needs $190\,\Myr$. After reaching the gravitational influence radius of the central SMBH at $R=72\,\pc$, the final decay takes $15\,\Myr$ (with Eq. \ref{taukep1}). If we compare these decay times with an isothermal model with the same enclosed mass at $R_0=1\,\kpc$ and adopting a constant Coulomb logarithm of $\ln\Lambda=6.9$, we find the corresponding times $792;\,200;\,4.1\,\Myr$, respectively. The total decay time is comparable, but in the final phase the decay is significantly slower in the Kepler potential compared to the standard isothermal approximation.

\section{Conclusions}\label{sec-concl}

Classical dynamical friction as derived by Chandrasekhar is still
often used for important astrophysical applications, such as
infalling small galaxies in dark matter halos or globular cluster
systems around galaxies. It is also important to understand the co-evolution
of super-massive black holes in galactic nuclei (after mergers) with their
host galaxies. These recent new interests in dynamical friction
problems have caused several new studies of dynamical friction in
higher order than originally by Chandrasekhar (e.g. regarding the velocity
distribution) or in a different physical framework (collective modes
in stellar or gaseous systems rather than particle-particle interactions).

Chandrasekhar's dynamical friction formula has been proved in many ways,
and agrees with the study of collective modes, but with notable 
exceptions, one which is being discussed in the literature now is the
possible absence of dynamical friction in harmonic cores, relevant for
cosmological structure formation, avoiding too strong infall of small
galaxies. Still in many papers today 
the classical Chandrasekhar formula is applied by using
a local isothermal approximation for the kinematics of the background
system (i.e. the $\chi$ function) and by fitting the Coulomb
logarithm $\ln\Lambda$ for each orbit.

In this article we tested quantitatively the effect of using the
self-consistent velocity distribution function in $\chi$ and a general
analytic formula for $\ln\Lambda$ derived by \citet{jus05}. We
performed high-resolution numerical simulations, with
particle-mesh and direct $N$-body codes, of the orbital
evolution of a massive black hole in a variety of stellar
distributions. We investigated circular and eccentric orbits in
self-gravitating cusps (Dehnen models) 
and in cusps in a central Kepler potential (so-called Bahcall-Wolf cusps) and in the Kepler limit of steep power law density profiles in the outskirts of stellar systems. The background distributions cover a large range of  power law indices
between $-1 \dots -5$ of the density profile.

The application of the self-consistent $\chi$ functions lead to correction factors in the orbital decay times in the range between $0.5\dots 3$ (Fig. \ref{figchitot}).
The main new feature in the improved general
form of the position and velocity dependent Coulomb logarithm (Eq. \ref{loglam}) is the local scale length $D_{\rm r}$ of the density profile as maximum impact parameter. 
In most applications the effect of the new $\ln\Lambda$ is a significant delay in the orbital decay.
A detailed comparison with orbital decay using the standard values $\ln\Lambda_\mathrm{s}$ and $\chi_\mathrm{s}$ shows that the corrections are very different in the different cases leading to a general improvement of the orbit approximation. In a few cases like circular orbits in the Bahcall-Wolf cusp with a resolved minimum impact parameter the standard formula can still be used. But already for eccentric orbits a measurable difference occurs.
We like to point out the generality of the new formula such that a fit of individual orbits is no longer necessary.
We find a general agreement in the orbital decay of the numerical simulations with the analytic predictions at the 10\% level. This holds for circular as well as eccentric orbits and in all background distributions in self-gravitating and in Kepler potentials.

Another more technical finding concerns the best choice of the minimum impact parameter $b_{\rm min}$ measuring the numerical resolution. It
should be reminded, that from the structure of $\ln\Lambda$, there is
implicitly one common scaling factor for $b_{\rm max},b_{\rm
  min},a_{\rm 90}$ free. So the normalisation of one of these
quantities must be fixed in order to determine the other two. In
\citet{jus05} we argued for $a_{\rm 90}$ and $b_{\rm max}$. Here we
fixed $b_{\rm max}=D_{\rm r}$ and proved that $a_{\rm 90}$ is the
correct effective  minimum impact parameter in the numerically
resolved cases. On that basis we determined the numerical resolution
for the different codes in terms of the softening length $\epsilon$ or grid cell
size $d_{\rm c}$. We find $b_{\rm min}=1.5\epsilon$ for the PP code
consistent with the results of other authors and 
$b_{\rm min}=d_{\rm c}/2$ for the PM code {\sc Superbox}.

Our result is very important for any conclusions about the rates
of massive black hole binary mergers and black hole ejections in the
course of galaxy formation in the hierarchical structure formation
picture (cf. e.g. \citet{vol03,vol07,dot10}). It helps also to interpret results
of recent numerical investigations of the problem of multiple black holes
in dense nuclei of galaxies \citep{mil01,hem02,mil03,ber05,ber06,ber09,ama10}.
We did not investigate in detail the feedback of the decaying BH on the background distribution. A general flattening of the central cusp is expected for mass ratios of order unity \citep{nak99a,nak99b,mer06}. The investigation of dynamical friction in shallow cusps is much more complicated, because the maximum impact parameter is not well defined and the outer boundary conditions influence the velocity distribution function deep into the cusp.

The new formula can be used for extensive parameter studies of orbital
decay, where a high numerical resolution is not possible to resolve
the dynamical friction force. Even if we did not test the formula for
extended objects like satellite galaxies or star clusters in this
article, it should work equally well in these cases. 
Since $\ln\Lambda$ is generally smaller in these applications, the correction due to the maximum impact parameter would be more significant.
Additional effects like mass loss of the satellite galaxies must be taken into account \citep{fuj06,fuj08}. 
The new Coulomb
logarithm works also for non-isotropic background distributions, as was
shown already in \citet{pen04}.

\section*{Acknowledgements}

We thank Ingo Berentzen for his support in performing and analysing the numerical simulations.

This research and the computer hardware used in Heidelberg
were supported by project ``GRACE'' I/80 041-043 of the
Volks\-wagen Foundation, by the Ministry of Science, Research
and the Arts of Baden-W\"urttemberg (Az: 823.219-439/30
and 823.219-439/36), and in part by the
German Science Foundation (DFG) under SFB 439 (sub-project
B11) "Galaxies in the Young Universe" at the University of
Heidelberg.

Part of the simulations were performed on the GPU
super-computer at NAOC funded by the "Silk Road Project"
of the Chinese Academy of Sciences.

FK is supported by a grant of the Higher Education Commission (HEC) of 
Pakistan administrated by the Deutscher Akademischer Austauschdienst (DAAD).

PB thanks for the special support of his work by the Ukrainian
National Academy of Sciences under the Main Astronomical
Observatory ``GRAPE/GRID'' computing cluster project. Computers
used in this project were linked by a special memorandum of
understanding between Astrogrid-D (German Astronomy
Community Grid, part of D-Grid) and the astronomical segment of
Ukrainian Academic GRID Network.

PB studies are also partially supported by a program
Cosmomicrophysics
of NAS Ukraine.

\appendix

\section[]{Distribution functions}\label{appfe}

The distribution function of a power law cusp (eqs. \ref{massy} and \ref{deny}) is given by
\bqn
f(E)&=&\left\{ \begin{array}{lll} 
        K |E|^p & \eta<2.5 & \mbox{Kepler}\\
        K |E|^p & 0\le\eta< 3 & \mbox{self-grav.}\\
        K \exp(-E/\sigma^2) & \eta= 1 & \mbox{self-grav.} \end{array}\right.\label{fE}
\eqn
with normalisation constant $K$ and energy
$E=\Phi+v^2/2$, where the zero point of the potential is at the centre for self-gravitating cusps with $\eta>1$ and at infinity for the Kepler case and self-grav. cusps with $\eta<1$. 
The dependence of the power law index $p$ on $\eta$ is different for the Kepler and the self-gravitating case (see below).
We consider the two cases where
$\Phi$ is given by the self-gravitating potential of the cusp or the Kepler case with  $\Phi$ dominated by the central mass $M_c$.
The natural normalisation of the velocity is $\sqrt{|2\Phi|}$, which corresponds to the escape velocity for vanishing potential at infinity. For practical use it is more comfortable to
normalise the velocities to the circular velocity ($u$ from Eq. \ref{eq-u}).
We introduce the normalised 1-dimensional distribution function $F(u)$ and the cumulative function $\chi(U)$ by
\bqn
4\pi v^2f(E)\dd v &=& \rho F(u)\dd u\\
        \chi(U)&=&\int_0^U F(u)\dd u\,.
\eqn
which can be written in the
form
\bqn
F(u)&=& \left\{ \begin{array}{lll} 
        K' u^2\left(1+\xi u^2\right)^p & \eta<2.5 & \mbox{Kepler}\\
        K' u^2\left(1+\xi u^2\right)^p & 0<\eta<3 & \mbox{self-grav.}\\
        K' u^2 \exp\left(-u^2\right)& \eta= 1 & \mbox{self-grav.}
        \end{array}\right.\label{eq-FU}\\
	&& \mbox{with}\quad \xi=\frac{V_c^2}{2\Phi}\nonumber
\eqn
Since $K'$ and $\xi$ are constant, $F(u)$ and $\chi(U)$ are independent of position $y$.

The velocity dispersion can be obtained by integrating the Jeans equation involving the second moment of the velocity distribution function \citep{bin87}
\bq
\sigma^2 (y) = \frac{-G}{\rho(y)R_{0}}
	\left[\int \frac{\rho(y') M(y')}{y'^2}\dd y' + C''\right] \label{je}
\eq
where the integration constant $C''$ depends on the inner and outer boundary conditions.

\subsection{Self-gravitating cusps}\label{sub-selfdf}

\begin{figure}
\centerline{
  \resizebox{0.98\hsize}{!}{\includegraphics[angle=270]{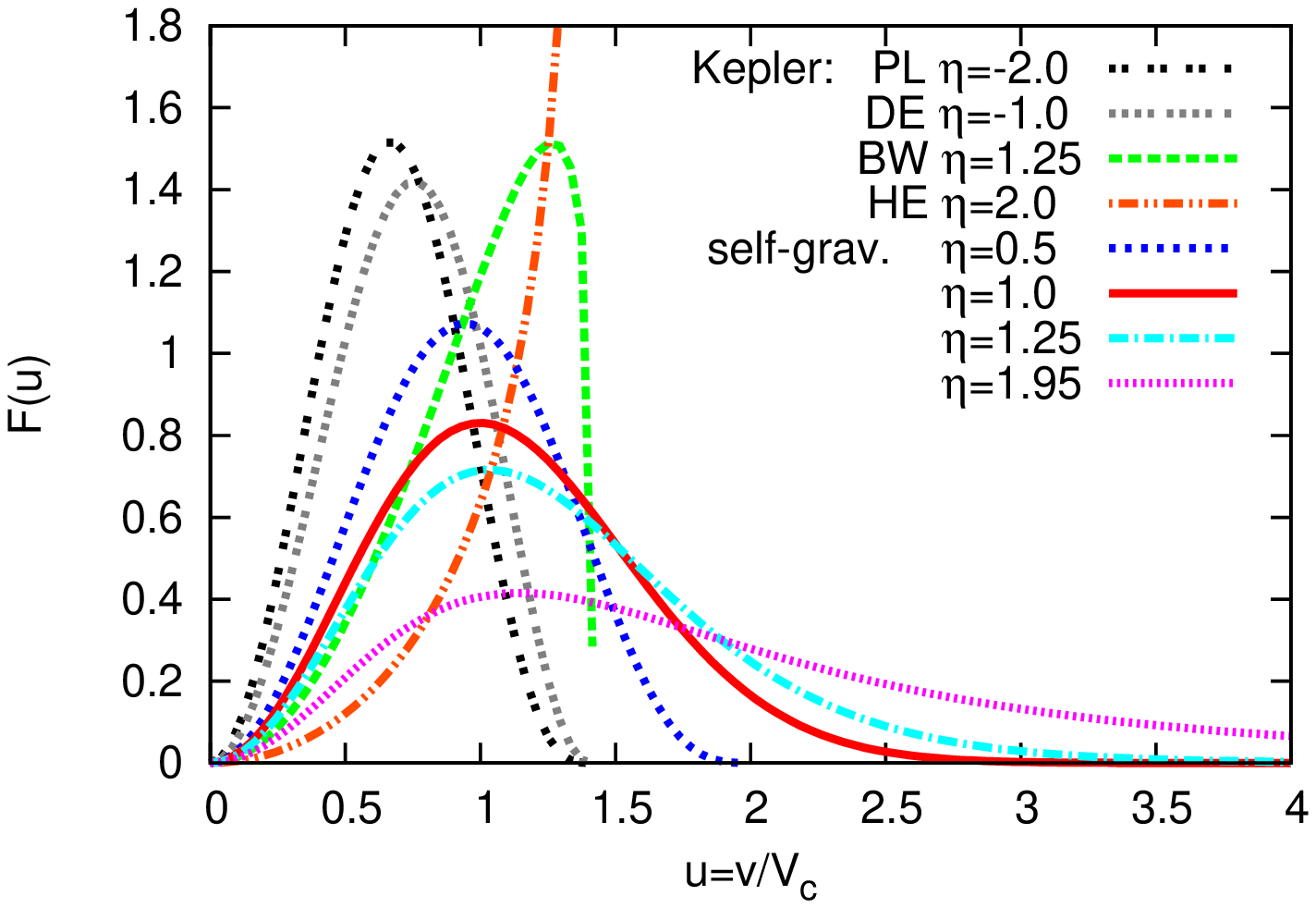}}
  }
\caption[]{
Normalised 1-dim distribution function $F(u)$ for
 different cases as a function of normalised velocity $u=v/V_{\rm c}$
  (see Eq. \ref{eq-FU}). The sequence with 
 $\eta=0.5,1.0,1.25,1.95$ shows a decreasing maximum.
  The full line is the Maxwellian ($\eta=1$). The Kepler potential cases are labelled by PL for Plummer, DE for Dehnen outskirts, BW for the Bahcall-Wolf cusp, HE for Hernquist cusp.
}\label{figfe}
\end{figure}

The constants in Eqs. \ref{fE} and \ref{eq-FU} are given by
\bq
\xi=\frac{\eta-1}{2}\qquad ;\qquad  p=\frac{3+\eta}{2(1-\eta)}
\eq
and
\bqn
K&=&\left\{ \begin{array}{ll}
	\frac{\rho_0}{4\pi\sqrt{2}B(\frac{3}{2},1+p)}
        \left(\frac{V_{\rm c0}^2}{\onema}\right)^{-p-\frac{3}{2}}
	& \eta<1 \\ \\
	\frac{\rho_0}{\pi\sqrt{\pi}}V_{\rm c0}^{-3} & \eta=1\\ \\
	\frac{\rho_0}{4\pi\sqrt{2}B(\frac{3}{2},-p-\frac{3}{2})}
        \left(\frac{V_{\rm c0}^2}{\amone}\right)^{-p-\frac{3}{2}}
	& 1<\eta<3 
        \end{array}\right. \\
K'&=&\left\{ \begin{array}{ll}
	\frac{2\left|\xi\right|^{3/2}}{B\left(\frac{3}{2},1+p\right)}
	& \eta<1 \\ \\
	\frac{4}{\sqrt{\pi}} & \eta=1\\ \\
\frac{2\left|\xi\right|^{3/2}}{B\left(\frac{3}{2},-p-\frac{3}{2}\right)} 
	& 1<\eta<3 
        \end{array}\right.
\eqn
 Here we used the
Beta function $B(x,y)=\Gamma(x)\Gamma(y)/\Gamma(x+y)$ (see \citet{gra80} (8.38)). In Fig. \ref{figfe} the distribution functions $F(u)$ are shown for different
 values of
$\eta=0.5,1.0,1.25,1.95$ (with decreasing maximum).
The thick line is the standard Maxwellian. For $\eta<1$ the energy range is
 finite with $u^2<|\xi|^{-1}$, whereas for
 $\eta\ge 1$ the potential is infinitely deep allowing for all velocities. 
\begin{figure}
\centerline{
  \resizebox{0.98\hsize}{!}{\includegraphics[angle=270]{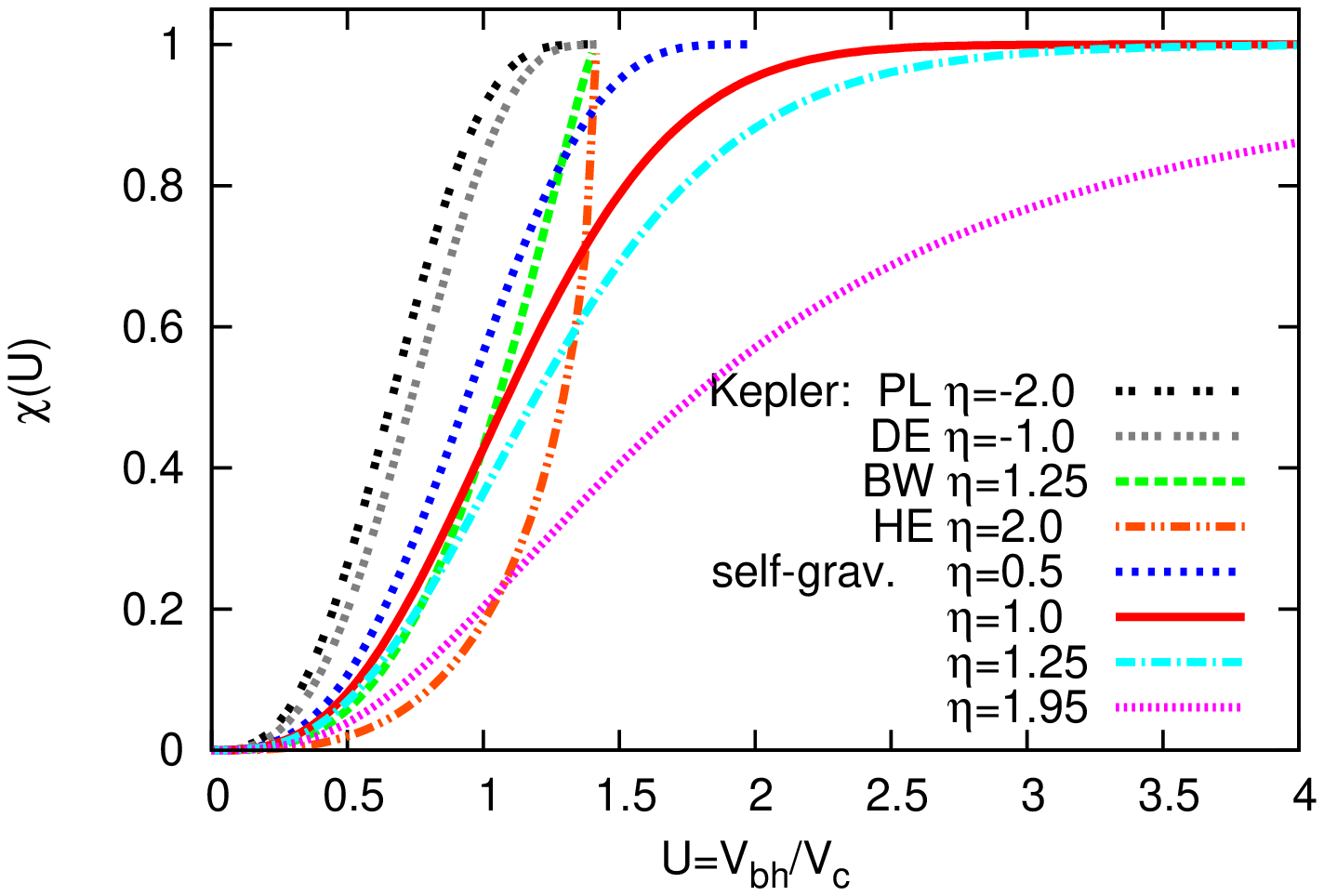}}
  }
\caption[]{
$\chi(U)$ for the same values of $\eta$ as in Fig. \ref{figfe}.
}\label{figchiU}
\end{figure}
\begin{figure}
\centerline{
  \resizebox{0.98\hsize}{!}{\includegraphics[angle=270]{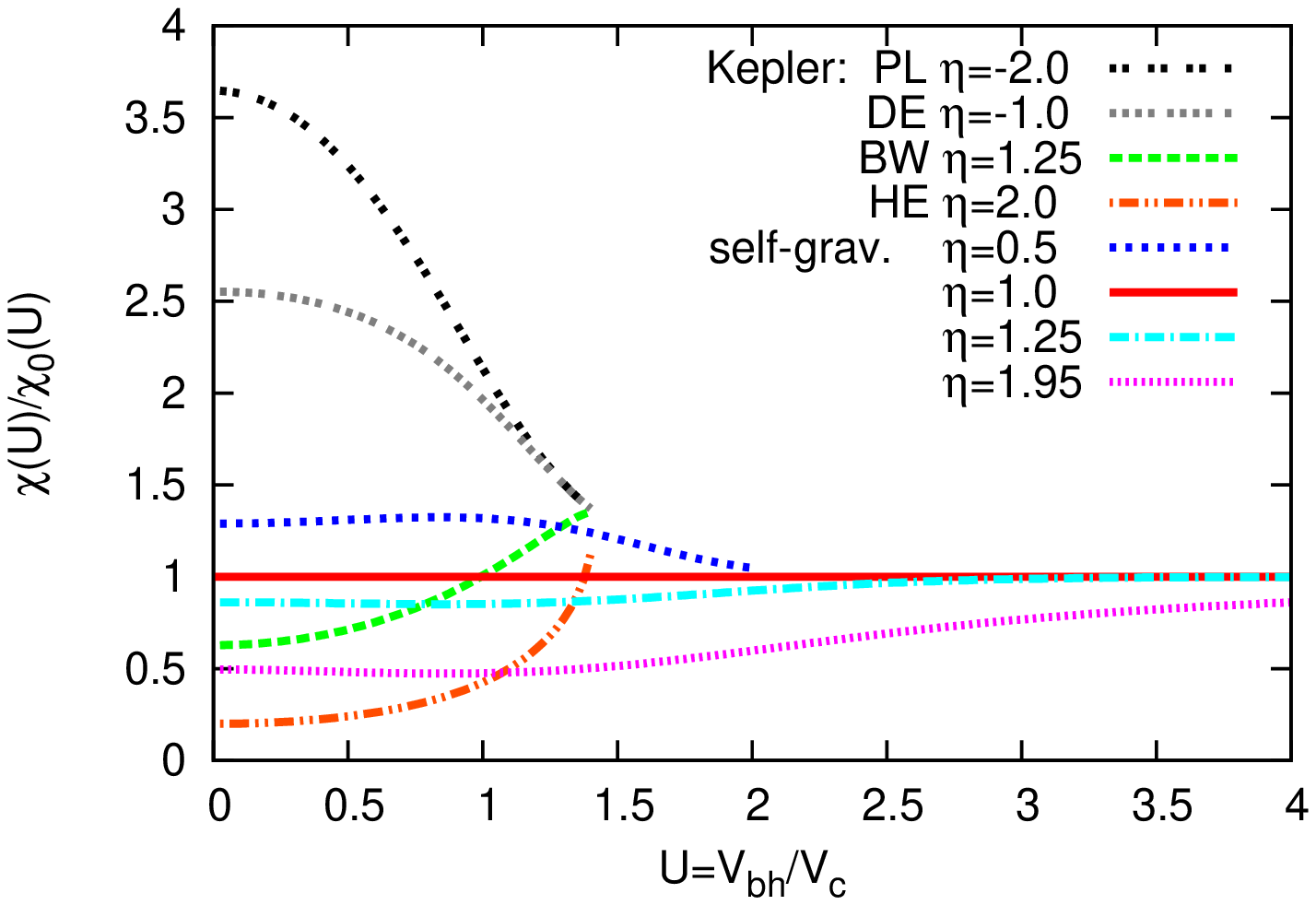}}
  }
\caption[]{
Same functions as in Fig. \ref{figchiU}, but here normalised to the Maxwellian:
$\chi(U)/\chi_{\rm s}(U)$ (increasing $\eta$ from top to bottom).
}\label{figchiR}
\end{figure}
\begin{figure}
\centerline{
  \resizebox{0.98\hsize}{!}{\includegraphics[angle=270]{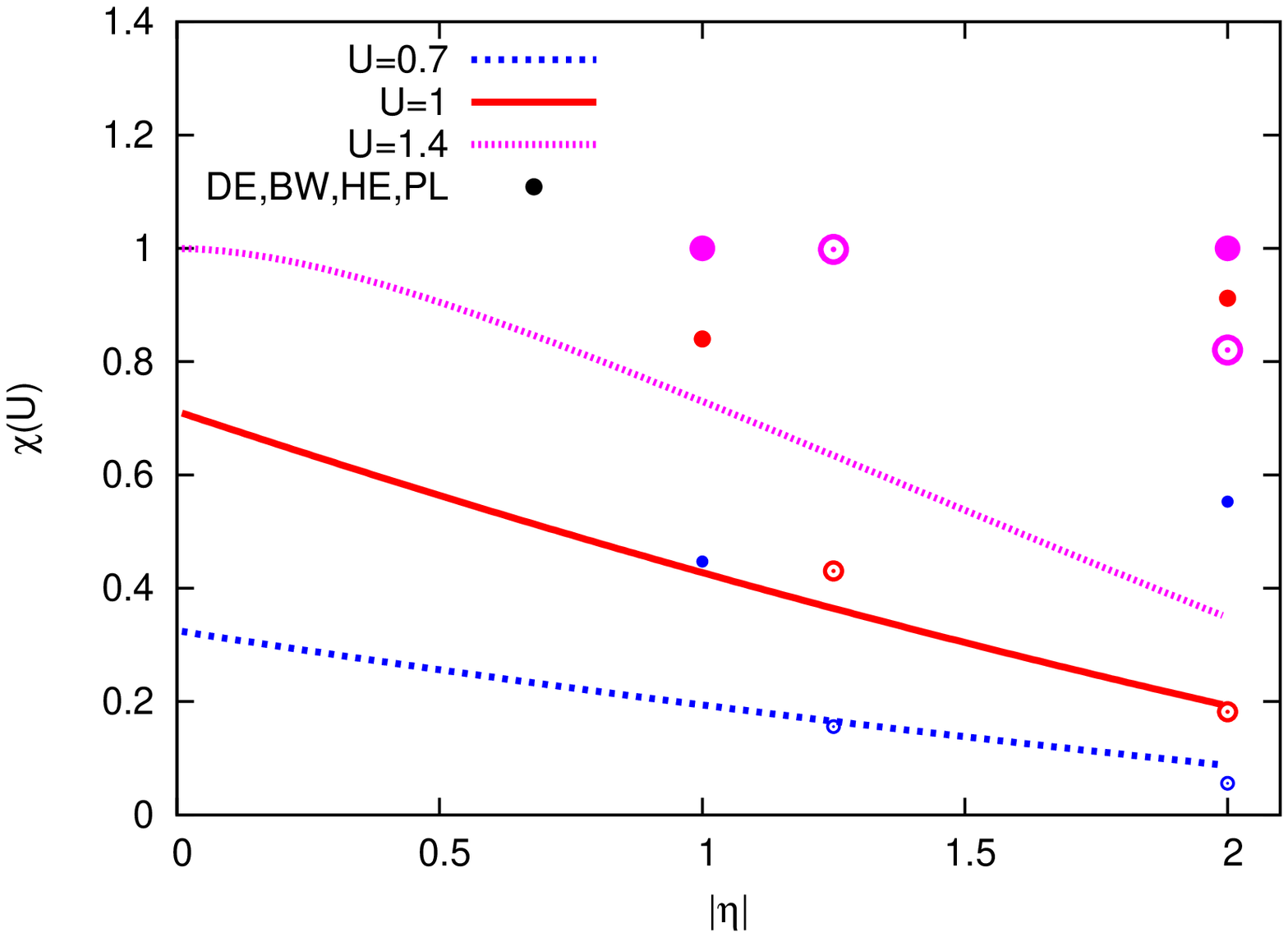}}
  }
\caption[]{
The functions
$\chi(U)$ as a function of  $\eta$ are shown for fixed $U=0.7,1.0,1.4$ (Dashed blue, full red,
dotted pink line, respectively). The circles give the corresponding values for the Kepler potential cases with increasing size for increasing $U$ ($\eta=-1$ for DE, $\eta=-2$ for PL, $\eta=1.25$ for BW, $\eta=2.0$ for HE). Open circles are for positive $\eta$ and full circles for negative $\eta$.
}\label{figchial}
\end{figure}
\begin{figure}
\centerline{
  \resizebox{0.98\hsize}{!}{\includegraphics[angle=270]{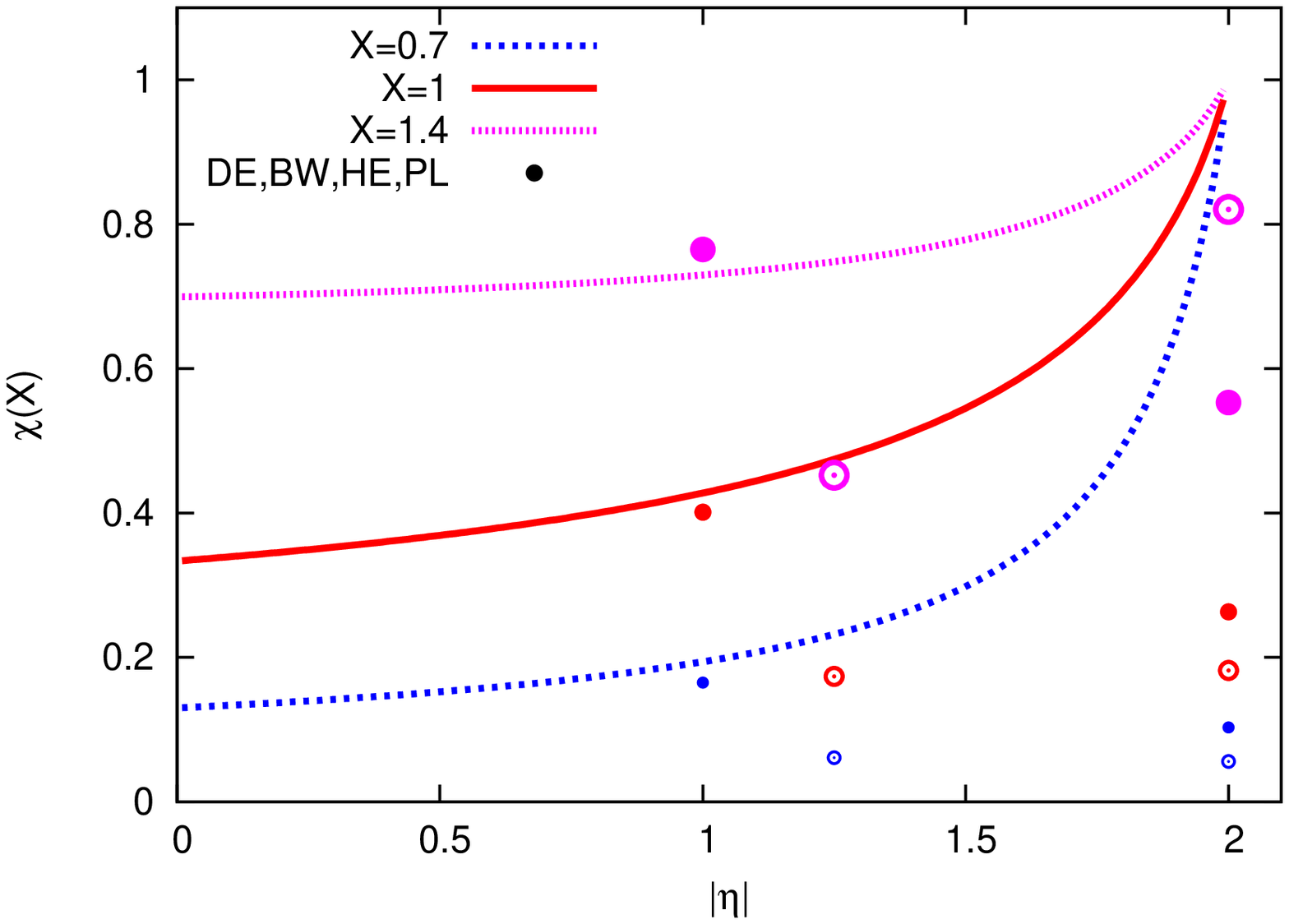}}
  }
\caption[]{
Same as in Fig. \ref{figchial} but for fixed $X=0.7,1.0,1.4$.
}\label{figchial2}
\end{figure}

For the dynamical friction force we need the cumulative function $\chi(U)$, which gives the
fraction of stars with velocity smaller than $U$. In Fig. \ref{figchiU} $\chi(U)$ is shown
for the same set of $\eta$-values as in Fig. \ref{figfe}. In order to see more clearly the
difference to the Maxwellian distribution function the ratios
$\chi(U)/\chi_{\rm s}(U)$ are plotted in Fig. \ref{figchiR}. 
For $\eta<1$ the $\chi$-values are larger and for 
shallower cusps with $\eta>1$ the values are smaller.
In Fig. \ref{figchial} we show $\chi(U)$ as a function of $\eta$ for fixed $U$. $U=1$
corresponds to the circular velocity, and $U=0.7,1.4$ are typical values for peri- and
apo-centre velocities with moderate eccentricities. Fig. \ref{figchial2} gives the same but
 for fixed $X=0.7,1.0,1.4$
showing the systematic difference due to the different $\eta$-dependence of $X$.

The velocity dispersion is needed for $X_\mathrm{c}$ (eq. \ref{Xc2}) used in $a_{90}$ (eq. \ref{a90}).  Substituting $M(y)$ and $\rho(y)$ from eq. \ref{massy} and eq. \ref{deny} yields
\bqn
\sigma^2 (y)&=&\left\{ \begin{array}{ll}
	 \frac{GM_{0}}{2R_{0}(2-\eta)}y^{\eta-1} 
	= \frac{V_{\rm c}^2}{2(2-\eta)} & 0<\eta < 2 \\ \\
	\frac{GM_{0}}{R_{0}}y\left(-\ln y + C \right) & \eta = 2 \\ \\
	 C'''y^{3-\eta} & 2<\eta <3 
        \end{array}\right.  \label{sig2}
\eqn
where the constants $C$ and $C'''$ depend on the outer boundary conditions.

\subsection{Kepler potential}\label{sub-keplerdf}

In the Kepler potential of a central SMBH 
(without the mean field contribution of the
 stellar cusp) we find
for the constants 
\bq
\xi=-\frac{1}{2}\qquad ;\qquad
        p=\frac{3}{2}\!-\!\eta\,. \label{constKep1}
\eq
and
\bqn
K&=&\frac{\rho_0}{4\pi\sqrt{2}B(3/2,1+p)}
        \left(V_{\rm c0}^2\right)^{-p-3/2}\\
K'&=&\frac{2\left|\xi\right|^{3/2}}{B\left(3/2,1+p\right)}\label{constKep2}
\eqn
In a pioneering work \citet{pee72} analysed the structure of a stellar cusp in a Kepler
potential of a central SMBH, which is stationary for times large compared to the relaxation time. Unfortunately he derived incorrect values for
$p$ and thus $\eta$. 
The correct derivation can be found in \citet{lig76} using scaling arguments and
in \citet{bah76} using a Fokker-Planck analysis. 
The resulting so-called Bahcall-Wolf cusp (BW) is given by $p=1/4$
leading to $\eta=5/4$ and the well-known density profile 
$\rho\propto y^{-7/4}$. In 
Figs. \ref{figfe} - \ref{figchial2} the corresponding functions or values are shown 
(dashed lines or full circles, respectively). 
In a Hernquist cusp (HE) with $\eta=2$ we have a shallow density profile $\rho\propto y^{-1}$ and find $p=-1/2$ leading to a diverging 1-D distribution function at the escape velocity.
The outskirts of Dehnen (DE) and Plummer (PL) distributions with densities $\rho\propto y^{-4}$ and $\rho\propto y^{-5}$ correspond to $\eta=-1$ and $\eta=-2$ with $p=5/2$ and $p=7/2$, respectively. The corresponding distribution functions and $\chi$ values are also plotted in Figs. \ref{figfe} -- \ref{figchial2}.

\section[]{Orbital decay}\label{apporb}

Here we derive explicitly the orbital decay of a massive object on a circular
orbit in a cuspy density distribution
with a position dependent Coulomb logarithm.  All values are normalised to the
values at the initial distance $R_0$. The distance to the centre is $y=R/R_0$.
The orbital decay can be computed by identifying the angular momentum loss due to the dynamical friction force (eq. \ref{dynfric})
\bq
\frac{\dd (y\,V_{\rm c})}{\dd y}\dot{y} =\dot{L} = y\, \dot{V}_{df} 
=-\frac{V_{c0}}{\tau_0}y^{\eta-1}\frac{M_0}{M(y)}
\frac{\ln\left(\Lambda_0 y^\beta\right)}{\ln\Lambda_0}
\label{ldot}
\eq
where we have used the definition of the decay timescale $\tau_0$ (eq. \ref{tau0}), the position dependence of the density (eq. \ref{deny}), replaced the square of the circular velocity by $GM(y)/R_0y$ with initial value $GM_0/R_0$, and the parametrisation of the Coulomb logarithm (eq. \ref{pdcl}). The enclosed mass $M(y)$ on the right hand side and the circular velocity $V_{\rm c}(y)$ on the left hand side are different for the Kepler case and the self-gravitating case.

\subsection{Kepler potential}\label{sub-keplerorbit}

In the Kepler potential with constant enclosed mass $M(y)=M_0=M_{\rm c}$ and
\bq
{V_{c}^2} = {V_{c0}^2}y^{-1} \label{vck}
\eq
we write eq. \ref{ldot} in the form
\bq
\dot{y} =-\frac{2}{\tau_0}
\frac{\ln\left(\Lambda_0 y^\beta\right)}{\ln\Lambda_0}\, y^{(\eta-\frac{1}{2})}
\label{34k}
\eq
We define for $\eta\neq 3/2$
\bq
z = \Lambda^{\frac{\kappa}{\beta}} = \Lambda_0^{\frac{\kappa}{\beta}} y^\kappa
, \qquad \kappa = \frac{3}{2} - \eta \label{z}
\eq
and find after a little mathematical manipulation
\bq
\frac{dz}{dt} = -\frac{2\kappa}{\tau_0}\frac{z_{0}}{\ln z_{0}}\ln z \label{36}
\eq
or
\bqn
t &=& -\frac{\tau_0}{2\kappa}\frac{\ln z_{0}}{z_{0}}\int \frac{1}{\ln z}  \dd z \nonumber\\
&=&\tau_{\rm df}\frac{\ln z_0}{z_0}\left[\Ei(\ln z_0)-\Ei(\ln z(y))\right] 
\qquad \beta,\kappa\ne 0\label{sol}
\eqn
with $\tau_{\rm df}$ from eq. \ref{taudf} and the exponential-integral function $\Ei(x)$ (see 2.2724.2 and 8.211.2 of \citet{gra80}).
For the special case of $\eta=3/2$ we can use $\Lambda$ as variable and find with
\bq
\frac{\dd}{\dd t}\ln(\ln\Lambda)=\frac{\beta}{\ln\Lambda}\frac{\dot y}{y}
\eq
the implicit solution
\bqn
t &=& \tau_{\rm df}\,\ln\left(\frac{\ln\Lambda_0}{\ln\Lambda(y)}\right)  
\qquad \beta\ne 0,\,\kappa=0\label{sol0}
\eqn
which can be inverted to the explicit solution given in eq. \ref{yt}.

For a point-like body the maximum and the minimum impact
parameter $D_{\rm r}$ and $a_{90}$ are both linear in $y$ leading to a constant Coulomb logarithm (i.e. $\beta=0$). For this case
direct integration of eq. \ref{34k} leads to the  explicit solutions given in eq. \ref{yt}.

For the Kepler potential the results can be easily generalised to expanding or contracting cusps. If the density varies proportional to a power of time, i.e. $\rho(y)=\rho(y,t=0)(1+t/t_{\rm a})^\nu$, then the differential equation \ref{34k} with time dependent $\tau_0$ can be converted back to the original form with initial $\rho_0$ in $\tau_0$ by
\bqn
\frac{\dd y}{\dd s}&=&\dot{y}\frac{\dd t}{\dd s} =-\frac{2}{\tau_0}
\frac{\ln\left(\Lambda_0 y^\beta\right)}{\ln\Lambda_0}\, y^{(\eta-\frac{1}{2})}\\
s&=&\left[(1+t/t_{\rm a})^{1+\nu}-1\right]\frac{t_{\rm a}}{1+\nu}
\label{34t}
\eqn
The implicit solution of the differential equation \ref{34k} is still given by Eq. \ref{sol} with the substitution $t\to s$. For a Plummer sphere with linear increasing Plummer radius $y_{\rm a}=y_{\rm a0}(1+t/t_{\rm a})$ we find for the outskirts $\nu=2$ (eq. \ref{mplumapp}).

\subsection{Self-gravitating cusps}\label{sub-selforbit}

In a  self-gravitating cusp 
we have $\beta=\eta,\,1,\,0$ for a point-like object, an extended
object, and a constant Coulomb logarithm,
respectively (Eq. \ref{pdcl}).
For the self-gravitating case we insert $M(y)=M_0y^{\eta}$ and
\bq
{V_{c}^2} = {V_{c0}^2}y^{\eta-1}
\eq
into equation \ref{ldot}. The resulting differential equation takes the form
\bq
\dot{y} =-\frac{2}{(1+\eta)\tau_0}
\frac{\ln\left(\Lambda_0 y^\beta\right)}{\ln\Lambda_0}\, y^{-\frac{1+\eta}{2}}
\label{34s}
\eq
With the definition
\bq
\kappa=\frac{3+\eta}{2} \label{kappas} \label{zself}
\eq
which is positive for all realistic cases, we find the same implicit solution for $\beta\ne 0$ and explicit solution for $\beta= 0$ as in the Kepler case but with a different $\tau_{\rm df}$ (eqs. \ref{ty1} and \ref{yt}). 

\bsp

\label{lastpage}

\end{document}